\documentclass[%
 reprint,
 amsmath,amssymb,
 aps,
]{revtex4-2}
\usepackage[utf8]{inputenc}%
\usepackage{multirow}
\usepackage{graphicx}
\usepackage{dcolumn}
\usepackage{bm}
\usepackage{color}

\begin{document}

\preprint{APS/123-QED}

\title{Layer-number and strain effects on the structural and electronic properties of PtSe$_2$ material}
\author{Rania Amairi}
 \email{rania.amairi@fsb.ucar.tn}
\affiliation {Faculté des Sciences de Bizerte, Laboratoire de Physique des Matériaux: Structure et Propriétés, Université de Carthage, 7021 Jarzouna, Tunisia }
\author{Adlen Smiri}%
 \email{smiriadlen4@gmail.com}
 \affiliation{Mathematics for Advanced Materials Open Innovation Laboratory (MathAM-OIL), National Institue of Advanced Industrial Science and Technology (AIST) c/o Advanced Institute for Materials Research, Tohoku University, 2-1-1, Katahira, Aoba-ku, Sendai, Miyagi 980-8577, Japan and Faculté des Sciences de Bizerte, Laboratoire de Physique des Matériaux: Structure et Propriétés, Université de Carthage, 7021 Jarzouna, Tunisia} 
\author{Sihem Jaziri}%
 \email{sihem.jaziri@fsb.rnu.tn}
\affiliation{Faculté des Sciences de Bizerte, Laboratoire de Physique des Matériaux: Structure et Propriétés, Université de Carthage, 7021 Jarzouna, Tunisia,\\
         and Faculté des Sciences de Tunis, Laboratoire de Physique de la Matiére Condensée and Département de Physique, Université Tunis el Manar, Campus Universitaire, 2092 Tunis, Tunisia }

\date{\today}

\begin{abstract}
\textbf{Abstract:} 
Bandgap engineering of low-dimensional materials forms a robust basis for advancements in optoelectronic technologies. Platinum diselenide (PtSe$_2$) material exhibits a transition from semi-metal to semiconductor (SM-SC) when going from bulk to monolayer (ML).  In this work, density functional theory (DFT) with various van der Waals (vdW) corrections has been tested to study the effect of the layer-number on the structural and electronic properties of the PtSe$_2$ material. The considered vdW corrections gave different results regarding the number of layers at which the SM-SC transition occurs. This variation is due to the different interlayer distances found for each correction, revealing the sensitivity of the bandgap to this distance in addition to the layer number. In fact, the bandgap increases with the increasing of the interlayer distance, due to the energy shift of conduction and valence bands dominated by Se-$p_z$ orbitals. According to the comparison with the available experimental data, the vdW corrections vdW-DF and rVV10 gave the most accurate results. Moreover, the control of the interlayer distance via vertical compressive strain led to the bandgap tuning of semiconductor PtSe$_2$ BL. Indeed, a semi-metal character of PtSe$_2$ BL can be obtained under 17\% vertical strain. Our work shows a deep understanding of the correlation between the structural and electronic properties, and thus a possibility to tune the bandgap by  strain means.


\end{abstract}

\maketitle


\section{\label{sec:level1}INTRODUCTION}

In recent years, two-dimensional (2D) materials have attracted great interest due to their exceptional properties, which differ significantly from bulk materials~\cite{choi2017recent}. 
The transition metal dichalcogenide (TMD) family represents an interesting and attractive group of 2D materials due to their rich electronic and optical properties~\cite{alfalasi2023designing,butler2013progress,wang2012electronics,mu2015graphene,liu2020mos,li2020recent,kang2019mos}.
The TMD monolayers (MLs) have the chemical formula MX$_2$ where M is a transition metal (Pt, Mo...) and X is a chalcogen (Se, S...).
In the last decade, the TMD of group 6 of the periodic table such as \textit{2H}-MoS$_2$, \textit{2H}-MoSe$_2$ and \textit{2H}-WSe$_2$ have attracted the attention of the condensed matter scientific community~\cite{mia2023two,shi2015recent}.  The important progress in the control of the electronic properties of  this group has stimulated the search for new two-dimensional (2D) materials. Group 10 TMD, such as  PtSe$_2$ and PdS$_2$, have been discovered as promising nanostructures for applications in electronics, optoelectronics, catalysis and sensors~\cite{pi2019recent,oyedele2017pdse2,gong2020two,setiyawati2019distinct,wagner2018highly,wang2021layered,gong2020two}. 
Among these noble TMDs, the \textit{1T} phase PtSe$_2$ material attracts considerable attention due to its unique properties~\cite{lei2023directional,hemmat2023layer,tharrault2023optical}. In particular, the \textit{1T}-PtSe$_2$ ML has high electronic mobility of 200 cm$^2$v$^{-1}$s$^{-1}$~\cite{zhao2017high}, that promotes efficient charge transport and paves the way for high-performance electronic devices such as transistors~\cite{almutairi2017ptse}, photodetectors~\cite{wu2018design}. The \textit{1T}-PtSe$_2$ material stands out from other TMD materials due to the unique correlation between its structural and electronic proprieties~\cite{li2021layer,zhang2021precise,island2016precise,kandemir2018structural}. Indeed, as the number of layers increases, the energy of the bandgap decreases~\cite{li2021layer,zhang2021precise,island2016precise,kandemir2018structural,yan2017high}. This peculiarity is attributed to the interlayer coupling and screening effects~\cite{ansari2019quantum}.  The semi-metallic nature of bulk \textit{1T}-PtSe$_2$ has been known for a long time~\cite{dai2003trends}. However, experimental studies  have shown that \textit{1T}-PtSe$_2$ ML exhibits semiconductor behavior~\cite{wang2015monolayer}. Additionally, a sharp decrease of the bandgap energy due to the increase of the number of layers was observed~\cite{yan2017high}.
However, the exact transition between the semi-metallic and semiconductor (SC-SM) behavior of \textit{1T}-PtSe$_2$ material as a function of the layer number has not been clearly determined. Neither numerical nor experimental studies have been able to provide a clear boundary where this transition occurs. Recently, in order to solve this problem, two experimental studies were conducted~\cite{li2021layer,zhang2021precise}. However, these two studies have shown two different results. In fact, $Zhang$ \textit{et} $al.$~\cite{zhang2021precise} showed that \textit{1T}-PtSe$_2$ material becomes a semi-metal from the fifth layer. Indeed, for a layer number that goes from one to four, the bandgap energy goes from 2.0$\pm$0.1, 1.1$\pm$0.1, 0.6$\pm$0.1 and 0.2$\pm$0.1 eV, respectively. However, $Li$ \textit{et} $al.$~\cite{li2021layer} proved that \textit{1T}-PtSe$_2$ material becomes a semi-metal from the third layer. In this case, the values of the bandgap energies are 1.8 eV for the  PtSe$_2$ ML and 0.6 eV for the Bilayer (BL). Furthermore, several\textit{ ab-initio }studies have been carried out and showed discrepancies in the results~\cite{ansari2019quantum,wang2015monolayer,zhao2017high,o2016raman,kandemir2018structural,zulfiqar2016tunable}. Therefore, to understand the effect of the layer number on the electronic proprieties of \textit{1T}-PtSe$_2$ material, a further insights on the origin of the SC-SM transition is needed.\\

For layered materials, especially TMDs, van der Waals (vdW) interactions play a crucial role, ensuring a stable connection between their layers and thereby facilitating the realisation of vdW heterostructures~\cite{ren2019electronic,selamneni2023mixed,khestanova2023robustness,thompson2023interlayer,zhen2023effect}. To incorporate these interactions into Density Functional Theory (DFT) calculations, various corrections have been proposed~\cite{berland2015van,lee2010higher,sabatini2013nonlocal,grimme2004accurate,moellmann2014dft}, typically grouped into two categories. The first category involves the vdW interaction by including non-local Coulomb interaction in the exchange-correlation functional, namely vdW-DF~\cite{berland2015van}, vdW-DF2~\cite{lee2010higher}, and rVV10~\cite{sabatini2013nonlocal} functionals. The second category encompasses semi-empirical corrections, where the vdW energy is including as a correction to  the total energy. For instance, DFT-D~\cite{grimme2004accurate} and DFT-D3~\cite{moellmann2014dft} are empirical methods. The performance of these corrections, in the two distinct categories, varies depending on the system and properties under investigation~\cite{peng2015scan+,chiter2016effect,freire2018comparison}. Indeed, some corrections prove more effective than others for specific types of systems or properties. For example, among various vdW corrections, the DFT+rVV10 method demonstrated the best agreement with experimental results of graphite material~\cite{sabatini2016phonons}. Therefore, in order to achieve more accurate and consistent results with experimental data, these various vdW corrections should be considered for \textit{1T}-PtSe$_2$ material.\\

Due to their crystal structure, few-layer TMDs emerge as prime candidates for strain engineering~\cite{dai2019strain,sortino2020dielectric,kansara2018effect,palepu2022comparative,wu2018giant,pandey2023straining,smiri2021optical,chaste2023straintronique,ippolito2021defect}. Indeed, this materials have the ability to undergo significant atomic displacements without nucleating defects~\cite{wu2018giant,pandey2023straining}.
Such materials can withstand mechanical strain up to 30\%, in contrast to silicon, which can only withstand about 1.5\%~\cite{pandey2023straining,pandey2022polymer,deng2018strain}. One of the strain engineering purpose is manipulating the electronic and optical properties of these materials~\cite{smiri2021optical,chaste2023straintronique,kansara2018effect,ippolito2021defect}. In fact, the control of bandgap energy paves the way for the development of new generations of optoelectronic devices endowed with desirable properties~\cite{chaste2023straintronique,kansara2018effect,ippolito2021defect,ccakirouglu2023automated}. For instance, the application of both uniaxial and biaxial strains to MoS$_2$ ML allows for the modifications of its electronic and optical properties~\cite{smiri2021optical,carrascoso2022biaxial,ren2022strain,lin2022regulation,wang2022first,le2020electronic,chen2023mobility}, such as the bandgap energy~\cite{wang2022first}, optical spectrum~\cite{yang2014lattice}, photoluminescence~\cite{le2020electronic}, and mobility~\cite{chen2023mobility}. These modifications are crucial for its integration into various technologies, including sensors, diodes, and field-effect transistors~\cite{carrascoso2022biaxial,kansara2018effect}. In order to broaden the applications of \textit{1T}-PtSe$_2$ material in nano-optoelectronic devices, both theoretical and experimental studies have been conducted to analyze the material response to strain, examining  ML~\cite{deng2018strain,guo2016biaxial,ge2023first,wang20232d,deng2018strain},  BL~\cite{li2016tuning,zhang2017mechanism,han2022deep} systems. However, for the  PtSe$_2$ BL under vertical compressive strain, to our knowledge while only one theoretical study was done~\cite{li2016tuning}, there are no existing experimental studies. In fact, the theoretical study conducted by $Li$ $et$ $al.$~\cite{li2016tuning} have shown a reducing in the bandgap energy and SC-SM transition at large vertical compressive strain of more than 40\%. This value is significantly larger than the in-plan strain that gave SC-SM transition in PtSe$_2$ ML~\cite{deng2018strain,hong2019tunable}, as well as for other TMDs~\cite{conley2013bandgap,armstrong2023directional,maniadaki2016strain,sharma2022tuning,su2021thermally}. Therefore, considering different approaches to further evaluate the vertical compressive strain and understand its impact on the PtSe$_2$ BL is of great interest for bandgap engeneerig.\\


In this article, we employ DFT plus vdW corrections to investigate the layer number and strain effects on the structural and electronic properties of PtSe$_2$ materials. To do this, five PtSe$_2$ systems, namely ML, BL, trilayer (TL), fourlayer (FL) and bulk were considered. In order to reveal the effect of layer number on the  structural parameters of PtSe$_2$ systems, We begin our work by performing a structural optimization using different vdW corrections. Then, a band structure calculations is conducted to understand the correlation between the electronic properties and the crystal structure including both layer number and crystal parameters. Finally, for a given vdW correction, the effect of a vertical compressive strain on the structural and electronic properties of PtSe$_2$ BL is studied. 

\section{\label{sec:level1} Computational details}
Our first principle calculations are based on the Density Functional Theory (DFT) using the Quantum Espresso packages~\cite{giannozzi2009quantum}. For the pseudopotential,  the projected augmented wave formalism (PAW) was considered~\cite{blochl1994projector},
The spin-orbit coupling (SOC) was consistently included in all Quantum Espresso calculations, using a fully relativistic pseudopotential. The structural optimization was performed using the Hellmann-Feynman Theorem~\cite{feynman1939forces}, where the convergence threshold for forces was set to $10^{-5}$ u.a. In this study, two structural optimization methods were used. For bulk PtSe$_2$ a fully structural optimization was considered while an atomic optimization was done for few layers systems. To avoid inter-layer interactions, a vacuum space larger than $19\AA$ was used.
The unit cells were visualized using the VESTA software~\cite{momma2011vesta}. The calculations was performed using the generalized gradient approximation (GGA) of the Perdew-Burke-Ernzerhof (PBE) functional for the treatment of the exchange-correlation interaction~\cite{perdew1996generalized}. However, GGA+PBE fails to describe the long-range coulomb interaction such as the inter-layer vdW interaction~\cite{giannozzi2017advanced,sabatini2016phonons}. To take into account these vdW interactions, two types of  vdW corrections in combination with the GGA exchange-correlation functional (GGA+PBE+vdW) were used.  In particular, the first type of corrections is based on empirical approaches as defined by \textit{Grimme}, namely DFT-D~\cite{grimme2004accurate} and DFT-D3~\cite{moellmann2014dft} methods. The second type is to consider non-local functional, such as vdW-DF~\cite{berland2015van}, vdW-DF2~\cite{lee2010higher}, and rVV10~\cite{sabatini2013nonlocal} methods. Each of these corrections has its own advantages in improving the accuracy of the calculations and getting closer results to the experimental results.
Convergence tests were performed on the various PtSe$_2$ systems for the kinetic energy cutoff and k-point sampling within the Brillouin zone. These tests were conducted for SCF calculation and structural optimization using vdW corrections. For the SCF calculation, the cutoff energy was set to $40$~Ry.  A uniform \textbf{k}-point sampling of $8\times$$8\times$$8$ was adopted for bulk PtSe$_2$, while a $12\times$$12\times$$1$ sampling was used for the few-layer structures, with a Gaussian broadening of $0.001$~Ry. For structural optimization, a higher energy cutoff of $60$~Ry  was employed with different corrections. \textbf{k}-point samplings of $4\times$$4\times$$4$ for the bulk and  $4\times$$4\times$$1$ for the few-layer structures were found to be sufficient for studying these systems.

 \section{ RESULTS AND DISCUSSION}
\subsection{Structural properties of PtSe$_2$ ML, BL, TL, FL and bulk}

\begin{figure}[h]
  \includegraphics[width=0.51\textwidth]{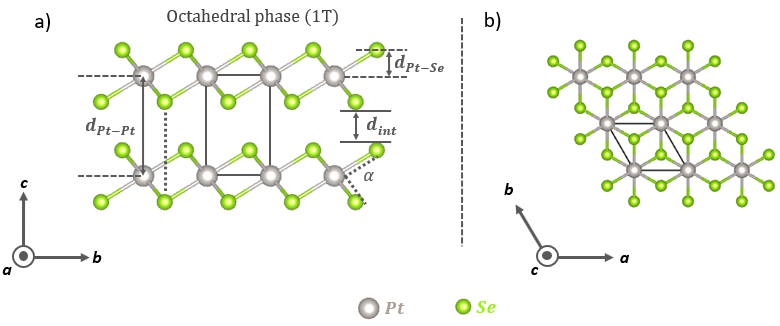}
   \caption{a) Side view of bulk \textit{1T}-PtSe$_2$. The interlayer distance is indicated by $d_{int}$ (distance between Se atoms of different layers), the interatomic distance between Pt and Se planes is shown as $d_{Pt-Se}$, and the distance between two Pt planes is indicated as $d_{Pt-Pt}$. b) Top view of the geometric structures of \textit{1T}-PtSe$_2$. }
  \label{fig1}
\end{figure}

The PtSe$_2$ material belongs to the centrosymmetric space group $D^3_{3d}$ ($P\bar{3}m1$) of the trigonal system~\cite{kandemir2018structural,zhang2017experimental,absor2020spin}. This material crystallizes in the characteristic \textit{1T} polytype, where the Pt and Se atoms form an octahedral coordination, as illustrated in the FIG.~\ref{fig1}. The primitive cell is defined by the lattice parameters \textit{a} (in-plane) and \textit{c} (out-of-plane), with primitive vectors defined as $\overrightarrow{a}$ = $\frac{1}{2}$\textit{a}($\sqrt{3}\overrightarrow{x}$-$\overrightarrow{y}$), $\overrightarrow{b}$ = $\frac{1}{2}$\textit{a}($\sqrt{3}\overrightarrow{x}$+$\overrightarrow{y}$), and $\overrightarrow{c}$ = \textit{z}$\overrightarrow{c}$ ~\cite{kandemir2018structural}. This primitive lattice contains three atoms where a Pt atom is located at the \textit{1a} Wyckoff site, while two Se atoms occupy the \textit{2d} Wyckoff sites~\cite{lei2017comparative}. By employing an internal parameter \textit{z}, the atomic positions can be expressed in fractional coordinates, namely (0,0,0) for the Pt atom, and ($\frac{2}{3}$,$\frac{1}{3}$,$\frac{1}{4}$+\textit{z}) and ($\frac{1}{3}$,$\frac{2}{3}$,$\frac{1}{4}$-\textit{z}) for the Se atoms. Furthermore, the Pt-Se bonding is of a covalent nature, while the PtSe$_2$ layers are connected by vdW forces. To take into account these vdW interactions, we employ different vdW corrections. Specifically, we employ  vdW-DF~\cite{berland2015van}, vdW-DF2~\cite{lee2010higher}, rVV10~\cite{sabatini2013nonlocal}, DFT-D~\cite{grimme2004accurate}, and DFT-D3~\cite{moellmann2014dft}.\\

\textbf{Bulk PtSe$_2$.} For bulk PtSe$_2$, a full structural optimization was performed. Indeed, the lattice vectors and atomic positions were fully relaxed for each vdW correction. The starting cell parameters were taken from the experiment~\cite{zhang2021precise}, where \textit{a} = 3.75~\AA ~and \textit{c} = 5.60~\AA. The results of the structural calculations, as well as those found in the literature and observed experimentally, are listed in TABLE~\ref{tab_bulk}. They include both \textit{a} and \textit{c} as well as the interlayer distance ($d_{int}$), the interatomic distance ($d_{Pt-Se}$), and the internal parameter  \textit{z} (see  FIG.~\ref{fig1}).

\begin{table}[h]
\caption{\label{tab:table3} Structural parameters of bulk PtSe$_2$: comparison of values obtained through literature and experimental observations with various GGA and GGA+vdW exchange-correlation potential approximations, including lattice constants \textit{a} and \textit{c}, the interlayer distance $d_{int}$, the interatomic distance $d_{Pt-Se}$ and internal parameter \textit{z}.}
\begin{ruledtabular}
\begin{tabular}{c c|c|c|c|c|c}
 & Approach & \textit{a} (\AA) & \textit{c} (\AA)& \textit{$d_{int}$} (\AA) & \textit{$d_{Pt-Se}$} (\AA)& \textit{z} \\ 
\hline
      & GGA & 3.79  & 5.16 & 2.59& 1.28& -0.001    \\
      & vdW-DF& 3.84  & 5.27 &2.65& 1.31& -0.001   \\
      & vdW-DF2& 3.90  & 5.28 & 2.61& 1.33&0.003   \\
      & rVV10 & 3.82  & 5.12 & 2.47&1.32 &0.008  \\
      & DFT-D & 3.76  & 4.69 & 2.14& 1.27&0.021  \\
      & DFT-D3 & 3.80  & 4.77 & 2.27& 1.25&0.012  \\
      & DFT-D3~\cite{zhang2021precise} & 3.77 & 4.80 & -& - &-  \\
      & DFT-D3~\cite{li2021layer}& 3.77 & 4.82 & - & - & -\\
      &DFT-D~\cite{kandemir2018structural} & 3.77 & 4.70 & - & - & -\\
      &GGA~\cite{zhang2021precise}& 3.75 & 6.12 & - &-&- \\
      & optB88-vdW~\cite{villaos2019thickness}& 3.78& 4.96 & - & -&- \\
      & Exp~\cite{zhang2021precise} & 3.75 &  5.60 & -&- &- \\
      &Exp~\cite{zhang2017experimental}& 3.72 & 5.07 & - &-&-\\
      & Exp~\cite{gronvold1960sulfides}& 3.82 & 5.08 & - &-&-\\
      &Exp~\cite{villaos2019thickness}& 3.72 &5.08&-&-&-
\label{tab_bulk}
\end{tabular}
\end{ruledtabular}
\end{table}
The lattice parameter \textit{a} is found around 3.8~\AA~for all the considered vdW corrections, with a fluctuation of less than 4\%. This value is in good agreement with previous experimental results~\cite{gronvold1960sulfides,zhang2021precise} and theoretical findings~\cite{zhang2021precise,kandemir2018structural,zhang2021precise,villaos2019thickness}.  Unlike \textit{a} parameter, \textit{c} is susceptible to change with the enhanced vdW interaction. This distance goes from 5.16~\AA~to 4.69~\AA~under vdW corrections, representing a decrease of 9.1\%. 
In fact, our results are in good agreement with both the experimental results and  previous \textit{ab initio} calculations~\cite{zhang2017experimental,gronvold1960sulfides,villaos2019thickness,zhang2021precise,kandemir2018structural}. 
Furthermore, the different vdW corrections also change the $d_{int}$ distance  from 2.59~\AA~to 2.14~\AA, representing a maximum decrease of 17.3\%.
Besides $d_{int}$, the $d_{Pt-Se}$ distance also exhibits a variation with the introduction of vdW corrections, reaching 6\%.
Therefore, the vdW corrections modify the interlayer vacuum and the covalent bond. 
These results show that vdW corrections play a fundamental role for the out-of-plane crystal features, namely \textit{c}, $d_{int}$, and $d_{Pt-Se}$.

Furthermore, the results given by DFT+vdW corrections are different from each other. According to the TABLE.~\ref{tab_bulk}, The empirical DFT-D and DFT-D3 corrections give the smallest values of the parameter \textit{c}, the $d_{int}$ and $d_{Pt-Se}$ distances. In particular, compared to experimental results~\cite{villaos2019thickness,gronvold1960sulfides,zhang2017experimental}, the \textit{c} parameter is underestimated by 7.6\% and 6.1\% for DFT-D and DFT-D3, respectively.  Additionally, the underestimation amounts to 16.2\% for DFT-D and 14.8\% for DFT-D3 compared to experimental observations~\cite{zhang2021precise}. However, the results obtained by these corrections are in good agreement with previous ab initio findings~\cite{zhang2021precise,kandemir2018structural}.
Compared to the other approximations, these two corrections show the largest variation from the GGA approximation. Indeed, the parameter \textit{c} decreases by 9.1\% and 7.5\% for DFT-D and DFT-D3, respectively. Additionally, the $d_{int}$ distance decreases by 17.3\% for DFT-D and 12.3\% for DFT-D3. Furthermore, besides the \textit{c} and $d_{int}$ parameters, the $d_{Pt-Se}$ distance  also undergoes a slight decrease compared to the GGA approximation.

\begin{figure}[h]
  \includegraphics[width=0.5\textwidth]{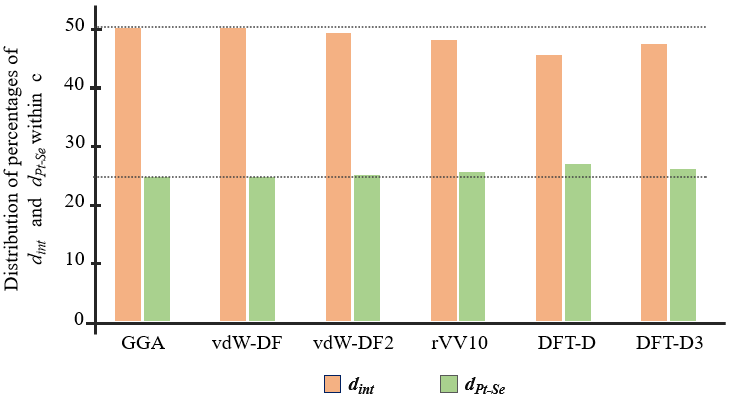}
   \caption{Diagram illustrating percentage of $d_{int}$ and $d_{Pt-Se}$ distances relative to lattice parameter \textit{c}. The dotted lines indicate the percentage of GGA.}
  \label{fig2}
\end{figure}

According to FIG.\ref{fig2}, in terms of the \textit{c} parameter, the values found for the $d_{Pt-Se}$ distance  represent 27\% of the \textit{c} parameter for the DFT-D correction and 26.2\% for the DFT-D3 correction. These values are higher than those obtained with GGA, which is 24.8\%.  With the introduction of these corrections, the Se atoms move away from the Pt atoms, as identified by the increase in the \textit{z} parameter, hence increasing the $d_{Pt-Se}$ distance. Therefore, the empirical corrections affect both the lattice parameters and atomic positions. In particular, they give the smaller unit cell volumes by reducing the lattice parameters. However, they cause the largest variation in the positions of Se atoms relative to Pt atoms, leading to higher values of the \textit{z} parameter.

The empirical corrections DFT-D and DFT-D3 often overestimate vdW interactions~\cite{moellmann2014dft, bucko2010improved}, which is also the case in bulk PtSe$_2$. However, the DFT-D3 correction shows an improvement compared to DFT-D. This correction has a lattice parameter \textit{c} of 4.77~\AA, closer to experimental results than that of DFT-D, which is 4.69~\AA. This improvement is explained by the fact that DFT-D3 takes into account eighth-order dispersion coefficients ($C^{ij}_8$), which are more sensitive and depend on the local geometry of the system (number of coordinations)~\cite{grimme2004accurate}.

On the other hand, corrections using non-local functionals, namely vdW-DF and vdW-DF2, tend to overestimate the \textit{c} parameter each in their own way. Indeed, the vdW-DF and vdW-DF2 corrections slightly overestimated the \textit{c} parameter by 3.6\% compared to experimental values reported in~\cite{villaos2019thickness,gronvold1960sulfides,zhang2017experimental}, and it underestimated by 5.7\% compared to experiment~\cite{zhang2021precise}. Additionally, the rVV10 correction yields results closer to the experimental observations reported in~\cite{villaos2019thickness,gronvold1960sulfides,zhang2017experimental}. Furthermore, the $d_{int}$ distance  found by the vdW-DF and vdW-DF2 corrections is slightly larger than that obtained with GGA. In contrast, rVV10 yields a smaller $d_{int}$ distance 
 compared to GGA, showing a decrease of about 4.6\%. 

As shown in FIG.\ref{fig2}, the non-local functionals result in larger $d_{int}$ and $d_{Pt-Se}$ distances  compared to empirical corrections. The $d_{Pt-Se}$ distance represents 24.8\% of the value of the \textit{c} parameter for vdW-DF, which is identical to that found by GGA. For vdW-DF2 and rVV10 corrections, the $d_{Pt-Se}$ distance represent approximately 25\% of the \textit{c} parameter of each correction.  In fact, these two corrections enhance vdW interactions between the layers more than vdW-DF, leading to a decrease in the $d_{int}$ distance  and an increase in the $d_{Pt-Se}$ distance  which represents over 25\% of the lattice parameter \textit{c}.  Unlike vdW-DF2 and rVV10, the Se atoms approach the Pt atoms in GGA and vdW-DF, resulting in a negative variation of the \textit{z} parameter. This observation in vdW-DF is attributed to the repulsive nature of the revPBE exchange functional, which underestimates the vdW interactions between the layers~\cite{lee2010higher}. Furthermore, this issue is addressed in the vdW-DF2 correction, which utilizes the PW86 exchange functional, thereby enhancing the vdW interactions between the layers~\cite{lee2010higher}.\\

\begin{table*}
\caption{\label{tab:table3} The calculated structural parameters for the BL, TL, and FL structures of PtSe$_2$ include, in particular, the  \textit{$d_{Pt-Pt}$} and \textit{$d_{int}$} distances, the interatomic distances \textit{$d_{Pt-Se}^{inner}$} and \textit{$d_{Pt-Se}^{outer}$}, as well as the angle $\alpha$ between the two Se planes, as schematically shown in FIG.\ref{fig1}, for different GGA and GGA+vdW functionals.}
\begin{ruledtabular}
\begin{tabular}{c c|c|c|c|c|c}
 System & Approach  & \textit{$d_{Pt-Pt}$} (\AA) & \textit{$d_{int}$} (\AA) &  \textit{$d_{Pt-Se}^{inner}$} (\AA) & \textit{$d_{Pt-Se}^{outer}$} (\AA) & $\alpha$ \\ 
\hline
\multirow{12}{*}{PtSe$_2$ BL}
      & GGA & 5.22 & 2.66 & 1.28 & 1.30 &83.50°   \\
      & vdW-DF & 5.49 & 2.87 & 1.31 & 1.33 & 83.81°  \\
      & vdW-DF2 & 5.46 & 2.78 & 1.34 & 1.35 & 83.99°   \\
      & rVV10 & 5.11 & 2.48 & 1.31 & 1.33 & 84.08°  \\
      & DFT-D & 4.64 & 2.11 & 1.26 & 1.32 & 83.80°  \\
      & DFT-D3 & 4.79 & 2.28 & 1.25 & 1.28 & 82.84° \\
      & DFT-D3~\cite{li2021layer} & 4.92 & - & - & - & -\\
      & DFT-D~\cite{kandemir2018structural} & 4.76  & 2.14 & - & - & - \\
      & optB86b-vdW~\cite{fang2019structural}  & 5.14 & - & - & - & -\\
      & Exp~\cite{li2021layer}&5$\pm$0.3&-& - & - &-\\
      &Exp~\cite{zhang2021precise} & 5.7 &- & - & - &-\\
      &Exp~\cite{shawkat2020thickness} &5.3&-& - & - &-\\
    
\hline

\multirow{8}{*}{PtSe$_2$ TL}
      & GGA & 5.21 & 2.66 & 1.27 & 1.29 &83.43°  \\  
      & vdW-DF & 5.50 & 2.88 & 1.31 & 1.33 & 83.81°   \\
      & vdW-DF2 & 5.47 & 2.78 & 1.34 & 1.35 &83.99° \\
      & rVV10 & 5.11 & 2.48 & 1.31 & 1.33 & 84.07°  \\
      & DFT-D & 4.64 & 2.10 & 1.26 & 1.32 & 83.76° \\
      & DFT-D3 & 4.77 &  2.25 & 1.25 & 1.28 & 82.82° \\
      & DFT-D~\cite{kandemir2018structural} & 4.72 & 2.14 & - & - &-\\
      &Exp~\cite{zhang2021precise} & 5.9 &-& - & - & -\\
         
\hline
\multirow{8}{*}{PtSe$_2$ FL}
      & GGA & 5.15 & 2.59 & 1.27 & 1.30 & 83.44°  \\  
      & vdW-DF & 5.49 & 2.88 & 1.31 & 1.33 & 83.81°   \\
      & vdW-DF2 & 5.46 & 2.78 & 1.34 & 1.35 & 83.98° \\
      & rVV10 & 5.10 & 2.47 & 1.31 & 1.33 & 84.08°  \\
      & DFT-D & 4.65 & 2.10 & 1.27 & 1.32 & 83.79° \\
      &DFT-D3 & 4.79 & 2.27 & 1.25 & 1.28 & 82.84 \\
      & DFT-D~\cite{kandemir2018structural} & -& 2.14 & -& -&-\\
      &Exp~\cite{zhang2021precise}  & 5.8&-&- & -&-
\label{tab_layer}
\end{tabular}
\end{ruledtabular}
\end{table*}
\textbf{Few-layer PtSe$_2$.} The 2D structures of PtSe$_2$, namely ML, BL, TL, and FL, were constructed from the optimized bulk structure (see TABLE~\ref{tab_bulk}).  These systems were created by manually setting the lattice parameters \textit{a} and \textit{c} specific to each correction. A 20~\AA vacuum was added along the \textit{z}-direction to simulate isolated layers. After creating these few-layer structures, an atomic optimization was performed. The results of this optimization are summarized in TABLE~\ref{tab_layer}. The broken perpendicular translation symmetry in the case of the BL, TL, and FL PtSe$_2$ systems can lead to changes in the out-of-plane parameters, namely the $d_{Pt-Pt}$,  $d_{int}$ and  $d_{Pt-Se}$ distances, for all exchange-correlation approximations. Indeed, the GGA and GGA+vdW functionals produce larger out-of-plane parameters compared to those found in bulk PtSe$_2$. The few-layer structures are more free to move along the \textit{oz} axis compared to bulk material. The latter is constrained by translational symmetry along this axis and thus affected by the continuous interaction between layers. Furthermore, as indicated in TABLE~\ref{tab_layer}, the  few-layer systems exhibit a negligible variation of $\pm$0.01~\AA~when transitioning between FL, TL, and BL for all out-of-plane parameters.

In the same way as for bulk PtSe$_2$, the vdW corrections affect the few-layer systems. Indeed, the $d_{Pt-Pt}$ and $d_{int}$ distances obtained with the rVV10, DFT-D, and DFT-D3 corrections show a decrease compared to the GGA method.  In contrast, the vdW-DF and vdW-DF2 corrections present larger $d_{Pt-Pt}$ and $d_{int}$  distances than those obtained with GGA.
Furthermore, according to TABLE~\ref{tab_layer}, although the $d_{Pt-Se}$ distance is equal across all layers of bulk PtSe$_2$, it varies for PtSe$_2$ BL, TL, and FL. Specifically, the outer $d_{Pt-Se}$ distances (\textit{$d_{Pt-Se}^{outer}$}) are slightly larger than the inner $d_{Pt-Se}$ distances (\textit{$d_{Pt-Se}^{inner}$}). This difference is approximately 0.02~\AA~for all corrections, except for DFT-D where it is around 0.06~\AA.  This means that the Se planes in each layer lose their symmetry relative to the Pt plane.
The inner Se atoms are more affected by enhanced vdW interactions between the layers when applying vdW corrections. In contrast, the outer Se atoms are less influenced by these corrections and therefore more free to move along the (\textit{oz}) axis during geometric optimization.

\textbf{Monolayer PtSe$_2$.} For the single PtSe$2$ ML, our calculations are limited to the GGA approximation due to the absence of vdW out-of-plane interactions. A lattice parameter of $a = 3.74~\AA$ and a monolayer thickness of $d_{\text{Se-Se}} = 2.62~\AA$ were found.
These values are in excellent agreement with previous theoretical results incorporating vdW corrections~\cite{kandemir2018structural,li2016tuning,zhang2021precise}. Consequently, for the PtSe$_2$ ML, the conventional GGA approximation for the exchange-correlation potential appears to be sufficient for accurately describing its crystal structure.\\

\subsection{Electronic structure of PtSe$_2$}
\subsubsection{Effect of the layer number on PtSe$_2$ band structure}
One of the most important characteristics of TMD is the variation of their electronic properties with respect to the number of layers. The PtSe$_2$ material stands out among the other TMD by a transition from a semi-metallic material in the bulk to a semiconductor material for a small number of layers. The challenge is to deeply understand the correlation between the number of layers and the SC-SM transition.  In this section, the band structures with and without SOC are calculated and presented for the different PtSe$_2$ systems, particularly for  ML, BL, TL, FL and bulk PtSe$_2$ materials. Our method uses the GGA and GGA plus vdW corrections in order to determine this SC-SM transition. Here, the energy dispersion of the different systems are calculated for the in-plane wave vector of the First Brillouin Zone (FBZ). Indeed, the considered high-symmetry path is K-$\Gamma$-M-K as illustrated in FIG.~\ref{fig_3}.\\

\begin{figure}[h]
\includegraphics[scale=0.35]{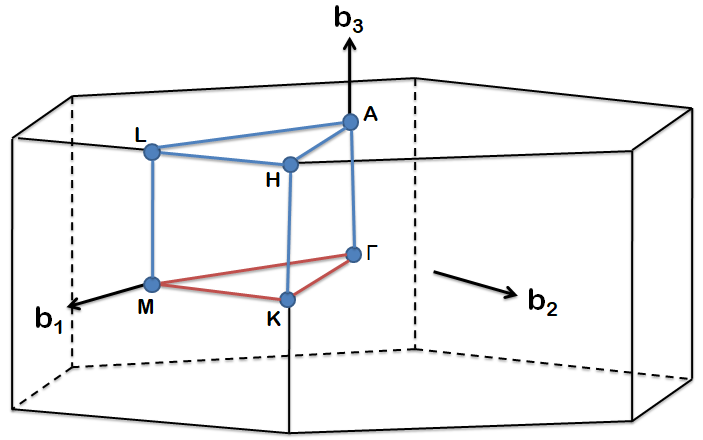}
\caption{\label{fig_3}
First Brillouin Zone in the reciprocal space of the PtSe$_2$. The line of high symmetry used for the study of the structure of the bands (red).}
\label{fig_3}
\end{figure}
As shown in FIG.~\ref{structure_bandes}, PtSe$_2$ ML exhibits a semiconductor character in all DFT and DFT+vdW approximations, with an indirect bandgap (E$_g$) of 1.35 eV. The vdW corrections, as discussed in the previous section, do not affect the properties of the PtSe$_2$ ML. The valence band of PtSe$_2$ ML reaches its maximum at the $\Gamma$ point, while the minimum of the conduction band is located between the $\Gamma$ and M points. The valence band shows an almost flat dispersion around the $\Gamma$ point, consistent with the structure found by Angle-Resolved Photoemission Spectroscopy (ARPES)~\cite{wang2015monolayer}.  The indirect bandgap has been observed in group 10 TMD monolayers (such as PtSe$_2$, PdSe$_2$...), unlike group 6 TMD monolayers (MoS$_2$, WSe$_2$...) which exhibit a direct bandgap~\cite{liu2018nature,chang2014ballistic,oyedele2017pdse2}. Consequently, the optical transition requires the involvement of phonons to ensure the conservation of electron momentum. This categorizes PtSe$_2$ ML as a poor emitter and absorber of light~\cite{allah2015regroupement,yoffe1993low}.\\

According to the FIG.~\ref{fig_4}, the bandgap of PtSe$_2$ BL, TL and FL depends on the used vdW correction. Under the vdW-DF approximation, $E_g$ reaches its maximum at 0.44 eV, 0.24 eV and 0.10 eV for PtSe$_2$ BL, TL, and FL systems, respectively. In contrast, for the DFT-D approximation,  the bandgap  is minimal, or even completely disappears for the same systems. Compared to GGA, the non-local functionals vdW-DF and vdW-DF2 show a wider bandgap. However, the corrections rVV10, DFT-D, and DFT-D3 produce a narrower bandgap. Despite this decrease in the bandgap, PtSe$_2$ BL retains its semiconductor character.

\begin{figure}[h]
\includegraphics[scale=0.33]{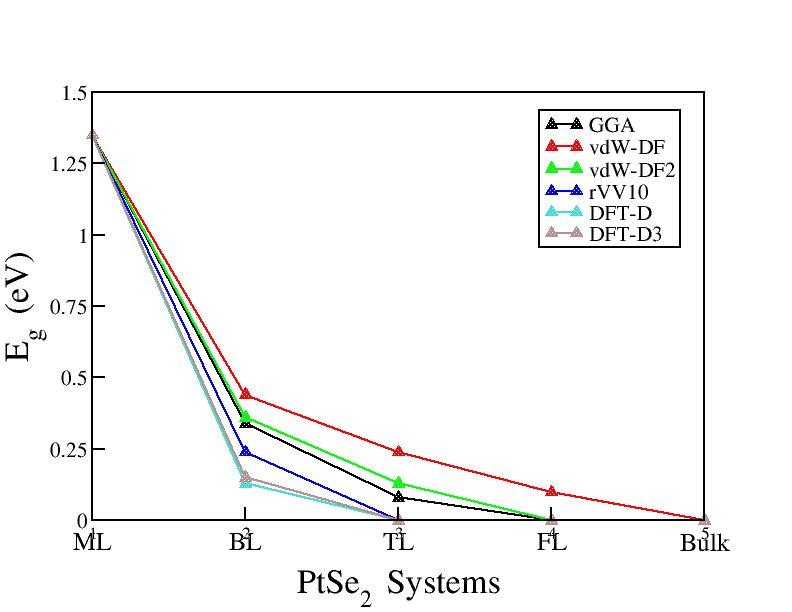}
\caption{\label{fig_4}
The bandgap values of the various PtSe$_2$ systems for all the considered exchange-correlation approximations.}
\label{fig_4}
\end{figure}

\begin{figure*}
\begin{center}
 \leavevmode
\includegraphics[width=18.8cm]{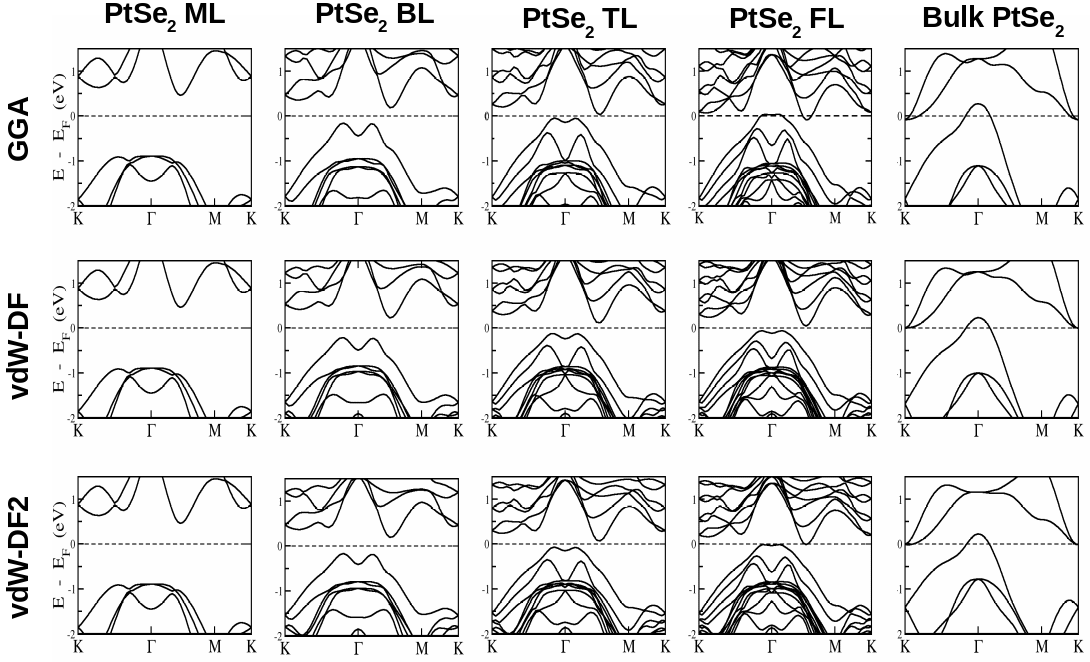}
\includegraphics[width=19cm]{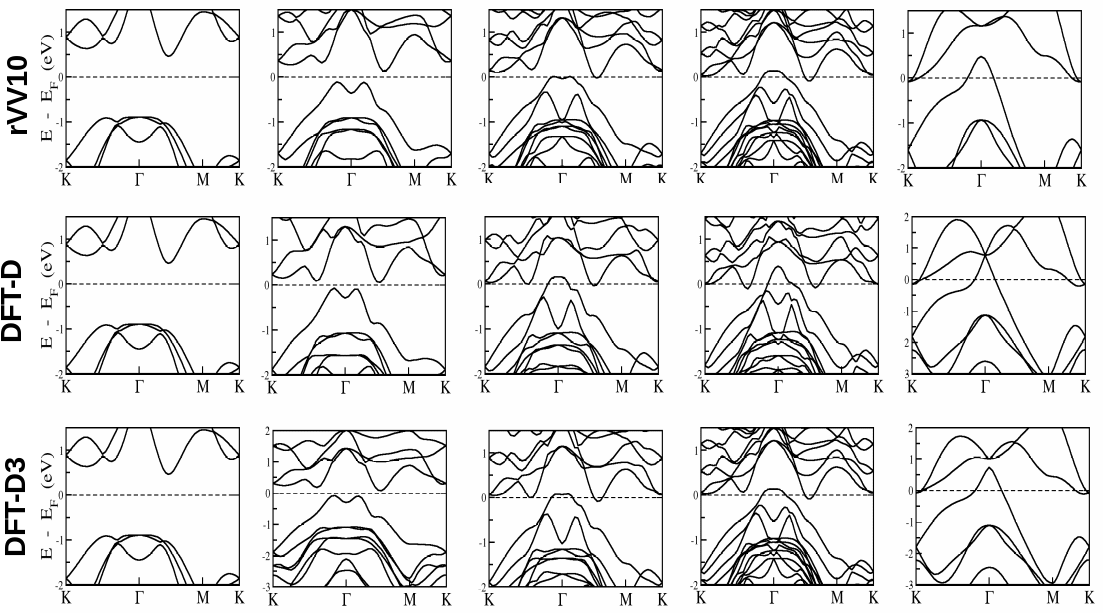}
\caption{Band structures of the different PtSe$_2$ systems (ML, BL, TL, FL and  bulk) with GGA approximation and vdW corrections without SOC. The Fermi level is considered as the origin of the energy.}
\label{structure_bandes}
\end{center}
\end{figure*}
Furthermore, FIG.~\ref{fig_4} shows that PtSe$_2$ TL exhibits a semi-metallic character for the rVV10, DFT-D, and DFT-D3 corrections, which is consistent with the experimental data from reference~\cite{li2021layer}. However, according to the vdW-DF and vdW-DF2 approximations, PtSe$_2$ TL shows a semiconductor behavior with a smaller bandgap compared to PtSe$_2$ BL. This result is in agreement with the experimental observations from reference~\cite{zhang2021precise}. This disparity stems mainly from the $d_{Pt-Pt}$ and $d_{int}$ distances, where a relatively larger value of the $d_{int}$ leads to a semiconductor character.

In Table~\ref{tab2}, we present the bandgap values obtained from our DFT calculations for various PtSe$_2$ systems, compared with previous theoretical and experimental results.  For PtSe$_2$ BL, our vdW-DF calculations show closer bandgap values to the experimental results~\cite{li2021layer} compared to all other corrections. However, while our calculations show that the PtSe$_2$ TL exhibits a semiconductor character, the experiment~\cite{li2021layer} demonstrates a semi-metallic nature.  Regarding the electronic transition, the vdW-DF correction is in good agreement with the experiment~\cite{zhang2021precise} and the previous theoretical study by HSE06~\cite{zhang2021precise}. Furthermore, the calculation obtained with the rVV10 correction agrees well with values found in theoretical studies using optB88-vdW~\cite{villaos2019thickness}, GGA~\cite{li2021layer}, and LDA~\cite{fang2019structural}. Our rVV10, DFT-D, and DFT-D3 approximations indicate a semi-metallic nature beginning from the third layer, consistent with the observations by \textit{Li} et \textit{al.}~\cite{li2021layer}.
\begin{table}[h]
\caption{\label{tab:table3}Comparaison of PtSe$_2$ ML, BL, TL, FL bandgap energies obtained by applying different theoretical functionals and observed in different experiments. A zero bandgap corresponds to the semi-metallic nature.}
\begin{ruledtabular}
\begin{tabular}{c c|c|c|c|c|c}
 & Approach &  ML & BL &  TL &  FL &  Bulk\\ 
\hline
      & GGA & 1.35 & 0.34 & 0.08 & 0 & 0 \\
      & vdW-DF& 1.35 & 0.44 &0.24 & 0.10 & 0\\
      & vdW-DF2 & 1.35 & 0.36 & 0.13 & 0 & 0\\
      & rVV10 & 1.35 & 0.24 & 0 & 0 & 0\\
      & DFT-D & 1.35 & 0.13 & 0 & 0 & 0\\
      & DFT-D3 & 1.35 & 0.15 & 0 & 0 & 0\\
      & DFT-D~\cite{kandemir2018structural} & 1.17 & 0.17 & 0 & 0 & 0\\
      &DFT-D3~\cite{li2016tuning} & 1.39 & 0.99 & - & - & 0\\
      &optB88-vdW~\cite{villaos2019thickness} & 1.18 & 0.21 & 0 & 0 & 0\\
      & GGA~\cite{li2021layer} & 1.20 & 0.22 & 0.01 & 0 & 0\\
      & GGA ~\cite{o2016raman} & 1.60 &  0.8 & - & - & 0\\
      &G$_0$W$_0$~\cite{zhuang2013computational} & 2.10 & - & - & -&0\\
      & G$_0$W$_0$~\cite{li2021layer} &2.44 & 1.174 & 0.30 & 0 & 0\\
    &HSE06~\cite{zhang2021precise}&1.837&0.796&0.417&0.148&0\\
      &LDA~\cite{fang2019structural}& 1.2&0.21&0&0&0\\
      & Exp~\cite{li2021layer} & 1.8 & 0.6 &0 &0&0\\
      &Exp~\cite{zhang2021precise} & 2.0 &1.1&0.6&0.2&0
\label{tab2}
\end{tabular}
\end{ruledtabular}
\end{table}
 However, all our GGA and GGA+vdW approximations, as well as the GGA and LDA approximations used in Ref.~\cite{li2021layer,fang2019structural}, underestimate the experimental bandgap values, while the G$_0$W$_0$ approximation overestimates them. For bulk PtSe$_2$, all corrections used in our calculations indicate a semi-metallic character, consistent with theoretical and experimental studies in the references~\cite{kandemir2018structural,li2016tuning,villaos2019thickness,li2021layer,o2016raman,zhuang2013computational,fang2019structural}. In fact, for PtSe$_2$ ML, the bandgap found by our calculation agrees with previous theoretical results~\cite{li2016tuning}.

\subsubsection{The effect of spin-orbit coupling on the bandgap of the PtSe$_2$ structure}
\begin{figure}[h]
\includegraphics[scale=0.8]{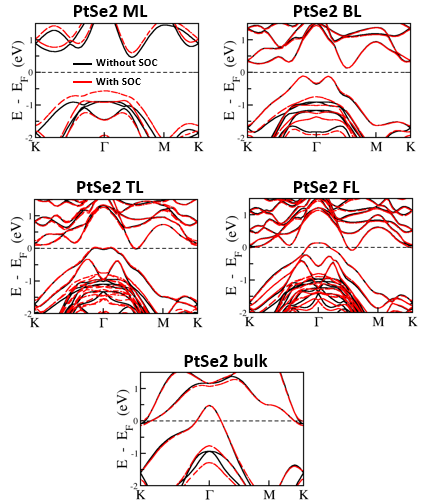}
\caption{\label{fig_7}
Band structures of ML, BL, TL, FL and bulk PtSe$_2$ systems, with and without SOC for the rVV10 correction.}
\label{fig_6}
\end{figure}

Given the high atomic mass of Pt and Se atoms, spin-orbit interaction is important in these 2D materials. In this section, we investigate the effect of SOC on the electrical properties of the different PtSe$_2$ systems. As shown in FIG.~\ref{fig_6}, the SOC alters the band structure of PtSe$_2$ ML by reducing the bandgap width from 1.35 eV without SOC to 1.16 eV with SOC. Moreover, the gap between the three valence bands increases  around the $\Gamma$ point, while the shape of the conduction band remains unchanged.  This result is consistent with the results reported by \textit{Kandemir et al.}~\cite{kandemir2018structural}. Furthermore,  for BL, TL, FL and bulk PtSe$_2$, the valence band splitting is also found. Specifically, the energy gap between the lower valence bands increases in these cases. However, the upmost valence band remains unaffected by SOC, thereby maintaining similar bandgap values as found without SOC. 

\subsubsection{The effect of the $d_{int}$ distance  on the bandgap of PtSe$_2$ structures}

Each of the vdW corrections yields different results regarding the SC-SM transition. This difference can be explained by the variation in $d_{int}$ distance. Indeed, according to Table~\ref{tab_layer} and FIG.~\ref{structure_bandes}, vdW-DF and vdW-DF2 corrections, which have larger $d_{int}$ distance compared to other approximations, exhibit the widest bandgap. Interestingly, for PtSe$_2$ TL, the 0.40 \AA~difference between vdW-DF and rVV10 corrections changes the electronic character of PtSe$_2$ TL. This highlights the sensitivity of the bandgap to $d_{int}$ distance. Furthermore, for a fixed vdW correction, the $d_{int}$ distance remains constant. This demonstrates that the bandgap width is also influenced by the number of layers. Indeed, the number of layers governs the transition, whereas the $d_{int}$ distance determines at which layer count it occurs.

Furthermore, the effect of $d_{int}$ distance on electronic properties can also explain the discrepancy between the results of the two experiments mentioned in Refs.~\cite{li2021layer,zhang2021precise}. In~\cite{li2021layer}, the SC-SM transition was observed at the third layer. However, in Ref.~\cite{zhang2021precise}, the transition happened at the fifth layer. This disparity between the two experiments can be attributed to the difference in $d_{int}$ distance. Specifically, while a distance $d_{Pt-Pt}$ of 5 $\pm$ 0.3 $\AA$ was reported in Ref.~\cite{li2021layer}, the experiment cited in Ref.~\cite{zhang2021precise} shows a distance of approximately around 5.6 $\AA$. The increase in bandgap is directly linked to the increase in $d_{int}$ distance, leading to transitions occurring in higher layers. The increasing in the bandgap  is directly related to the increase in the $d_{int}$ distance, leading to transitions occurring in higher layers. The observed difference in the $d_{Pt-Pt}$ distance  between the two experiments may due to the growth protocols, experimental conditions. The hight sensitivity to the $d_{int}$ distance suggest that the SC-SM transition in PtSe$_2$ material is not unique for a given layer number.

The sensitivity of the bandgap to interlayer separation suggests that the d$_{int}$ distance  can be adjusted through vertical compressive strain along the ($oz$) axis. This approach can be applied to the case of PtSe$_2$ BL which the best candidate to illustrate the effect of this strain. A detailed analysis of the strain effect on PtSe$_2$ BL is provided in the third part of this article.
\subsubsection{Origin of the SC-SM transition}
\begin{figure*}
\includegraphics[width=18cm]{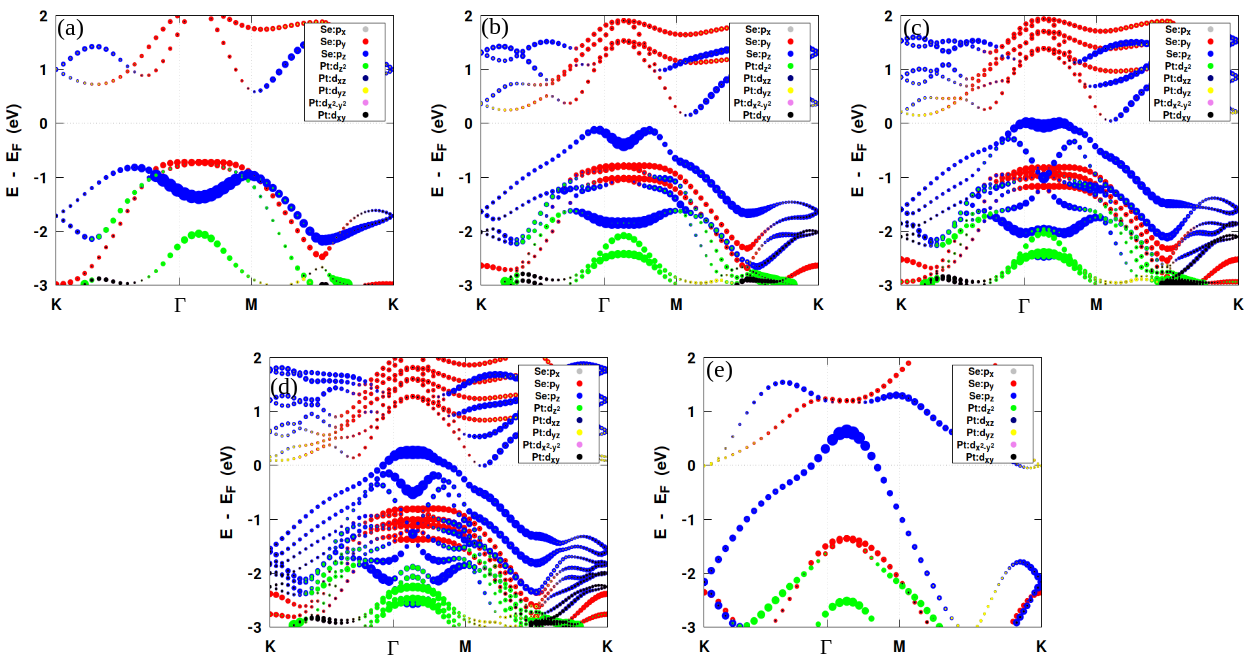}
\caption{\label{proj_band}Projected Band Structures for different PtSe$_2$ Systems (ML(a), BL(b), TL(c), FL(d) and Bulk(e)), using rVV10 correction}.
\end{figure*}

In order to further explore the origin of the SC-SM transition, we calculated the projected band structures for the different layered systems (see FIG.~\ref{proj_band}), using GGA+rVV10 approximation. In PtSe$_2$ ML, the maximum of the valence band (VBM) is mainly constituted by the $p_y$ orbitals of the chalcogen atom Se. Besides, the conduction band (CB) is essentially constructed by both $p_y$ and $p_z$ orbitals of the Se atom. As the number of layers increases, moving from PtSe$_2$ ML to bulk material, an interesting change occurs. The $p_z$ orbitals rises above the $p_y$ orbitals and takes on a higher energy, while the $p_y$ orbitals retains almost the same position and energy. For the conduction band, the doublet of the $p_y$ and $p_z$ orbitals moves to a lower energy level, until the overlap with the $p_z$ orbitals of the valence band. This highlights the sensitivity of the $p_z$ orbitals to interactions between the layers in the material. Therefore, the origin behind the SC-SM transition is mainly the evolution of $p_z$ orbitals.

\subsection{Strain effect on PtSe$_2$ BL}
The sensitivity to the $d_{int}$ distance open the possibility of tuning the bandgap of PtSe$_2$ BL. In fact, a way of controlling the $d_{int}$ distance is by applying a vertical compressive strain. Indeed, in Ref.~\cite{li2016tuning}, this strain was applied by changing and constraining the $d_{int}$ distance. Here, the strain was defined as $\epsilon=\frac{(d_0-d)}{d_0}$, where $d$ and $d_0$ are the $d_{int}$ distance with and without strain, respectively. In this picture, the in-plane parameters are kept unchanged. In fact, in order to optimize the BL structure, the supercell approach should be considered to treat the non-periodicity along $z$ direction. However, this fact make impossible to control $c$ parameter of the BL. In our work, we present a fully optimized structure method. Indeed, the BL structure is optimized after the application of strain $\epsilon$ which is defined as,
\begin{align}
\epsilon&=\frac{(c-c')}{c'}
\end{align}
where $c$ and $c'$ are the BL lattice parameter without and with strain, respectively.  In order to optimize the BL structure, a two step approach was adopted. First, an in-plane optimization is performed for the bulk PtSe$_2$ material. Here, we consider the following approximation,
\begin{align}
c'&=c^{Bulk} (xy)\simeq c^{BL} (L\rightarrow \infty)
\label{formula1}
\end{align}
Here, $c'$ is equal to the lattice parameter of the bulk $c^{Bulk} (xy)$ for an in-plane optimization, where $L$ is the vacuum width of BL supercell. For each strain the $c^{Bulk}$ is fixed only the in-plane parameters such as $d_{int}$ and $a$ are allowed to move. For instance, in the case of unstrained BL, by fixing $c^{Bulk}$ to $c^{BL}=5.18$ $\AA$,  the calculated parameter of the bulk after an  in-plane optimization, $a^{Bulk}= 3.8$ $\AA$ and $d^{Bulk}_{int} = 2.51$ $\AA$ 
 are almost equal to the parameters $a^{BL}= 3.79$ $\AA$ and $d^{BL}_{int} = 2.50$ $\AA$ of the BL. In the second step, for a given strain,  the in-plane optimized parameters, $a^{Bulk}$ and $d^{Bulk}_{int}$, used together with the supercell approach, we calculate the band structure of the strained BL.\\

\begin{figure}[h]
\includegraphics[scale=0.34]{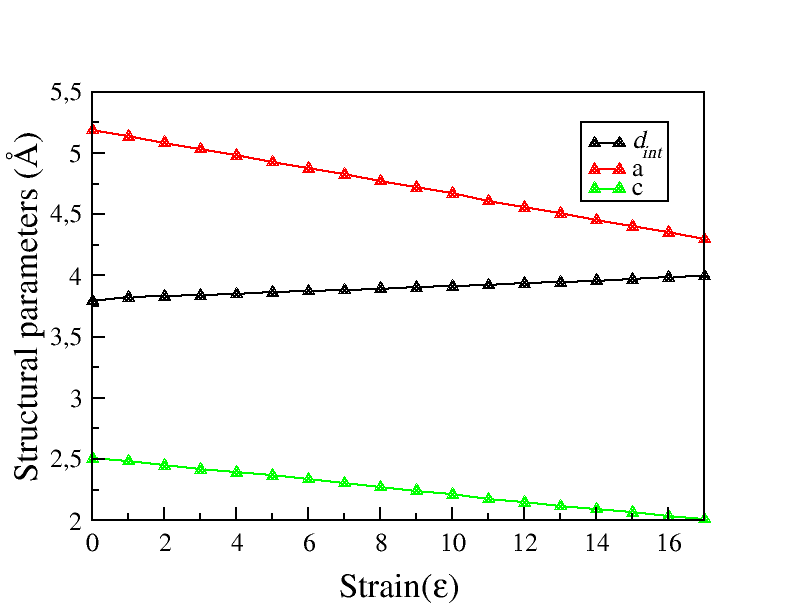}
\caption{Evolution of structural parameters $a$, $c$, and $d_{int}$ as a function of vertical compressive strain for PtSe$_2$ BL.}
\label{structural_parameters}
\end{figure}
By applying vertical compressive strain to the PtSe$_2$ BL, a variations in structural parameters is induced, namely in the $a$ and the $d_{int}$ distances, as depicted in FIG.~\ref{structural_parameters}. The $d_{int}$ distance  consistently decreases as the applied strain increases, with a reduction exceeding 0.40 $\AA$ for a compression of $\epsilon=$ 15\%. This corresponds to a reduction of 0.77 $\AA$ for the $c$ distance. In contrast to $d_{int}$, the parameter $a$ experiences a relatively modest increase. For an unstrained structure, the parameter $a$ is 3.79 $\AA$, while with a strain of $\epsilon=$ 15\%, it reaches 3.96 $\AA$. This slight increase in parameter $a$ is attributed to the system  ability to freely relax within the ($xy$) plane, constantly seeking the ground state of energy.\\

\begin{figure}[h]
\includegraphics[scale=0.335]{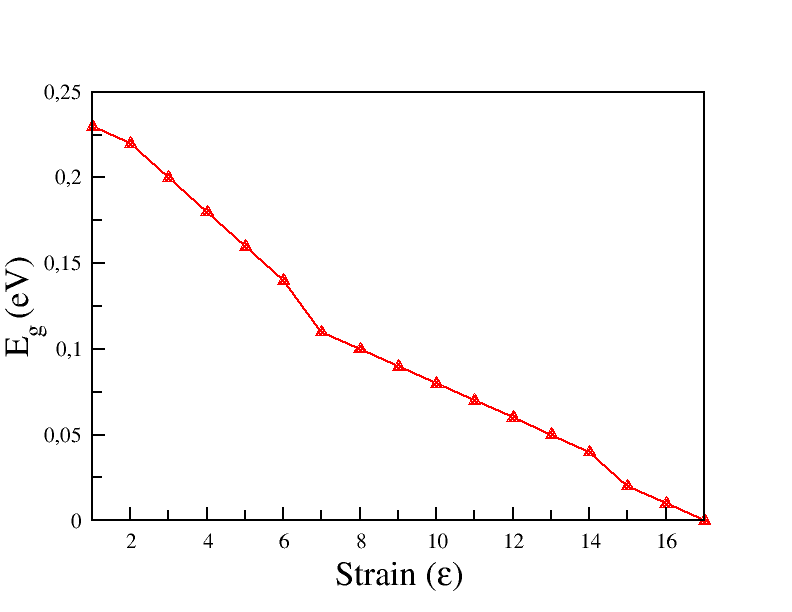}
\caption{bandgap of the PtSe$_2$ BL as a function of the vertical compressive strain.}
\label{energy_strain}
\end{figure}

In order to reveal the influence of the vertical compressive strain on the bandgap and the band structure,  we plot FIG.~\ref{energy_strain} and~\ref{strain_tot}, respectively.  In particular,  according to FIG.~\ref{energy_strain}, the increase in $\epsilon$ leads to the decrease of the bandgap energy. Indeed, in the range of $\epsilon=$  1\% to 16\% , PtSe$_2$ BL maintains its semiconductor character, with a decrease in the bandgap energy ranging from 0.25 eV to 0.01 eV. The bandgap varies with a step of approximately -0.02 eV/1\%, which is of the same order of magnitude as the experimental results obtained for the monolayer MoSe$_2$ under vertical compressive strain, approximately -27$\pm$2/1\%~\cite{island2016precise}. Note that, for a vertical strain lower than 8\%, the bandgap becomes narrow and less than 0.1 eV. From $\epsilon$  = 17\%, the PtSe$_2$ BL adopts a semi-metallic character, exhibiting a negative bandgap. Therefore, the strain offer the possibility of reversibly tuning the BL bandgap by about 0.25 eV through application. The reduction of the bandgap energy under vertical compressive strain was also found by $Li$ et $al.$~\cite{li2016tuning}. However, in this reference, the SC-SM transition was assigned to a large strain of over 40\%, where the bandgap goes from 1 to 0 eV. This discrepancy is due to the use of different vdW correction and the in-plane optimization in our calculations of strained BL.\\

\begin{figure}[h]
\includegraphics[scale=0.32]{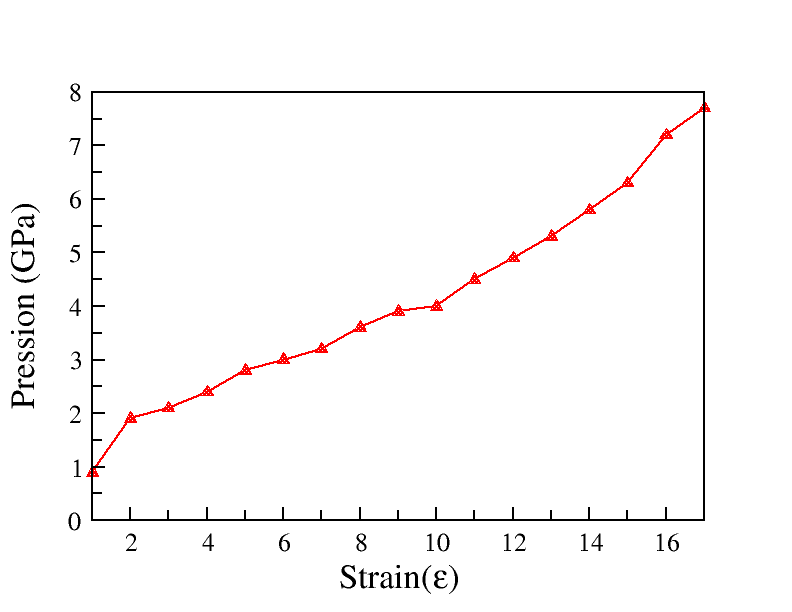}
\caption{ Applied pressure per unit cell of PtSe$_2$ BL as a function of the vertical compressive strain ($\epsilon$).}
\label{pression}
\end{figure}

\begin{figure}[h]
\includegraphics[scale=0.55]{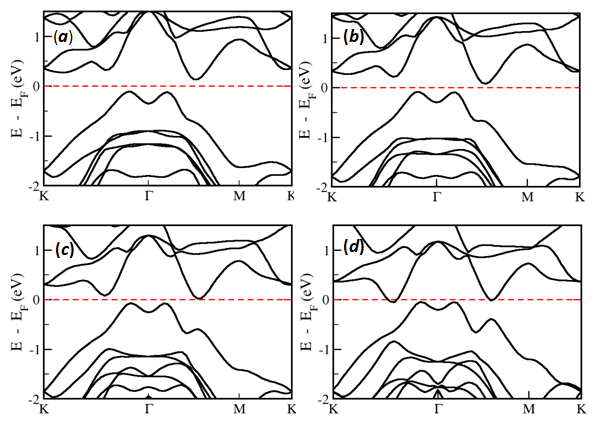}
\caption{Band structures of PtSe$_2$ BL for Strain $\epsilon$ = 1\% (a), $\epsilon$ = 5\% (b), $\epsilon$ = 10\% (c), and $\epsilon$ = 17\% (d), using GGA+rVV10 functional. }
\label{strain_tot}
\end{figure}

To evaluate the applied pressure to the PtSe$_2$ BL under strain, we present in the FIG.~\ref{pression} the calculated pressure for each level of strain, using the formula $P= (E^{tot}_0 - E^{tot})/((c - c')\times A$))~\cite{li2016tuning,kou2013graphene,kou2014robust}. The terms $E^{tot}_0$ and $E^{tot}$ represent the total energies of the systems without and under strain, respectively,  and $A$ represents the surface of the primitive cell of the strained system. Our results indicate that the pressure is ranging from 1 GPa  to 7.7 GPa. The pressure needed for the SC-SM transition is 7.7 GPa which larger than the values obtained in Ref.~\cite{li2016tuning,lei2023directional}. In fact, this pressure is lower than that found in the case of  PtSe$_2$ ML to achieve its semi-metallic character under in-plane strain~\cite{deng2018strain}. Indeed, for PtSe$_2$ ML~\cite{deng2018strain}, a strain of 8\% was found to induce an SC-SM transition, requiring a pressure exceeding 10 GPa.\\

\begin{figure}[h]
      \includegraphics[width=0.239\textwidth]{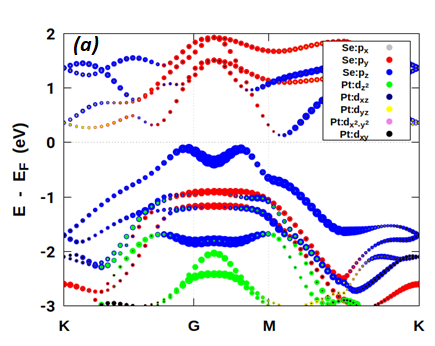}
      \label{ptse2_1}
\includegraphics[width=0.237\textwidth]{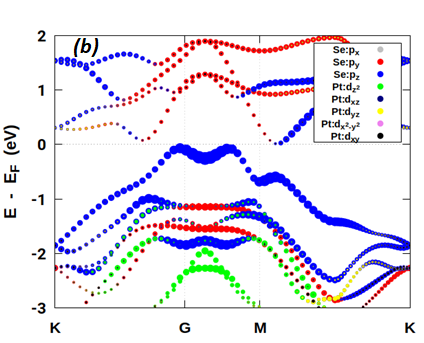}
      \label{ptse2_10}                      
\caption{Projected band structures for the PtSe$_2$ BL under a 1\% strain (a) and a 10\% strain (b).}
  \label{fig_1.10}
\end{figure}

Furthermore, the band structures of strained BL are shown in FIG.~\ref{strain_tot}. The increase of the strain leads to energy shift in the valence and conduction bands. Besides, under the effect of the strain, a new band parabolic dispersion appears, in the  valence band near \textbf{M} point.  Moreover, the lower valence band loses its flat form. To further understand the effect of the strain on the electronic bands, we plot the projected band structures in FIG.~\ref{fig_1.10}. The indicated modifications in the band structure under vertical compressive strain, stem mainly from  $p_z$ orbitals of Se atom. In fact, the interlayer interaction is enhanced due to the applied strain leading the valence band dominated by $p_z$ orbital to shift toward higher energy. Meanwhile, the bands dominated in-plane $p_x$ and  $p_y$ orbital contribution of Se atom remain almost unchanged. \\

\section{CONCLUSIONS}
In this work, we presented an investigation of the structural and electronic properties of the PtSe$_2$ ML, BL, TL, FL, bulk and strained BL using first principle calculations.  In order to treat the vdW interactions between PtSe$_2$ layers, the functionals GGA, GGA+vdW-DF, GGA+vdW-DF2, GGA+rVV10, GGA+DFT-D and GGA+DFT-D3 were adopted. Each of these corrections uniquely modifies the crystalline and electronic structure of PtSe$_2$. Indeed, the vdW-DF and vdW-DF2 corrections result in larger $d_{Pt-Pt}$ and $d_{int}$ distances compared to those obtained with GGA. This contrasts with the rVV10, DFT-D, and DFT-D3 corrections, which yield smaller distances. The $d_{Pt-Pt}$ and $d_{int}$ distances remain unchanged for each correction during the transition from bulk to ML.\\

Furthermore, regarding the electronic properties of PtSe$_2$, the SM-SC transition is found depending on the considered vdW corrections. Indeed, each correction results in a unique SC-SM transition. This dependence stems from the high sensitivity to changes in the $d_{int}$ distance. A 0.4 $\AA$ change in the $d_{int}$ distance between vdW-DF and rVV10 corrections in the PtSe$_2$ TL was sufficient to change its nature from semiconductor to semi-metal. Additionally, our findings suggest that the layer-number dictates the SC-SM transition for a specific correction. However, the $d_{int}$ distance determines at which layer count this transition occurs. Therefore, this distance plays a key role for the electronic properties of PtSe$_2$ material. More precisely, the SC-SM transition in PtSe$_2$ material originates from the energy shift of Se-$p_z$ dominated the conduction and valence bands. Our results reveal that the exact determination of the SC-SM transition in this material requires careful attention to the out-of-plane interaction. \\
Due to the $d_{int}$ distance effect, applying a vertical compressive strain represent an effective tool to tune the bandgap of PtSe$_2$ material. In particular, the PtSe$_2$ BL show the highest sensitivity to the $d_{int}$ among the other considered systems. Hence, by applying a vertical compressive strain from 1 to 16\%, the bandgap decreased from 0.25 to 0.01 eV. The SC-SM transition was obtained at 17\% due to the decrease of the $d_{int}$ distance. The bandgap tuning of PtSe$_2$ BL under vertical compressive strain is of particular interest in the development and fabrication of optoelectronics nanodevices.

\section*{DATA AVAILABILITY}
The data that support the findings of this study are available from the corresponding author upon reasonable request.









\nocite{*}

\bibliography{apssamp.bib}

\begin{thebibliography}{102}%
\makeatletter
\providecommand \@ifxundefined [1]{%
 \@ifx{#1\undefined}
}%
\providecommand \@ifnum [1]{%
 \ifnum #1\expandafter \@firstoftwo
 \else \expandafter \@secondoftwo
 \fi
}%
\providecommand \@ifx [1]{%
 \ifx #1\expandafter \@firstoftwo
 \else \expandafter \@secondoftwo
 \fi
}%
\providecommand \natexlab [1]{#1}%
\providecommand \enquote  [1]{``#1''}%
\providecommand \bibnamefont  [1]{#1}%
\providecommand \bibfnamefont [1]{#1}%
\providecommand \citenamefont [1]{#1}%
\providecommand \href@noop [0]{\@secondoftwo}%
\providecommand \href [0]{\begingroup \@sanitize@url \@href}%
\providecommand \@href[1]{\@@startlink{#1}\@@href}%
\providecommand \@@href[1]{\endgroup#1\@@endlink}%
\providecommand \@sanitize@url [0]{\catcode `\\12\catcode `\$12\catcode `\&12\catcode `\#12\catcode `\^12\catcode `\_12\catcode `\%12\relax}%
\providecommand \@@startlink[1]{}%
\providecommand \@@endlink[0]{}%
\providecommand \url  [0]{\begingroup\@sanitize@url \@url }%
\providecommand \@url [1]{\endgroup\@href {#1}{\urlprefix }}%
\providecommand \urlprefix  [0]{URL }%
\providecommand \Eprint [0]{\href }%
\providecommand \doibase [0]{https://doi.org/}%
\providecommand \selectlanguage [0]{\@gobble}%
\providecommand \bibinfo  [0]{\@secondoftwo}%
\providecommand \bibfield  [0]{\@secondoftwo}%
\providecommand \translation [1]{[#1]}%
\providecommand \BibitemOpen [0]{}%
\providecommand \bibitemStop [0]{}%
\providecommand \bibitemNoStop [0]{.\EOS\space}%
\providecommand \EOS [0]{\spacefactor3000\relax}%
\providecommand \BibitemShut  [1]{\csname bibitem#1\endcsname}%
\let\auto@bib@innerbib\@empty
\bibitem [{\citenamefont {Choi}\ \emph {et~al.}(2017)\citenamefont {Choi}, \citenamefont {Choudhary}, \citenamefont {Han}, \citenamefont {Park}, \citenamefont {Akinwande},\ and\ \citenamefont {Lee}}]{choi2017recent}%
  \BibitemOpen
  \bibfield  {author} {\bibinfo {author} {\bibfnamefont {W.}~\bibnamefont {Choi}}, \bibinfo {author} {\bibfnamefont {N.}~\bibnamefont {Choudhary}}, \bibinfo {author} {\bibfnamefont {G.~H.}\ \bibnamefont {Han}}, \bibinfo {author} {\bibfnamefont {J.}~\bibnamefont {Park}}, \bibinfo {author} {\bibfnamefont {D.}~\bibnamefont {Akinwande}},\ and\ \bibinfo {author} {\bibfnamefont {Y.~H.}\ \bibnamefont {Lee}},\ }\bibfield  {title} {\bibinfo {title} {Recent development of two-dimensional transition metal dichalcogenides and their applications},\ }\href@noop {} {\bibfield  {journal} {\bibinfo  {journal} {Materials Today}\ }\textbf {\bibinfo {volume} {20}},\ \bibinfo {pages} {116} (\bibinfo {year} {2017})}\BibitemShut {NoStop}%
\bibitem [{\citenamefont {Alfalasi}\ \emph {et~al.}(2023)\citenamefont {Alfalasi}, \citenamefont {Feng},\ and\ \citenamefont {Tit}}]{alfalasi2023designing}%
  \BibitemOpen
  \bibfield  {author} {\bibinfo {author} {\bibfnamefont {W.}~\bibnamefont {Alfalasi}}, \bibinfo {author} {\bibfnamefont {Y.~P.}\ \bibnamefont {Feng}},\ and\ \bibinfo {author} {\bibfnamefont {N.}~\bibnamefont {Tit}},\ }\bibfield  {title} {\bibinfo {title} {Designing a functionalized 2d-tmd (mox$_2$, x= s, se) hosting half-metallicity for selective gas-sensing applications: Atomic-scale study},\ }\href@noop {} {\bibfield  {journal} {\bibinfo  {journal} {Acta Materialia}\ }\textbf {\bibinfo {volume} {246}},\ \bibinfo {pages} {118655} (\bibinfo {year} {2023})}\BibitemShut {NoStop}%
\bibitem [{\citenamefont {Butler}\ \emph {et~al.}(2013)\citenamefont {Butler}, \citenamefont {Hollen}, \citenamefont {Cao}, \citenamefont {Cui}, \citenamefont {Gupta}, \citenamefont {Guti{\'e}rrez}, \citenamefont {Heinz}, \citenamefont {Hong}, \citenamefont {Huang}, \citenamefont {Ismach} \emph {et~al.}}]{butler2013progress}%
  \BibitemOpen
  \bibfield  {author} {\bibinfo {author} {\bibfnamefont {S.~Z.}\ \bibnamefont {Butler}}, \bibinfo {author} {\bibfnamefont {S.~M.}\ \bibnamefont {Hollen}}, \bibinfo {author} {\bibfnamefont {L.}~\bibnamefont {Cao}}, \bibinfo {author} {\bibfnamefont {Y.}~\bibnamefont {Cui}}, \bibinfo {author} {\bibfnamefont {J.~A.}\ \bibnamefont {Gupta}}, \bibinfo {author} {\bibfnamefont {H.~R.}\ \bibnamefont {Guti{\'e}rrez}}, \bibinfo {author} {\bibfnamefont {T.~F.}\ \bibnamefont {Heinz}}, \bibinfo {author} {\bibfnamefont {S.~S.}\ \bibnamefont {Hong}}, \bibinfo {author} {\bibfnamefont {J.}~\bibnamefont {Huang}}, \bibinfo {author} {\bibfnamefont {A.~F.}\ \bibnamefont {Ismach}}, \emph {et~al.},\ }\bibfield  {title} {\bibinfo {title} {Progress, challenges, and opportunities in two-dimensional materials beyond graphene},\ }\href@noop {} {\bibfield  {journal} {\bibinfo  {journal} {ACS nano}\ }\textbf {\bibinfo {volume} {7}},\ \bibinfo {pages} {2898} (\bibinfo {year} {2013})}\BibitemShut {NoStop}%
\bibitem [{\citenamefont {Wang}\ \emph {et~al.}(2012)\citenamefont {Wang}, \citenamefont {Kalantar-Zadeh}, \citenamefont {Kis}, \citenamefont {Coleman},\ and\ \citenamefont {Strano}}]{wang2012electronics}%
  \BibitemOpen
  \bibfield  {author} {\bibinfo {author} {\bibfnamefont {Q.~H.}\ \bibnamefont {Wang}}, \bibinfo {author} {\bibfnamefont {K.}~\bibnamefont {Kalantar-Zadeh}}, \bibinfo {author} {\bibfnamefont {A.}~\bibnamefont {Kis}}, \bibinfo {author} {\bibfnamefont {J.~N.}\ \bibnamefont {Coleman}},\ and\ \bibinfo {author} {\bibfnamefont {M.~S.}\ \bibnamefont {Strano}},\ }\bibfield  {title} {\bibinfo {title} {Electronics and optoelectronics of two-dimensional transition metal dichalcogenides},\ }\href@noop {} {\bibfield  {journal} {\bibinfo  {journal} {Nature nanotechnology}\ }\textbf {\bibinfo {volume} {7}},\ \bibinfo {pages} {699} (\bibinfo {year} {2012})}\BibitemShut {NoStop}%
\bibitem [{\citenamefont {Mu}\ \emph {et~al.}(2015)\citenamefont {Mu}, \citenamefont {Wang}, \citenamefont {Yuan}, \citenamefont {Xiao}, \citenamefont {Chen}, \citenamefont {Chen}, \citenamefont {Chen}, \citenamefont {Song}, \citenamefont {Wang}, \citenamefont {Xue} \emph {et~al.}}]{mu2015graphene}%
  \BibitemOpen
  \bibfield  {author} {\bibinfo {author} {\bibfnamefont {H.}~\bibnamefont {Mu}}, \bibinfo {author} {\bibfnamefont {Z.}~\bibnamefont {Wang}}, \bibinfo {author} {\bibfnamefont {J.}~\bibnamefont {Yuan}}, \bibinfo {author} {\bibfnamefont {S.}~\bibnamefont {Xiao}}, \bibinfo {author} {\bibfnamefont {C.}~\bibnamefont {Chen}}, \bibinfo {author} {\bibfnamefont {Y.}~\bibnamefont {Chen}}, \bibinfo {author} {\bibfnamefont {Y.}~\bibnamefont {Chen}}, \bibinfo {author} {\bibfnamefont {J.}~\bibnamefont {Song}}, \bibinfo {author} {\bibfnamefont {Y.}~\bibnamefont {Wang}}, \bibinfo {author} {\bibfnamefont {Y.}~\bibnamefont {Xue}}, \emph {et~al.},\ }\bibfield  {title} {\bibinfo {title} {Graphene-bi2te3 heterostructure as saturable absorber for short pulse generation},\ }\href@noop {} {\bibfield  {journal} {\bibinfo  {journal} {Acs Photonics}\ }\textbf {\bibinfo {volume} {2}},\ \bibinfo {pages} {832} (\bibinfo {year} {2015})}\BibitemShut {NoStop}%
\bibitem [{\citenamefont {Liu}\ \emph {et~al.}(2020)\citenamefont {Liu}, \citenamefont {Hu}, \citenamefont {Zhang}, \citenamefont {Li}, \citenamefont {Gao}, \citenamefont {Tian}, \citenamefont {Zhou}, \citenamefont {Zhang}, \citenamefont {Tang}, \citenamefont {Zhang} \emph {et~al.}}]{liu2020mos}%
  \BibitemOpen
  \bibfield  {author} {\bibinfo {author} {\bibfnamefont {J.}~\bibnamefont {Liu}}, \bibinfo {author} {\bibfnamefont {Z.}~\bibnamefont {Hu}}, \bibinfo {author} {\bibfnamefont {Y.}~\bibnamefont {Zhang}}, \bibinfo {author} {\bibfnamefont {H.-Y.}\ \bibnamefont {Li}}, \bibinfo {author} {\bibfnamefont {N.}~\bibnamefont {Gao}}, \bibinfo {author} {\bibfnamefont {Z.}~\bibnamefont {Tian}}, \bibinfo {author} {\bibfnamefont {L.}~\bibnamefont {Zhou}}, \bibinfo {author} {\bibfnamefont {B.}~\bibnamefont {Zhang}}, \bibinfo {author} {\bibfnamefont {J.}~\bibnamefont {Tang}}, \bibinfo {author} {\bibfnamefont {J.}~\bibnamefont {Zhang}}, \emph {et~al.},\ }\bibfield  {title} {\bibinfo {title} {Mos$_2$ nanosheets sensitized with quantum dots for room-temperature gas sensors},\ }\href@noop {} {\bibfield  {journal} {\bibinfo  {journal} {Nano-Micro Letters}\ }\textbf {\bibinfo {volume} {12}},\ \bibinfo {pages} {1} (\bibinfo {year} {2020})}\BibitemShut {NoStop}%
\bibitem [{\citenamefont {Li}\ \emph {et~al.}(2020)\citenamefont {Li}, \citenamefont {Gong}, \citenamefont {Chen}, \citenamefont {Lin}, \citenamefont {Khan}, \citenamefont {Zhang}, \citenamefont {Li}, \citenamefont {Zhang},\ and\ \citenamefont {Xie}}]{li2020recent}%
  \BibitemOpen
  \bibfield  {author} {\bibinfo {author} {\bibfnamefont {D.}~\bibnamefont {Li}}, \bibinfo {author} {\bibfnamefont {Y.}~\bibnamefont {Gong}}, \bibinfo {author} {\bibfnamefont {Y.}~\bibnamefont {Chen}}, \bibinfo {author} {\bibfnamefont {J.}~\bibnamefont {Lin}}, \bibinfo {author} {\bibfnamefont {Q.}~\bibnamefont {Khan}}, \bibinfo {author} {\bibfnamefont {Y.}~\bibnamefont {Zhang}}, \bibinfo {author} {\bibfnamefont {Y.}~\bibnamefont {Li}}, \bibinfo {author} {\bibfnamefont {H.}~\bibnamefont {Zhang}},\ and\ \bibinfo {author} {\bibfnamefont {H.}~\bibnamefont {Xie}},\ }\bibfield  {title} {\bibinfo {title} {Recent progress of two-dimensional thermoelectric materials},\ }\href@noop {} {\bibfield  {journal} {\bibinfo  {journal} {Nano-Micro Letters}\ }\textbf {\bibinfo {volume} {12}},\ \bibinfo {pages} {1} (\bibinfo {year} {2020})}\BibitemShut {NoStop}%
\bibitem [{\citenamefont {Kang}\ \emph {et~al.}(2019)\citenamefont {Kang}, \citenamefont {Cheng}, \citenamefont {Zheng}, \citenamefont {Cheng}, \citenamefont {Chen}, \citenamefont {Li}, \citenamefont {Tan}, \citenamefont {Xiong}, \citenamefont {Zhai},\ and\ \citenamefont {Gao}}]{kang2019mos}%
  \BibitemOpen
  \bibfield  {author} {\bibinfo {author} {\bibfnamefont {Z.}~\bibnamefont {Kang}}, \bibinfo {author} {\bibfnamefont {Y.}~\bibnamefont {Cheng}}, \bibinfo {author} {\bibfnamefont {Z.}~\bibnamefont {Zheng}}, \bibinfo {author} {\bibfnamefont {F.}~\bibnamefont {Cheng}}, \bibinfo {author} {\bibfnamefont {Z.}~\bibnamefont {Chen}}, \bibinfo {author} {\bibfnamefont {L.}~\bibnamefont {Li}}, \bibinfo {author} {\bibfnamefont {X.}~\bibnamefont {Tan}}, \bibinfo {author} {\bibfnamefont {L.}~\bibnamefont {Xiong}}, \bibinfo {author} {\bibfnamefont {T.}~\bibnamefont {Zhai}},\ and\ \bibinfo {author} {\bibfnamefont {Y.}~\bibnamefont {Gao}},\ }\bibfield  {title} {\bibinfo {title} {Mos$_2$-based photodetectors powered by asymmetric contact structure with large work function difference},\ }\href@noop {} {\bibfield  {journal} {\bibinfo  {journal} {Nano-micro letters}\ }\textbf {\bibinfo {volume} {11}},\ \bibinfo {pages} {1} (\bibinfo {year} {2019})}\BibitemShut {NoStop}%
\bibitem [{\citenamefont {Mia}\ \emph {et~al.}(2023)\citenamefont {Mia}, \citenamefont {Meyyappan},\ and\ \citenamefont {Giri}}]{mia2023two}%
  \BibitemOpen
  \bibfield  {author} {\bibinfo {author} {\bibfnamefont {A.~K.}\ \bibnamefont {Mia}}, \bibinfo {author} {\bibfnamefont {M.}~\bibnamefont {Meyyappan}},\ and\ \bibinfo {author} {\bibfnamefont {P.}~\bibnamefont {Giri}},\ }\bibfield  {title} {\bibinfo {title} {Two-dimensional transition metal dichalcogenide based biosensors: From fundamentals to healthcare applications},\ }\href@noop {} {\bibfield  {journal} {\bibinfo  {journal} {Biosensors}\ }\textbf {\bibinfo {volume} {13}},\ \bibinfo {pages} {169} (\bibinfo {year} {2023})}\BibitemShut {NoStop}%
\bibitem [{\citenamefont {Shi}\ \emph {et~al.}(2015)\citenamefont {Shi}, \citenamefont {Li},\ and\ \citenamefont {Li}}]{shi2015recent}%
  \BibitemOpen
  \bibfield  {author} {\bibinfo {author} {\bibfnamefont {Y.}~\bibnamefont {Shi}}, \bibinfo {author} {\bibfnamefont {H.}~\bibnamefont {Li}},\ and\ \bibinfo {author} {\bibfnamefont {L.-J.}\ \bibnamefont {Li}},\ }\bibfield  {title} {\bibinfo {title} {Recent advances in controlled synthesis of two-dimensional transition metal dichalcogenides via vapour deposition techniques},\ }\href@noop {} {\bibfield  {journal} {\bibinfo  {journal} {Chemical Society Reviews}\ }\textbf {\bibinfo {volume} {44}},\ \bibinfo {pages} {2744} (\bibinfo {year} {2015})}\BibitemShut {NoStop}%
\bibitem [{\citenamefont {Pi}\ \emph {et~al.}(2019)\citenamefont {Pi}, \citenamefont {Li}, \citenamefont {Liu}, \citenamefont {Zhang}, \citenamefont {Li},\ and\ \citenamefont {Zhai}}]{pi2019recent}%
  \BibitemOpen
  \bibfield  {author} {\bibinfo {author} {\bibfnamefont {L.}~\bibnamefont {Pi}}, \bibinfo {author} {\bibfnamefont {L.}~\bibnamefont {Li}}, \bibinfo {author} {\bibfnamefont {K.}~\bibnamefont {Liu}}, \bibinfo {author} {\bibfnamefont {Q.}~\bibnamefont {Zhang}}, \bibinfo {author} {\bibfnamefont {H.}~\bibnamefont {Li}},\ and\ \bibinfo {author} {\bibfnamefont {T.}~\bibnamefont {Zhai}},\ }\bibfield  {title} {\bibinfo {title} {Recent progress on 2d noble-transition-metal dichalcogenides},\ }\href@noop {} {\bibfield  {journal} {\bibinfo  {journal} {Advanced Functional Materials}\ }\textbf {\bibinfo {volume} {29}},\ \bibinfo {pages} {1904932} (\bibinfo {year} {2019})}\BibitemShut {NoStop}%
\bibitem [{\citenamefont {Oyedele}\ \emph {et~al.}(2017)\citenamefont {Oyedele}, \citenamefont {Yang}, \citenamefont {Liang}, \citenamefont {Puretzky}, \citenamefont {Wang}, \citenamefont {Zhang}, \citenamefont {Yu}, \citenamefont {Pudasaini}, \citenamefont {Ghosh}, \citenamefont {Liu} \emph {et~al.}}]{oyedele2017pdse2}%
  \BibitemOpen
  \bibfield  {author} {\bibinfo {author} {\bibfnamefont {A.~D.}\ \bibnamefont {Oyedele}}, \bibinfo {author} {\bibfnamefont {S.}~\bibnamefont {Yang}}, \bibinfo {author} {\bibfnamefont {L.}~\bibnamefont {Liang}}, \bibinfo {author} {\bibfnamefont {A.~A.}\ \bibnamefont {Puretzky}}, \bibinfo {author} {\bibfnamefont {K.}~\bibnamefont {Wang}}, \bibinfo {author} {\bibfnamefont {J.}~\bibnamefont {Zhang}}, \bibinfo {author} {\bibfnamefont {P.}~\bibnamefont {Yu}}, \bibinfo {author} {\bibfnamefont {P.~R.}\ \bibnamefont {Pudasaini}}, \bibinfo {author} {\bibfnamefont {A.~W.}\ \bibnamefont {Ghosh}}, \bibinfo {author} {\bibfnamefont {Z.}~\bibnamefont {Liu}}, \emph {et~al.},\ }\bibfield  {title} {\bibinfo {title} {Pdse$_2$: pentagonal two-dimensional layers with high air stability for electronics},\ }\href@noop {} {\bibfield  {journal} {\bibinfo  {journal} {Journal of the American Chemical Society}\ }\textbf {\bibinfo {volume} {139}},\ \bibinfo {pages} {14090} (\bibinfo {year} {2017})}\BibitemShut {NoStop}%
\bibitem [{\citenamefont {Gong}\ \emph {et~al.}(2020)\citenamefont {Gong}, \citenamefont {Lin}, \citenamefont {Chen}, \citenamefont {Khan}, \citenamefont {Wang}, \citenamefont {Zhang}, \citenamefont {Nie}, \citenamefont {Xie},\ and\ \citenamefont {Li}}]{gong2020two}%
  \BibitemOpen
  \bibfield  {author} {\bibinfo {author} {\bibfnamefont {Y.}~\bibnamefont {Gong}}, \bibinfo {author} {\bibfnamefont {Z.}~\bibnamefont {Lin}}, \bibinfo {author} {\bibfnamefont {Y.-X.}\ \bibnamefont {Chen}}, \bibinfo {author} {\bibfnamefont {Q.}~\bibnamefont {Khan}}, \bibinfo {author} {\bibfnamefont {C.}~\bibnamefont {Wang}}, \bibinfo {author} {\bibfnamefont {B.}~\bibnamefont {Zhang}}, \bibinfo {author} {\bibfnamefont {G.}~\bibnamefont {Nie}}, \bibinfo {author} {\bibfnamefont {N.}~\bibnamefont {Xie}},\ and\ \bibinfo {author} {\bibfnamefont {D.}~\bibnamefont {Li}},\ }\bibfield  {title} {\bibinfo {title} {Two-dimensional platinum diselenide: synthesis, emerging applications, and future challenges},\ }\href@noop {} {\bibfield  {journal} {\bibinfo  {journal} {Nano-Micro Letters}\ }\textbf {\bibinfo {volume} {12}},\ \bibinfo {pages} {1} (\bibinfo {year} {2020})}\BibitemShut {NoStop}%
\bibitem [{\citenamefont {Setiyawati}\ \emph {et~al.}(2019)\citenamefont {Setiyawati}, \citenamefont {Chiang}, \citenamefont {Ho},\ and\ \citenamefont {Tang}}]{setiyawati2019distinct}%
  \BibitemOpen
  \bibfield  {author} {\bibinfo {author} {\bibfnamefont {I.}~\bibnamefont {Setiyawati}}, \bibinfo {author} {\bibfnamefont {K.-R.}\ \bibnamefont {Chiang}}, \bibinfo {author} {\bibfnamefont {H.-M.}\ \bibnamefont {Ho}},\ and\ \bibinfo {author} {\bibfnamefont {Y.-H.}\ \bibnamefont {Tang}},\ }\bibfield  {title} {\bibinfo {title} {Distinct electronic and transport properties between 1t-hfse$_2$ and 1t-ptse$_2$},\ }\href@noop {} {\bibfield  {journal} {\bibinfo  {journal} {Chinese Journal of Physics}\ }\textbf {\bibinfo {volume} {62}},\ \bibinfo {pages} {151} (\bibinfo {year} {2019})}\BibitemShut {NoStop}%
\bibitem [{\citenamefont {Wagner}\ \emph {et~al.}(2018)\citenamefont {Wagner}, \citenamefont {Yim}, \citenamefont {McEvoy}, \citenamefont {Kataria}, \citenamefont {Yokaribas}, \citenamefont {Kuc}, \citenamefont {Pindl}, \citenamefont {Fritzen}, \citenamefont {Heine}, \citenamefont {Duesberg} \emph {et~al.}}]{wagner2018highly}%
  \BibitemOpen
  \bibfield  {author} {\bibinfo {author} {\bibfnamefont {S.}~\bibnamefont {Wagner}}, \bibinfo {author} {\bibfnamefont {C.}~\bibnamefont {Yim}}, \bibinfo {author} {\bibfnamefont {N.}~\bibnamefont {McEvoy}}, \bibinfo {author} {\bibfnamefont {S.}~\bibnamefont {Kataria}}, \bibinfo {author} {\bibfnamefont {V.}~\bibnamefont {Yokaribas}}, \bibinfo {author} {\bibfnamefont {A.}~\bibnamefont {Kuc}}, \bibinfo {author} {\bibfnamefont {S.}~\bibnamefont {Pindl}}, \bibinfo {author} {\bibfnamefont {C.-P.}\ \bibnamefont {Fritzen}}, \bibinfo {author} {\bibfnamefont {T.}~\bibnamefont {Heine}}, \bibinfo {author} {\bibfnamefont {G.~S.}\ \bibnamefont {Duesberg}}, \emph {et~al.},\ }\bibfield  {title} {\bibinfo {title} {Highly sensitive electromechanical piezoresistive pressure sensors based on large-area layered ptse$_2$ films},\ }\href@noop {} {\bibfield  {journal} {\bibinfo  {journal} {Nano letters}\ }\textbf {\bibinfo {volume} {18}},\ \bibinfo {pages} {3738} (\bibinfo {year} {2018})}\BibitemShut {NoStop}%
\bibitem [{\citenamefont {Wang}\ \emph {et~al.}(2021)\citenamefont {Wang}, \citenamefont {Wang}, \citenamefont {McEvoy}, \citenamefont {Fan},\ and\ \citenamefont {Blau}}]{wang2021layered}%
  \BibitemOpen
  \bibfield  {author} {\bibinfo {author} {\bibfnamefont {G.}~\bibnamefont {Wang}}, \bibinfo {author} {\bibfnamefont {Z.}~\bibnamefont {Wang}}, \bibinfo {author} {\bibfnamefont {N.}~\bibnamefont {McEvoy}}, \bibinfo {author} {\bibfnamefont {P.}~\bibnamefont {Fan}},\ and\ \bibinfo {author} {\bibfnamefont {W.~J.}\ \bibnamefont {Blau}},\ }\bibfield  {title} {\bibinfo {title} {Layered ptse$_2$ for sensing, photonic, and (opto-)electronic applications},\ }\href@noop {} {\bibfield  {journal} {\bibinfo  {journal} {Advanced Materials}\ }\textbf {\bibinfo {volume} {33}},\ \bibinfo {pages} {2004070} (\bibinfo {year} {2021})}\BibitemShut {NoStop}%
\bibitem [{\citenamefont {Lei}\ \emph {et~al.}(2023)\citenamefont {Lei}, \citenamefont {Hu}, \citenamefont {Han}, \citenamefont {Yuan}, \citenamefont {Jiao}, \citenamefont {Luo},\ and\ \citenamefont {Liu}}]{lei2023directional}%
  \BibitemOpen
  \bibfield  {author} {\bibinfo {author} {\bibfnamefont {W.}~\bibnamefont {Lei}}, \bibinfo {author} {\bibfnamefont {R.}~\bibnamefont {Hu}}, \bibinfo {author} {\bibfnamefont {S.}~\bibnamefont {Han}}, \bibinfo {author} {\bibfnamefont {H.}~\bibnamefont {Yuan}}, \bibinfo {author} {\bibfnamefont {W.}~\bibnamefont {Jiao}}, \bibinfo {author} {\bibfnamefont {Y.}~\bibnamefont {Luo}},\ and\ \bibinfo {author} {\bibfnamefont {H.}~\bibnamefont {Liu}},\ }\bibfield  {title} {\bibinfo {title} {Directional control of the electronic and phonon transport properties in the ferroelastic ptse$_2$},\ }\href@noop {} {\bibfield  {journal} {\bibinfo  {journal} {The Journal of Physical Chemistry C}\ } (\bibinfo {year} {2023})}\BibitemShut {NoStop}%
\bibitem [{\citenamefont {Hemmat}\ \emph {et~al.}(2023)\citenamefont {Hemmat}, \citenamefont {Ayari}, \citenamefont {Mi{\v{c}}ica}, \citenamefont {Vergnet}, \citenamefont {Guo}, \citenamefont {Arfaoui}, \citenamefont {Yu}, \citenamefont {Vala}, \citenamefont {Wright}, \citenamefont {Postava} \emph {et~al.}}]{hemmat2023layer}%
  \BibitemOpen
  \bibfield  {author} {\bibinfo {author} {\bibfnamefont {M.}~\bibnamefont {Hemmat}}, \bibinfo {author} {\bibfnamefont {S.}~\bibnamefont {Ayari}}, \bibinfo {author} {\bibfnamefont {M.}~\bibnamefont {Mi{\v{c}}ica}}, \bibinfo {author} {\bibfnamefont {H.}~\bibnamefont {Vergnet}}, \bibinfo {author} {\bibfnamefont {S.}~\bibnamefont {Guo}}, \bibinfo {author} {\bibfnamefont {M.}~\bibnamefont {Arfaoui}}, \bibinfo {author} {\bibfnamefont {X.}~\bibnamefont {Yu}}, \bibinfo {author} {\bibfnamefont {D.}~\bibnamefont {Vala}}, \bibinfo {author} {\bibfnamefont {A.}~\bibnamefont {Wright}}, \bibinfo {author} {\bibfnamefont {K.}~\bibnamefont {Postava}}, \emph {et~al.},\ }\bibfield  {title} {\bibinfo {title} {Layer-controlled nonlinear terahertz valleytronics in two-dimensional semimetal and semiconductor ptse$_2$},\ }\href@noop {} {\bibfield  {journal} {\bibinfo  {journal} {InfoMat}\ ,\ \bibinfo {pages} {e12468}} (\bibinfo {year} {2023})}\BibitemShut {NoStop}%
\bibitem [{\citenamefont {Tharrault}\ \emph {et~al.}(2023)\citenamefont {Tharrault}, \citenamefont {Ayari}, \citenamefont {Desgu{\'e}}, \citenamefont {Arfaoui}, \citenamefont {Goff}, \citenamefont {Morfin}, \citenamefont {Palomo}, \citenamefont {Rosticher}, \citenamefont {Jaziri}, \citenamefont {Pla{\c{c}}ais} \emph {et~al.}}]{tharrault2023optical}%
  \BibitemOpen
  \bibfield  {author} {\bibinfo {author} {\bibfnamefont {M.}~\bibnamefont {Tharrault}}, \bibinfo {author} {\bibfnamefont {S.}~\bibnamefont {Ayari}}, \bibinfo {author} {\bibfnamefont {E.}~\bibnamefont {Desgu{\'e}}}, \bibinfo {author} {\bibfnamefont {M.}~\bibnamefont {Arfaoui}}, \bibinfo {author} {\bibfnamefont {R.~L.}\ \bibnamefont {Goff}}, \bibinfo {author} {\bibfnamefont {P.}~\bibnamefont {Morfin}}, \bibinfo {author} {\bibfnamefont {J.}~\bibnamefont {Palomo}}, \bibinfo {author} {\bibfnamefont {M.}~\bibnamefont {Rosticher}}, \bibinfo {author} {\bibfnamefont {S.}~\bibnamefont {Jaziri}}, \bibinfo {author} {\bibfnamefont {B.}~\bibnamefont {Pla{\c{c}}ais}}, \emph {et~al.},\ }\bibfield  {title} {\bibinfo {title} {The optical absorption in indirect semiconductor to semimetal ptse$_2$ arises from direct transitions},\ }\href@noop {} {\bibfield  {journal} {\bibinfo  {journal} {arXiv preprint arXiv:2311.01847}\ } (\bibinfo {year} {2023})}\BibitemShut {NoStop}%
\bibitem [{\citenamefont {Zhao}\ \emph {et~al.}(2017)\citenamefont {Zhao}, \citenamefont {Qiao}, \citenamefont {Yu}, \citenamefont {Yu}, \citenamefont {Xu}, \citenamefont {Lau}, \citenamefont {Zhou}, \citenamefont {Liu}, \citenamefont {Wang}, \citenamefont {Ji} \emph {et~al.}}]{zhao2017high}%
  \BibitemOpen
  \bibfield  {author} {\bibinfo {author} {\bibfnamefont {Y.}~\bibnamefont {Zhao}}, \bibinfo {author} {\bibfnamefont {J.}~\bibnamefont {Qiao}}, \bibinfo {author} {\bibfnamefont {Z.}~\bibnamefont {Yu}}, \bibinfo {author} {\bibfnamefont {P.}~\bibnamefont {Yu}}, \bibinfo {author} {\bibfnamefont {K.}~\bibnamefont {Xu}}, \bibinfo {author} {\bibfnamefont {S.~P.}\ \bibnamefont {Lau}}, \bibinfo {author} {\bibfnamefont {W.}~\bibnamefont {Zhou}}, \bibinfo {author} {\bibfnamefont {Z.}~\bibnamefont {Liu}}, \bibinfo {author} {\bibfnamefont {X.}~\bibnamefont {Wang}}, \bibinfo {author} {\bibfnamefont {W.}~\bibnamefont {Ji}}, \emph {et~al.},\ }\bibfield  {title} {\bibinfo {title} {High-electron-mobility and air-stable 2d layered ptse$_2$ fets},\ }\href@noop {} {\bibfield  {journal} {\bibinfo  {journal} {Advanced Materials}\ }\textbf {\bibinfo {volume} {29}},\ \bibinfo {pages} {1604230} (\bibinfo {year} {2017})}\BibitemShut {NoStop}%
\bibitem [{\citenamefont {AlMutairi}\ \emph {et~al.}(2017)\citenamefont {AlMutairi}, \citenamefont {Yin},\ and\ \citenamefont {Yoon}}]{almutairi2017ptse}%
  \BibitemOpen
  \bibfield  {author} {\bibinfo {author} {\bibfnamefont {A.}~\bibnamefont {AlMutairi}}, \bibinfo {author} {\bibfnamefont {D.}~\bibnamefont {Yin}},\ and\ \bibinfo {author} {\bibfnamefont {Y.}~\bibnamefont {Yoon}},\ }\bibfield  {title} {\bibinfo {title} {Ptse$_2$ field-effect transistors: New opportunities for electronic devices},\ }\href@noop {} {\bibfield  {journal} {\bibinfo  {journal} {IEEE Electron Device Letters}\ }\textbf {\bibinfo {volume} {39}},\ \bibinfo {pages} {151} (\bibinfo {year} {2017})}\BibitemShut {NoStop}%
\bibitem [{\citenamefont {Wu}\ \emph {et~al.}(2018{\natexlab{a}})\citenamefont {Wu}, \citenamefont {Wang}, \citenamefont {Zeng}, \citenamefont {Jia}, \citenamefont {Wu}, \citenamefont {Xu}, \citenamefont {Shi}, \citenamefont {Tian}, \citenamefont {Li},\ and\ \citenamefont {Tsang}}]{wu2018design}%
  \BibitemOpen
  \bibfield  {author} {\bibinfo {author} {\bibfnamefont {D.}~\bibnamefont {Wu}}, \bibinfo {author} {\bibfnamefont {Y.}~\bibnamefont {Wang}}, \bibinfo {author} {\bibfnamefont {L.}~\bibnamefont {Zeng}}, \bibinfo {author} {\bibfnamefont {C.}~\bibnamefont {Jia}}, \bibinfo {author} {\bibfnamefont {E.}~\bibnamefont {Wu}}, \bibinfo {author} {\bibfnamefont {T.}~\bibnamefont {Xu}}, \bibinfo {author} {\bibfnamefont {Z.}~\bibnamefont {Shi}}, \bibinfo {author} {\bibfnamefont {Y.}~\bibnamefont {Tian}}, \bibinfo {author} {\bibfnamefont {X.}~\bibnamefont {Li}},\ and\ \bibinfo {author} {\bibfnamefont {Y.~H.}\ \bibnamefont {Tsang}},\ }\bibfield  {title} {\bibinfo {title} {Design of 2d layered ptse$_2$ heterojunction for the high-performance, room-temperature, broadband, infrared photodetector},\ }\href@noop {} {\bibfield  {journal} {\bibinfo  {journal} {Acs Photonics}\ }\textbf {\bibinfo {volume} {5}},\ \bibinfo {pages} {3820} (\bibinfo {year} {2018}{\natexlab{a}})}\BibitemShut {NoStop}%
\bibitem [{\citenamefont {Li}\ \emph {et~al.}(2021)\citenamefont {Li}, \citenamefont {Kolekar}, \citenamefont {Ghorbani-Asl}, \citenamefont {Lehnert}, \citenamefont {Biskupek}, \citenamefont {Kaiser}, \citenamefont {Krasheninnikov},\ and\ \citenamefont {Batzill}}]{li2021layer}%
  \BibitemOpen
  \bibfield  {author} {\bibinfo {author} {\bibfnamefont {J.}~\bibnamefont {Li}}, \bibinfo {author} {\bibfnamefont {S.}~\bibnamefont {Kolekar}}, \bibinfo {author} {\bibfnamefont {M.}~\bibnamefont {Ghorbani-Asl}}, \bibinfo {author} {\bibfnamefont {T.}~\bibnamefont {Lehnert}}, \bibinfo {author} {\bibfnamefont {J.}~\bibnamefont {Biskupek}}, \bibinfo {author} {\bibfnamefont {U.}~\bibnamefont {Kaiser}}, \bibinfo {author} {\bibfnamefont {A.~V.}\ \bibnamefont {Krasheninnikov}},\ and\ \bibinfo {author} {\bibfnamefont {M.}~\bibnamefont {Batzill}},\ }\bibfield  {title} {\bibinfo {title} {Layer-dependent band gaps of platinum dichalcogenides},\ }\href@noop {} {\bibfield  {journal} {\bibinfo  {journal} {ACS Nano}\ }\textbf {\bibinfo {volume} {15}},\ \bibinfo {pages} {13249} (\bibinfo {year} {2021})}\BibitemShut {NoStop}%
\bibitem [{\citenamefont {Zhang}\ \emph {et~al.}(2021)\citenamefont {Zhang}, \citenamefont {Yang}, \citenamefont {Sahdan}, \citenamefont {Arramel}, \citenamefont {Xu}, \citenamefont {Xing}, \citenamefont {Feng}, \citenamefont {Zhang}, \citenamefont {Wang},\ and\ \citenamefont {Wee}}]{zhang2021precise}%
  \BibitemOpen
  \bibfield  {author} {\bibinfo {author} {\bibfnamefont {L.}~\bibnamefont {Zhang}}, \bibinfo {author} {\bibfnamefont {T.}~\bibnamefont {Yang}}, \bibinfo {author} {\bibfnamefont {M.~F.}\ \bibnamefont {Sahdan}}, \bibinfo {author} {\bibnamefont {Arramel}}, \bibinfo {author} {\bibfnamefont {W.}~\bibnamefont {Xu}}, \bibinfo {author} {\bibfnamefont {K.}~\bibnamefont {Xing}}, \bibinfo {author} {\bibfnamefont {Y.~P.}\ \bibnamefont {Feng}}, \bibinfo {author} {\bibfnamefont {W.}~\bibnamefont {Zhang}}, \bibinfo {author} {\bibfnamefont {Z.}~\bibnamefont {Wang}},\ and\ \bibinfo {author} {\bibfnamefont {A.~T.}\ \bibnamefont {Wee}},\ }\bibfield  {title} {\bibinfo {title} {Precise layer-dependent electronic structure of mbe-grown ptse$_2$},\ }\href@noop {} {\bibfield  {journal} {\bibinfo  {journal} {Advanced Electronic Materials}\ }\textbf {\bibinfo {volume} {7}},\ \bibinfo {pages} {2100559} (\bibinfo {year} {2021})}\BibitemShut {NoStop}%
\bibitem [{\citenamefont {Island}\ \emph {et~al.}(2016)\citenamefont {Island}, \citenamefont {Kuc}, \citenamefont {Diependaal}, \citenamefont {Bratschitsch}, \citenamefont {Van Der~Zant}, \citenamefont {Heine},\ and\ \citenamefont {Castellanos-Gomez}}]{island2016precise}%
  \BibitemOpen
  \bibfield  {author} {\bibinfo {author} {\bibfnamefont {J.~O.}\ \bibnamefont {Island}}, \bibinfo {author} {\bibfnamefont {A.}~\bibnamefont {Kuc}}, \bibinfo {author} {\bibfnamefont {E.~H.}\ \bibnamefont {Diependaal}}, \bibinfo {author} {\bibfnamefont {R.}~\bibnamefont {Bratschitsch}}, \bibinfo {author} {\bibfnamefont {H.~S.}\ \bibnamefont {Van Der~Zant}}, \bibinfo {author} {\bibfnamefont {T.}~\bibnamefont {Heine}},\ and\ \bibinfo {author} {\bibfnamefont {A.}~\bibnamefont {Castellanos-Gomez}},\ }\bibfield  {title} {\bibinfo {title} {Precise and reversible band gap tuning in single-layer mose$_2$ by uniaxial strain},\ }\href@noop {} {\bibfield  {journal} {\bibinfo  {journal} {Nanoscale}\ }\textbf {\bibinfo {volume} {8}},\ \bibinfo {pages} {2589} (\bibinfo {year} {2016})}\BibitemShut {NoStop}%
\bibitem [{\citenamefont {Kandemir}\ \emph {et~al.}(2018)\citenamefont {Kandemir}, \citenamefont {Akbali}, \citenamefont {Kahraman}, \citenamefont {Badalov}, \citenamefont {Ozcan}, \citenamefont {{\.I}yikanat},\ and\ \citenamefont {Sahin}}]{kandemir2018structural}%
  \BibitemOpen
  \bibfield  {author} {\bibinfo {author} {\bibfnamefont {A.}~\bibnamefont {Kandemir}}, \bibinfo {author} {\bibfnamefont {B.}~\bibnamefont {Akbali}}, \bibinfo {author} {\bibfnamefont {Z.}~\bibnamefont {Kahraman}}, \bibinfo {author} {\bibfnamefont {S.}~\bibnamefont {Badalov}}, \bibinfo {author} {\bibfnamefont {M.}~\bibnamefont {Ozcan}}, \bibinfo {author} {\bibfnamefont {F.}~\bibnamefont {{\.I}yikanat}},\ and\ \bibinfo {author} {\bibfnamefont {H.}~\bibnamefont {Sahin}},\ }\bibfield  {title} {\bibinfo {title} {Structural, electronic and phononic properties of ptse$_2$: from monolayer to bulk},\ }\href@noop {} {\bibfield  {journal} {\bibinfo  {journal} {Semiconductor Science and Technology}\ }\textbf {\bibinfo {volume} {33}},\ \bibinfo {pages} {085002} (\bibinfo {year} {2018})}\BibitemShut {NoStop}%
\bibitem [{\citenamefont {Yan}\ \emph {et~al.}(2017)\citenamefont {Yan}, \citenamefont {Wang}, \citenamefont {Zhou}, \citenamefont {Zhang}, \citenamefont {Zhang}, \citenamefont {Zhang}, \citenamefont {Yao}, \citenamefont {Lu}, \citenamefont {Yang}, \citenamefont {Wu} \emph {et~al.}}]{yan2017high}%
  \BibitemOpen
  \bibfield  {author} {\bibinfo {author} {\bibfnamefont {M.}~\bibnamefont {Yan}}, \bibinfo {author} {\bibfnamefont {E.}~\bibnamefont {Wang}}, \bibinfo {author} {\bibfnamefont {X.}~\bibnamefont {Zhou}}, \bibinfo {author} {\bibfnamefont {G.}~\bibnamefont {Zhang}}, \bibinfo {author} {\bibfnamefont {H.}~\bibnamefont {Zhang}}, \bibinfo {author} {\bibfnamefont {K.}~\bibnamefont {Zhang}}, \bibinfo {author} {\bibfnamefont {W.}~\bibnamefont {Yao}}, \bibinfo {author} {\bibfnamefont {N.}~\bibnamefont {Lu}}, \bibinfo {author} {\bibfnamefont {S.}~\bibnamefont {Yang}}, \bibinfo {author} {\bibfnamefont {S.}~\bibnamefont {Wu}}, \emph {et~al.},\ }\bibfield  {title} {\bibinfo {title} {High quality atomically thin ptse$_2$ films grown by molecular beam epitaxy},\ }\href@noop {} {\bibfield  {journal} {\bibinfo  {journal} {2D Materials}\ }\textbf {\bibinfo {volume} {4}},\ \bibinfo {pages} {045015} (\bibinfo {year} {2017})}\BibitemShut {NoStop}%
\bibitem [{\citenamefont {Ansari}\ \emph {et~al.}(2019)\citenamefont {Ansari}, \citenamefont {Monaghan}, \citenamefont {McEvoy}, \citenamefont {Coile{\'a}in}, \citenamefont {Cullen}, \citenamefont {Lin}, \citenamefont {Siris}, \citenamefont {Stimpel-Lindner}, \citenamefont {Burke}, \citenamefont {Mirabelli} \emph {et~al.}}]{ansari2019quantum}%
  \BibitemOpen
  \bibfield  {author} {\bibinfo {author} {\bibfnamefont {L.}~\bibnamefont {Ansari}}, \bibinfo {author} {\bibfnamefont {S.}~\bibnamefont {Monaghan}}, \bibinfo {author} {\bibfnamefont {N.}~\bibnamefont {McEvoy}}, \bibinfo {author} {\bibfnamefont {C.~{\'O}.}\ \bibnamefont {Coile{\'a}in}}, \bibinfo {author} {\bibfnamefont {C.~P.}\ \bibnamefont {Cullen}}, \bibinfo {author} {\bibfnamefont {J.}~\bibnamefont {Lin}}, \bibinfo {author} {\bibfnamefont {R.}~\bibnamefont {Siris}}, \bibinfo {author} {\bibfnamefont {T.}~\bibnamefont {Stimpel-Lindner}}, \bibinfo {author} {\bibfnamefont {K.~F.}\ \bibnamefont {Burke}}, \bibinfo {author} {\bibfnamefont {G.}~\bibnamefont {Mirabelli}}, \emph {et~al.},\ }\bibfield  {title} {\bibinfo {title} {Quantum confinement-induced semimetal-to-semiconductor evolution in large-area ultra-thin ptse$_2$ films grown at 400° c},\ }\href@noop {} {\bibfield  {journal} {\bibinfo  {journal} {npj 2D Materials and Applications}\ }\textbf {\bibinfo {volume} {3}},\ \bibinfo {pages} {33} (\bibinfo {year}
  {2019})}\BibitemShut {NoStop}%
\bibitem [{\citenamefont {Dai}\ \emph {et~al.}(2003)\citenamefont {Dai}, \citenamefont {Koo}, \citenamefont {Whangbo}, \citenamefont {Soulard}, \citenamefont {Rocquefelte},\ and\ \citenamefont {Jobic}}]{dai2003trends}%
  \BibitemOpen
  \bibfield  {author} {\bibinfo {author} {\bibfnamefont {D.}~\bibnamefont {Dai}}, \bibinfo {author} {\bibfnamefont {H.-J.}\ \bibnamefont {Koo}}, \bibinfo {author} {\bibfnamefont {M.-H.}\ \bibnamefont {Whangbo}}, \bibinfo {author} {\bibfnamefont {C.}~\bibnamefont {Soulard}}, \bibinfo {author} {\bibfnamefont {X.}~\bibnamefont {Rocquefelte}},\ and\ \bibinfo {author} {\bibfnamefont {S.}~\bibnamefont {Jobic}},\ }\bibfield  {title} {\bibinfo {title} {Trends in the structure and bonding in the layered platinum dioxide and dichalcogenides ptq$_2$ (q= o, s, se, te)},\ }\href@noop {} {\bibfield  {journal} {\bibinfo  {journal} {Journal of Solid State Chemistry}\ }\textbf {\bibinfo {volume} {173}},\ \bibinfo {pages} {114} (\bibinfo {year} {2003})}\BibitemShut {NoStop}%
\bibitem [{\citenamefont {Wang}\ \emph {et~al.}(2015)\citenamefont {Wang}, \citenamefont {Li}, \citenamefont {Yao}, \citenamefont {Song}, \citenamefont {Sun}, \citenamefont {Pan}, \citenamefont {Ren}, \citenamefont {Li}, \citenamefont {Okunishi}, \citenamefont {Wang} \emph {et~al.}}]{wang2015monolayer}%
  \BibitemOpen
  \bibfield  {author} {\bibinfo {author} {\bibfnamefont {Y.}~\bibnamefont {Wang}}, \bibinfo {author} {\bibfnamefont {L.}~\bibnamefont {Li}}, \bibinfo {author} {\bibfnamefont {W.}~\bibnamefont {Yao}}, \bibinfo {author} {\bibfnamefont {S.}~\bibnamefont {Song}}, \bibinfo {author} {\bibfnamefont {J.}~\bibnamefont {Sun}}, \bibinfo {author} {\bibfnamefont {J.}~\bibnamefont {Pan}}, \bibinfo {author} {\bibfnamefont {X.}~\bibnamefont {Ren}}, \bibinfo {author} {\bibfnamefont {C.}~\bibnamefont {Li}}, \bibinfo {author} {\bibfnamefont {E.}~\bibnamefont {Okunishi}}, \bibinfo {author} {\bibfnamefont {Y.-Q.}\ \bibnamefont {Wang}}, \emph {et~al.},\ }\bibfield  {title} {\bibinfo {title} {Monolayer ptse$_2$, a new semiconducting transition-metal-dichalcogenide, epitaxially grown by direct selenization of pt},\ }\href@noop {} {\bibfield  {journal} {\bibinfo  {journal} {Nano letters}\ }\textbf {\bibinfo {volume} {15}},\ \bibinfo {pages} {4013} (\bibinfo {year} {2015})}\BibitemShut {NoStop}%
\bibitem [{\citenamefont {O’Brien}\ \emph {et~al.}(2016)\citenamefont {O’Brien}, \citenamefont {McEvoy}, \citenamefont {Motta}, \citenamefont {Zheng}, \citenamefont {Berner}, \citenamefont {Kotakoski}, \citenamefont {Elibol}, \citenamefont {Pennycook}, \citenamefont {Meyer}, \citenamefont {Yim} \emph {et~al.}}]{o2016raman}%
  \BibitemOpen
  \bibfield  {author} {\bibinfo {author} {\bibfnamefont {M.}~\bibnamefont {O’Brien}}, \bibinfo {author} {\bibfnamefont {N.}~\bibnamefont {McEvoy}}, \bibinfo {author} {\bibfnamefont {C.}~\bibnamefont {Motta}}, \bibinfo {author} {\bibfnamefont {J.-Y.}\ \bibnamefont {Zheng}}, \bibinfo {author} {\bibfnamefont {N.~C.}\ \bibnamefont {Berner}}, \bibinfo {author} {\bibfnamefont {J.}~\bibnamefont {Kotakoski}}, \bibinfo {author} {\bibfnamefont {K.}~\bibnamefont {Elibol}}, \bibinfo {author} {\bibfnamefont {T.~J.}\ \bibnamefont {Pennycook}}, \bibinfo {author} {\bibfnamefont {J.~C.}\ \bibnamefont {Meyer}}, \bibinfo {author} {\bibfnamefont {C.}~\bibnamefont {Yim}}, \emph {et~al.},\ }\bibfield  {title} {\bibinfo {title} {Raman characterization of platinum diselenide thin films},\ }\href@noop {} {\bibfield  {journal} {\bibinfo  {journal} {2D Materials}\ }\textbf {\bibinfo {volume} {3}},\ \bibinfo {pages} {021004} (\bibinfo {year} {2016})}\BibitemShut {NoStop}%
\bibitem [{\citenamefont {Zulfiqar}\ \emph {et~al.}(2016)\citenamefont {Zulfiqar}, \citenamefont {Zhao}, \citenamefont {Li}, \citenamefont {Nazir},\ and\ \citenamefont {Ni}}]{zulfiqar2016tunable}%
  \BibitemOpen
  \bibfield  {author} {\bibinfo {author} {\bibfnamefont {M.}~\bibnamefont {Zulfiqar}}, \bibinfo {author} {\bibfnamefont {Y.}~\bibnamefont {Zhao}}, \bibinfo {author} {\bibfnamefont {G.}~\bibnamefont {Li}}, \bibinfo {author} {\bibfnamefont {S.}~\bibnamefont {Nazir}},\ and\ \bibinfo {author} {\bibfnamefont {J.}~\bibnamefont {Ni}},\ }\bibfield  {title} {\bibinfo {title} {Tunable conductivity and half metallic ferromagnetism in monolayer platinum diselenide: a first-principles study},\ }\href@noop {} {\bibfield  {journal} {\bibinfo  {journal} {The Journal of Physical Chemistry C}\ }\textbf {\bibinfo {volume} {120}},\ \bibinfo {pages} {25030} (\bibinfo {year} {2016})}\BibitemShut {NoStop}%
\bibitem [{\citenamefont {Ren}\ \emph {et~al.}(2019)\citenamefont {Ren}, \citenamefont {Sun}, \citenamefont {Luo}, \citenamefont {Wang}, \citenamefont {Xu}, \citenamefont {Yu},\ and\ \citenamefont {Tang}}]{ren2019electronic}%
  \BibitemOpen
  \bibfield  {author} {\bibinfo {author} {\bibfnamefont {K.}~\bibnamefont {Ren}}, \bibinfo {author} {\bibfnamefont {M.}~\bibnamefont {Sun}}, \bibinfo {author} {\bibfnamefont {Y.}~\bibnamefont {Luo}}, \bibinfo {author} {\bibfnamefont {S.}~\bibnamefont {Wang}}, \bibinfo {author} {\bibfnamefont {Y.}~\bibnamefont {Xu}}, \bibinfo {author} {\bibfnamefont {J.}~\bibnamefont {Yu}},\ and\ \bibinfo {author} {\bibfnamefont {W.}~\bibnamefont {Tang}},\ }\bibfield  {title} {\bibinfo {title} {Electronic and optical properties of van der waals vertical heterostructures based on two-dimensional transition metal dichalcogenides: First-principles calculations},\ }\href@noop {} {\bibfield  {journal} {\bibinfo  {journal} {Physics Letters A}\ }\textbf {\bibinfo {volume} {383}},\ \bibinfo {pages} {1487} (\bibinfo {year} {2019})}\BibitemShut {NoStop}%
\bibitem [{\citenamefont {Selamneni}\ and\ \citenamefont {Sahatiya}(2023)}]{selamneni2023mixed}%
  \BibitemOpen
  \bibfield  {author} {\bibinfo {author} {\bibfnamefont {V.}~\bibnamefont {Selamneni}}\ and\ \bibinfo {author} {\bibfnamefont {P.}~\bibnamefont {Sahatiya}},\ }\bibfield  {title} {\bibinfo {title} {Mixed dimensional transition metal dichalcogenides (tmds) vdw heterostructure based photodetectors: a review},\ }\href@noop {} {\bibfield  {journal} {\bibinfo  {journal} {Microelectronic Engineering}\ }\textbf {\bibinfo {volume} {269}},\ \bibinfo {pages} {111926} (\bibinfo {year} {2023})}\BibitemShut {NoStop}%
\bibitem [{\citenamefont {Khestanova}\ \emph {et~al.}(2023)\citenamefont {Khestanova}, \citenamefont {Ivanova}, \citenamefont {Gillen}, \citenamefont {D’Elia}, \citenamefont {Gallego~Lacey}, \citenamefont {Wysocki}, \citenamefont {Gruneis}, \citenamefont {Kravtsov}, \citenamefont {Strupinski}, \citenamefont {Maultzsch} \emph {et~al.}}]{khestanova2023robustness}%
  \BibitemOpen
  \bibfield  {author} {\bibinfo {author} {\bibfnamefont {E.}~\bibnamefont {Khestanova}}, \bibinfo {author} {\bibfnamefont {T.}~\bibnamefont {Ivanova}}, \bibinfo {author} {\bibfnamefont {R.}~\bibnamefont {Gillen}}, \bibinfo {author} {\bibfnamefont {A.}~\bibnamefont {D’Elia}}, \bibinfo {author} {\bibfnamefont {O.~N.}\ \bibnamefont {Gallego~Lacey}}, \bibinfo {author} {\bibfnamefont {L.}~\bibnamefont {Wysocki}}, \bibinfo {author} {\bibfnamefont {A.}~\bibnamefont {Gruneis}}, \bibinfo {author} {\bibfnamefont {V.}~\bibnamefont {Kravtsov}}, \bibinfo {author} {\bibfnamefont {W.}~\bibnamefont {Strupinski}}, \bibinfo {author} {\bibfnamefont {J.}~\bibnamefont {Maultzsch}}, \emph {et~al.},\ }\bibfield  {title} {\bibinfo {title} {Robustness of momentum-indirect interlayer excitons in mos$_2$/wse$_2$ heterostructure against charge carrier doping},\ }\href@noop {} {\bibfield  {journal} {\bibinfo  {journal} {ACS Photonics}\ }\textbf {\bibinfo {volume} {10}},\ \bibinfo {pages} {1159} (\bibinfo {year} {2023})}\BibitemShut
  {NoStop}%
\bibitem [{\citenamefont {Thompson}\ \emph {et~al.}(2023)\citenamefont {Thompson}, \citenamefont {Lumsargis}, \citenamefont {Feierabend}, \citenamefont {Zhao}, \citenamefont {Wang}, \citenamefont {Dou}, \citenamefont {Huang},\ and\ \citenamefont {Malic}}]{thompson2023interlayer}%
  \BibitemOpen
  \bibfield  {author} {\bibinfo {author} {\bibfnamefont {J.~J.}\ \bibnamefont {Thompson}}, \bibinfo {author} {\bibfnamefont {V.}~\bibnamefont {Lumsargis}}, \bibinfo {author} {\bibfnamefont {M.}~\bibnamefont {Feierabend}}, \bibinfo {author} {\bibfnamefont {Q.}~\bibnamefont {Zhao}}, \bibinfo {author} {\bibfnamefont {K.}~\bibnamefont {Wang}}, \bibinfo {author} {\bibfnamefont {L.}~\bibnamefont {Dou}}, \bibinfo {author} {\bibfnamefont {L.}~\bibnamefont {Huang}},\ and\ \bibinfo {author} {\bibfnamefont {E.}~\bibnamefont {Malic}},\ }\bibfield  {title} {\bibinfo {title} {Interlayer exciton landscape in ws$_2$/tetracene heterostructures},\ }\href@noop {} {\bibfield  {journal} {\bibinfo  {journal} {Nanoscale}\ }\textbf {\bibinfo {volume} {15}},\ \bibinfo {pages} {1730} (\bibinfo {year} {2023})}\BibitemShut {NoStop}%
\bibitem [{\citenamefont {Zhen}\ \emph {et~al.}(2023)\citenamefont {Zhen}, \citenamefont {Liu}, \citenamefont {Liu}, \citenamefont {Zhang}, \citenamefont {Zhong},\ and\ \citenamefont {Huang}}]{zhen2023effect}%
  \BibitemOpen
  \bibfield  {author} {\bibinfo {author} {\bibfnamefont {X.}~\bibnamefont {Zhen}}, \bibinfo {author} {\bibfnamefont {H.}~\bibnamefont {Liu}}, \bibinfo {author} {\bibfnamefont {F.}~\bibnamefont {Liu}}, \bibinfo {author} {\bibfnamefont {S.}~\bibnamefont {Zhang}}, \bibinfo {author} {\bibfnamefont {J.}~\bibnamefont {Zhong}},\ and\ \bibinfo {author} {\bibfnamefont {Z.}~\bibnamefont {Huang}},\ }\bibfield  {title} {\bibinfo {title} {Effect of s vacancy and interlayer interaction on the electronic and optical properties of mos$_2$/wse$_2$ heterostructure},\ }\href@noop {} {\bibfield  {journal} {\bibinfo  {journal} {Journal of Electronic Materials}\ }\textbf {\bibinfo {volume} {52}},\ \bibinfo {pages} {1186} (\bibinfo {year} {2023})}\BibitemShut {NoStop}%
\bibitem [{\citenamefont {Berland}\ \emph {et~al.}(2015)\citenamefont {Berland}, \citenamefont {Cooper}, \citenamefont {Lee}, \citenamefont {Schr{\"o}der}, \citenamefont {Thonhauser}, \citenamefont {Hyldgaard},\ and\ \citenamefont {Lundqvist}}]{berland2015van}%
  \BibitemOpen
  \bibfield  {author} {\bibinfo {author} {\bibfnamefont {K.}~\bibnamefont {Berland}}, \bibinfo {author} {\bibfnamefont {V.~R.}\ \bibnamefont {Cooper}}, \bibinfo {author} {\bibfnamefont {K.}~\bibnamefont {Lee}}, \bibinfo {author} {\bibfnamefont {E.}~\bibnamefont {Schr{\"o}der}}, \bibinfo {author} {\bibfnamefont {T.}~\bibnamefont {Thonhauser}}, \bibinfo {author} {\bibfnamefont {P.}~\bibnamefont {Hyldgaard}},\ and\ \bibinfo {author} {\bibfnamefont {B.~I.}\ \bibnamefont {Lundqvist}},\ }\bibfield  {title} {\bibinfo {title} {van der waals forces in density functional theory: a review of the vdw-df method},\ }\href@noop {} {\bibfield  {journal} {\bibinfo  {journal} {Reports on Progress in Physics}\ }\textbf {\bibinfo {volume} {78}},\ \bibinfo {pages} {066501} (\bibinfo {year} {2015})}\BibitemShut {NoStop}%
\bibitem [{\citenamefont {Lee}\ \emph {et~al.}(2010)\citenamefont {Lee}, \citenamefont {Murray}, \citenamefont {Kong}, \citenamefont {Lundqvist},\ and\ \citenamefont {Langreth}}]{lee2010higher}%
  \BibitemOpen
  \bibfield  {author} {\bibinfo {author} {\bibfnamefont {K.}~\bibnamefont {Lee}}, \bibinfo {author} {\bibfnamefont {{\'E}.~D.}\ \bibnamefont {Murray}}, \bibinfo {author} {\bibfnamefont {L.}~\bibnamefont {Kong}}, \bibinfo {author} {\bibfnamefont {B.~I.}\ \bibnamefont {Lundqvist}},\ and\ \bibinfo {author} {\bibfnamefont {D.~C.}\ \bibnamefont {Langreth}},\ }\bibfield  {title} {\bibinfo {title} {Higher-accuracy van der waals density functional},\ }\href@noop {} {\bibfield  {journal} {\bibinfo  {journal} {Physical Review B}\ }\textbf {\bibinfo {volume} {82}},\ \bibinfo {pages} {081101} (\bibinfo {year} {2010})}\BibitemShut {NoStop}%
\bibitem [{\citenamefont {Sabatini}\ \emph {et~al.}(2013)\citenamefont {Sabatini}, \citenamefont {Gorni},\ and\ \citenamefont {De~Gironcoli}}]{sabatini2013nonlocal}%
  \BibitemOpen
  \bibfield  {author} {\bibinfo {author} {\bibfnamefont {R.}~\bibnamefont {Sabatini}}, \bibinfo {author} {\bibfnamefont {T.}~\bibnamefont {Gorni}},\ and\ \bibinfo {author} {\bibfnamefont {S.}~\bibnamefont {De~Gironcoli}},\ }\bibfield  {title} {\bibinfo {title} {Nonlocal van der waals density functional made simple and efficient},\ }\href@noop {} {\bibfield  {journal} {\bibinfo  {journal} {Physical Review B}\ }\textbf {\bibinfo {volume} {87}},\ \bibinfo {pages} {041108} (\bibinfo {year} {2013})}\BibitemShut {NoStop}%
\bibitem [{\citenamefont {Grimme}(2004)}]{grimme2004accurate}%
  \BibitemOpen
  \bibfield  {author} {\bibinfo {author} {\bibfnamefont {S.}~\bibnamefont {Grimme}},\ }\bibfield  {title} {\bibinfo {title} {Accurate description of van der waals complexes by density functional theory including empirical corrections},\ }\href@noop {} {\bibfield  {journal} {\bibinfo  {journal} {Journal of computational chemistry}\ }\textbf {\bibinfo {volume} {25}},\ \bibinfo {pages} {1463} (\bibinfo {year} {2004})}\BibitemShut {NoStop}%
\bibitem [{\citenamefont {Moellmann}\ and\ \citenamefont {Grimme}(2014)}]{moellmann2014dft}%
  \BibitemOpen
  \bibfield  {author} {\bibinfo {author} {\bibfnamefont {J.}~\bibnamefont {Moellmann}}\ and\ \bibinfo {author} {\bibfnamefont {S.}~\bibnamefont {Grimme}},\ }\bibfield  {title} {\bibinfo {title} {Dft-d3 study of some molecular crystals},\ }\href@noop {} {\bibfield  {journal} {\bibinfo  {journal} {The Journal of Physical Chemistry C}\ }\textbf {\bibinfo {volume} {118}},\ \bibinfo {pages} {7615} (\bibinfo {year} {2014})}\BibitemShut {NoStop}%
\bibitem [{\citenamefont {Peng}\ \emph {et~al.}(2015)\citenamefont {Peng}, \citenamefont {Yang}, \citenamefont {Sun},\ and\ \citenamefont {Perdew}}]{peng2015scan+}%
  \BibitemOpen
  \bibfield  {author} {\bibinfo {author} {\bibfnamefont {H.}~\bibnamefont {Peng}}, \bibinfo {author} {\bibfnamefont {Z.-H.}\ \bibnamefont {Yang}}, \bibinfo {author} {\bibfnamefont {J.}~\bibnamefont {Sun}},\ and\ \bibinfo {author} {\bibfnamefont {J.~P.}\ \bibnamefont {Perdew}},\ }\bibfield  {title} {\bibinfo {title} {Scan+ rvv10: A promising van der waals density functional},\ }\href@noop {} {\bibfield  {journal} {\bibinfo  {journal} {arXiv preprint arXiv:1510.05712}\ } (\bibinfo {year} {2015})}\BibitemShut {NoStop}%
\bibitem [{\citenamefont {Chiter}\ \emph {et~al.}(2016)\citenamefont {Chiter}, \citenamefont {Nguyen}, \citenamefont {Tarrat}, \citenamefont {Benoit}, \citenamefont {Tang},\ and\ \citenamefont {Lacaze-Dufaure}}]{chiter2016effect}%
  \BibitemOpen
  \bibfield  {author} {\bibinfo {author} {\bibfnamefont {F.}~\bibnamefont {Chiter}}, \bibinfo {author} {\bibfnamefont {V.~B.}\ \bibnamefont {Nguyen}}, \bibinfo {author} {\bibfnamefont {N.}~\bibnamefont {Tarrat}}, \bibinfo {author} {\bibfnamefont {M.}~\bibnamefont {Benoit}}, \bibinfo {author} {\bibfnamefont {H.}~\bibnamefont {Tang}},\ and\ \bibinfo {author} {\bibfnamefont {C.}~\bibnamefont {Lacaze-Dufaure}},\ }\bibfield  {title} {\bibinfo {title} {Effect of van der waals corrections on dft-computed metallic surface properties},\ }\href@noop {} {\bibfield  {journal} {\bibinfo  {journal} {Materials Research Express}\ }\textbf {\bibinfo {volume} {3}},\ \bibinfo {pages} {046501} (\bibinfo {year} {2016})}\BibitemShut {NoStop}%
\bibitem [{\citenamefont {Freire}\ \emph {et~al.}(2018)\citenamefont {Freire}, \citenamefont {Guedes-Sobrinho}, \citenamefont {Kiejna},\ and\ \citenamefont {Da~Silva}}]{freire2018comparison}%
  \BibitemOpen
  \bibfield  {author} {\bibinfo {author} {\bibfnamefont {R.~L.}\ \bibnamefont {Freire}}, \bibinfo {author} {\bibfnamefont {D.}~\bibnamefont {Guedes-Sobrinho}}, \bibinfo {author} {\bibfnamefont {A.}~\bibnamefont {Kiejna}},\ and\ \bibinfo {author} {\bibfnamefont {J.~L.}\ \bibnamefont {Da~Silva}},\ }\bibfield  {title} {\bibinfo {title} {Comparison of the performance of van der waals dispersion functionals in the description of water and ethanol on transition metal surfaces},\ }\href@noop {} {\bibfield  {journal} {\bibinfo  {journal} {The Journal of Physical Chemistry C}\ }\textbf {\bibinfo {volume} {122}},\ \bibinfo {pages} {1577} (\bibinfo {year} {2018})}\BibitemShut {NoStop}%
\bibitem [{\citenamefont {Sabatini}\ \emph {et~al.}(2016)\citenamefont {Sabatini}, \citenamefont {K{\"u}{\c{c}}{\"u}kbenli}, \citenamefont {Pham},\ and\ \citenamefont {de~Gironcoli}}]{sabatini2016phonons}%
  \BibitemOpen
  \bibfield  {author} {\bibinfo {author} {\bibfnamefont {R.}~\bibnamefont {Sabatini}}, \bibinfo {author} {\bibfnamefont {E.}~\bibnamefont {K{\"u}{\c{c}}{\"u}kbenli}}, \bibinfo {author} {\bibfnamefont {C.~H.}\ \bibnamefont {Pham}},\ and\ \bibinfo {author} {\bibfnamefont {S.}~\bibnamefont {de~Gironcoli}},\ }\bibfield  {title} {\bibinfo {title} {Phonons in nonlocal van der waals density functional theory},\ }\href@noop {} {\bibfield  {journal} {\bibinfo  {journal} {Physical Review B}\ }\textbf {\bibinfo {volume} {93}},\ \bibinfo {pages} {235120} (\bibinfo {year} {2016})}\BibitemShut {NoStop}%
\bibitem [{\citenamefont {Dai}\ \emph {et~al.}(2019)\citenamefont {Dai}, \citenamefont {Liu},\ and\ \citenamefont {Zhang}}]{dai2019strain}%
  \BibitemOpen
  \bibfield  {author} {\bibinfo {author} {\bibfnamefont {Z.}~\bibnamefont {Dai}}, \bibinfo {author} {\bibfnamefont {L.}~\bibnamefont {Liu}},\ and\ \bibinfo {author} {\bibfnamefont {Z.}~\bibnamefont {Zhang}},\ }\bibfield  {title} {\bibinfo {title} {Strain engineering of 2d materials: issues and opportunities at the interface},\ }\href@noop {} {\bibfield  {journal} {\bibinfo  {journal} {Advanced Materials}\ }\textbf {\bibinfo {volume} {31}},\ \bibinfo {pages} {1805417} (\bibinfo {year} {2019})}\BibitemShut {NoStop}%
\bibitem [{\citenamefont {Sortino}\ \emph {et~al.}(2020)\citenamefont {Sortino}, \citenamefont {Brooks}, \citenamefont {Zotev}, \citenamefont {Genco}, \citenamefont {Cambiasso}, \citenamefont {Mignuzzi}, \citenamefont {Maier}, \citenamefont {Burkard}, \citenamefont {Sapienza},\ and\ \citenamefont {Tartakovskii}}]{sortino2020dielectric}%
  \BibitemOpen
  \bibfield  {author} {\bibinfo {author} {\bibfnamefont {L.}~\bibnamefont {Sortino}}, \bibinfo {author} {\bibfnamefont {M.}~\bibnamefont {Brooks}}, \bibinfo {author} {\bibfnamefont {P.~G.}\ \bibnamefont {Zotev}}, \bibinfo {author} {\bibfnamefont {A.}~\bibnamefont {Genco}}, \bibinfo {author} {\bibfnamefont {J.}~\bibnamefont {Cambiasso}}, \bibinfo {author} {\bibfnamefont {S.}~\bibnamefont {Mignuzzi}}, \bibinfo {author} {\bibfnamefont {S.~A.}\ \bibnamefont {Maier}}, \bibinfo {author} {\bibfnamefont {G.}~\bibnamefont {Burkard}}, \bibinfo {author} {\bibfnamefont {R.}~\bibnamefont {Sapienza}},\ and\ \bibinfo {author} {\bibfnamefont {A.~I.}\ \bibnamefont {Tartakovskii}},\ }\bibfield  {title} {\bibinfo {title} {Dielectric nanoantennas for strain engineering in atomically thin two-dimensional semiconductors},\ }\href@noop {} {\bibfield  {journal} {\bibinfo  {journal} {ACS Photonics}\ }\textbf {\bibinfo {volume} {7}},\ \bibinfo {pages} {2413} (\bibinfo {year} {2020})}\BibitemShut {NoStop}%
\bibitem [{\citenamefont {Kansara}\ \emph {et~al.}(2018)\citenamefont {Kansara}, \citenamefont {Gupta},\ and\ \citenamefont {Sonvane}}]{kansara2018effect}%
  \BibitemOpen
  \bibfield  {author} {\bibinfo {author} {\bibfnamefont {S.}~\bibnamefont {Kansara}}, \bibinfo {author} {\bibfnamefont {S.~K.}\ \bibnamefont {Gupta}},\ and\ \bibinfo {author} {\bibfnamefont {Y.}~\bibnamefont {Sonvane}},\ }\bibfield  {title} {\bibinfo {title} {Effect of strain engineering on 2d dichalcogenides transition metal: a dft study},\ }\href@noop {} {\bibfield  {journal} {\bibinfo  {journal} {Computational Materials Science}\ }\textbf {\bibinfo {volume} {141}},\ \bibinfo {pages} {235} (\bibinfo {year} {2018})}\BibitemShut {NoStop}%
\bibitem [{\citenamefont {Palepu}\ \emph {et~al.}(2022)\citenamefont {Palepu}, \citenamefont {Anand}, \citenamefont {Parshi}, \citenamefont {Jain}, \citenamefont {Tiwari}, \citenamefont {Bhattacharya}, \citenamefont {Chakraborty},\ and\ \citenamefont {Kanungo}}]{palepu2022comparative}%
  \BibitemOpen
  \bibfield  {author} {\bibinfo {author} {\bibfnamefont {J.}~\bibnamefont {Palepu}}, \bibinfo {author} {\bibfnamefont {P.~P.}\ \bibnamefont {Anand}}, \bibinfo {author} {\bibfnamefont {P.}~\bibnamefont {Parshi}}, \bibinfo {author} {\bibfnamefont {V.}~\bibnamefont {Jain}}, \bibinfo {author} {\bibfnamefont {A.}~\bibnamefont {Tiwari}}, \bibinfo {author} {\bibfnamefont {S.}~\bibnamefont {Bhattacharya}}, \bibinfo {author} {\bibfnamefont {S.}~\bibnamefont {Chakraborty}},\ and\ \bibinfo {author} {\bibfnamefont {S.}~\bibnamefont {Kanungo}},\ }\bibfield  {title} {\bibinfo {title} {Comparative analysis of strain engineering on the electronic properties of homogenous and heterostructure bilayers of mox$_2$ (x= s, se, te)},\ }\href@noop {} {\bibfield  {journal} {\bibinfo  {journal} {Micro and Nanostructures}\ }\textbf {\bibinfo {volume} {168}},\ \bibinfo {pages} {207334} (\bibinfo {year} {2022})}\BibitemShut {NoStop}%
\bibitem [{\citenamefont {Wu}\ \emph {et~al.}(2018{\natexlab{b}})\citenamefont {Wu}, \citenamefont {Wang}, \citenamefont {Ercius}, \citenamefont {Wright}, \citenamefont {Leppert-Simenauer}, \citenamefont {Burke}, \citenamefont {Dubey}, \citenamefont {Dogare},\ and\ \citenamefont {Pettes}}]{wu2018giant}%
  \BibitemOpen
  \bibfield  {author} {\bibinfo {author} {\bibfnamefont {W.}~\bibnamefont {Wu}}, \bibinfo {author} {\bibfnamefont {J.}~\bibnamefont {Wang}}, \bibinfo {author} {\bibfnamefont {P.}~\bibnamefont {Ercius}}, \bibinfo {author} {\bibfnamefont {N.~C.}\ \bibnamefont {Wright}}, \bibinfo {author} {\bibfnamefont {D.~M.}\ \bibnamefont {Leppert-Simenauer}}, \bibinfo {author} {\bibfnamefont {R.~A.}\ \bibnamefont {Burke}}, \bibinfo {author} {\bibfnamefont {M.}~\bibnamefont {Dubey}}, \bibinfo {author} {\bibfnamefont {A.~M.}\ \bibnamefont {Dogare}},\ and\ \bibinfo {author} {\bibfnamefont {M.~T.}\ \bibnamefont {Pettes}},\ }\bibfield  {title} {\bibinfo {title} {Giant mechano-optoelectronic effect in an atomically thin semiconductor},\ }\href@noop {} {\bibfield  {journal} {\bibinfo  {journal} {Nano letters}\ }\textbf {\bibinfo {volume} {18}},\ \bibinfo {pages} {2351} (\bibinfo {year} {2018}{\natexlab{b}})}\BibitemShut {NoStop}%
\bibitem [{\citenamefont {Pandey}\ \emph {et~al.}(2023)\citenamefont {Pandey}, \citenamefont {Pandey}, \citenamefont {Ahuja},\ and\ \citenamefont {Kumar}}]{pandey2023straining}%
  \BibitemOpen
  \bibfield  {author} {\bibinfo {author} {\bibfnamefont {M.}~\bibnamefont {Pandey}}, \bibinfo {author} {\bibfnamefont {C.}~\bibnamefont {Pandey}}, \bibinfo {author} {\bibfnamefont {R.}~\bibnamefont {Ahuja}},\ and\ \bibinfo {author} {\bibfnamefont {R.}~\bibnamefont {Kumar}},\ }\bibfield  {title} {\bibinfo {title} {Straining techniques for strain engineering of 2d materials towards flexible straintronic applications},\ }\href@noop {} {\bibfield  {journal} {\bibinfo  {journal} {Nano Energy}\ ,\ \bibinfo {pages} {108278}} (\bibinfo {year} {2023})}\BibitemShut {NoStop}%
\bibitem [{\citenamefont {Smiri}\ \emph {et~al.}(2021)\citenamefont {Smiri}, \citenamefont {Amand},\ and\ \citenamefont {Jaziri}}]{smiri2021optical}%
  \BibitemOpen
  \bibfield  {author} {\bibinfo {author} {\bibfnamefont {A.}~\bibnamefont {Smiri}}, \bibinfo {author} {\bibfnamefont {T.}~\bibnamefont {Amand}},\ and\ \bibinfo {author} {\bibfnamefont {S.}~\bibnamefont {Jaziri}},\ }\bibfield  {title} {\bibinfo {title} {Optical properties of excitons in two-dimensional transition metal dichalcogenide nanobubbles},\ }\href@noop {} {\bibfield  {journal} {\bibinfo  {journal} {The Journal of Chemical Physics}\ }\textbf {\bibinfo {volume} {154}} (\bibinfo {year} {2021})}\BibitemShut {NoStop}%
\bibitem [{\citenamefont {Chaste}(2023)}]{chaste2023straintronique}%
  \BibitemOpen
  \bibfield  {author} {\bibinfo {author} {\bibfnamefont {J.}~\bibnamefont {Chaste}},\ }\emph {\bibinfo {title} {Straintronique et transport thermique dans les mat{\'e}riaux 2D}},\ \href@noop {} {Ph.D. thesis},\ \bibinfo  {school} {Universit{\'e} Paris Saclay} (\bibinfo {year} {2023})\BibitemShut {NoStop}%
\bibitem [{\citenamefont {Ippolito}(2021)}]{ippolito2021defect}%
  \BibitemOpen
  \bibfield  {author} {\bibinfo {author} {\bibfnamefont {S.}~\bibnamefont {Ippolito}},\ }\emph {\bibinfo {title} {Defect engineering in 2D semiconductors: fabrication of hybrid multifunctional devices}},\ \href@noop {} {Ph.D. thesis},\ \bibinfo  {school} {Universit{\'e} de Strasbourg} (\bibinfo {year} {2021})\BibitemShut {NoStop}%
\bibitem [{\citenamefont {Pandey}\ and\ \citenamefont {Kumar}(2022)}]{pandey2022polymer}%
  \BibitemOpen
  \bibfield  {author} {\bibinfo {author} {\bibfnamefont {M.}~\bibnamefont {Pandey}}\ and\ \bibinfo {author} {\bibfnamefont {R.}~\bibnamefont {Kumar}},\ }\bibfield  {title} {\bibinfo {title} {Polymer curing assisted formation of optically visible sub-micron blisters of multilayer graphene for local strain engineering},\ }\href@noop {} {\bibfield  {journal} {\bibinfo  {journal} {Journal of Physics: Condensed Matter}\ }\textbf {\bibinfo {volume} {34}},\ \bibinfo {pages} {245401} (\bibinfo {year} {2022})}\BibitemShut {NoStop}%
\bibitem [{\citenamefont {Deng}\ \emph {et~al.}(2018)\citenamefont {Deng}, \citenamefont {Li},\ and\ \citenamefont {Zhang}}]{deng2018strain}%
  \BibitemOpen
  \bibfield  {author} {\bibinfo {author} {\bibfnamefont {S.}~\bibnamefont {Deng}}, \bibinfo {author} {\bibfnamefont {L.}~\bibnamefont {Li}},\ and\ \bibinfo {author} {\bibfnamefont {Y.}~\bibnamefont {Zhang}},\ }\bibfield  {title} {\bibinfo {title} {Strain modulated electronic, mechanical, and optical properties of the monolayer pds$_2$, pdse$_2$, and ptse$_2$ for tunable devices},\ }\href@noop {} {\bibfield  {journal} {\bibinfo  {journal} {ACS Applied Nano Materials}\ }\textbf {\bibinfo {volume} {1}},\ \bibinfo {pages} {1932} (\bibinfo {year} {2018})}\BibitemShut {NoStop}%
\bibitem [{\citenamefont {{\c{C}}ak{\i}ro{\u{g}}lu}\ \emph {et~al.}(2023)\citenamefont {{\c{C}}ak{\i}ro{\u{g}}lu}, \citenamefont {Island}, \citenamefont {Xie}, \citenamefont {Frisenda},\ and\ \citenamefont {Castellanos-Gomez}}]{ccakirouglu2023automated}%
  \BibitemOpen
  \bibfield  {author} {\bibinfo {author} {\bibfnamefont {O.}~\bibnamefont {{\c{C}}ak{\i}ro{\u{g}}lu}}, \bibinfo {author} {\bibfnamefont {J.~O.}\ \bibnamefont {Island}}, \bibinfo {author} {\bibfnamefont {Y.}~\bibnamefont {Xie}}, \bibinfo {author} {\bibfnamefont {R.}~\bibnamefont {Frisenda}},\ and\ \bibinfo {author} {\bibfnamefont {A.}~\bibnamefont {Castellanos-Gomez}},\ }\bibfield  {title} {\bibinfo {title} {An automated system for strain engineering and straintronics of 2d materials},\ }\href@noop {} {\bibfield  {journal} {\bibinfo  {journal} {Advanced Materials Technologies}\ }\textbf {\bibinfo {volume} {8}},\ \bibinfo {pages} {2201091} (\bibinfo {year} {2023})}\BibitemShut {NoStop}%
\bibitem [{\citenamefont {Carrascoso}\ \emph {et~al.}(2022)\citenamefont {Carrascoso}, \citenamefont {Frisenda},\ and\ \citenamefont {Castellanos-Gomez}}]{carrascoso2022biaxial}%
  \BibitemOpen
  \bibfield  {author} {\bibinfo {author} {\bibfnamefont {F.}~\bibnamefont {Carrascoso}}, \bibinfo {author} {\bibfnamefont {R.}~\bibnamefont {Frisenda}},\ and\ \bibinfo {author} {\bibfnamefont {A.}~\bibnamefont {Castellanos-Gomez}},\ }\bibfield  {title} {\bibinfo {title} {Biaxial versus uniaxial strain tuning of single-layer mos$_2$},\ }\href@noop {} {\bibfield  {journal} {\bibinfo  {journal} {Nano Materials Science}\ }\textbf {\bibinfo {volume} {4}},\ \bibinfo {pages} {44} (\bibinfo {year} {2022})}\BibitemShut {NoStop}%
\bibitem [{\citenamefont {Ren}\ and\ \citenamefont {Xiang}(2022)}]{ren2022strain}%
  \BibitemOpen
  \bibfield  {author} {\bibinfo {author} {\bibfnamefont {H.}~\bibnamefont {Ren}}\ and\ \bibinfo {author} {\bibfnamefont {G.}~\bibnamefont {Xiang}},\ }\bibfield  {title} {\bibinfo {title} {Strain-modulated magnetism in mos$_2$},\ }\href@noop {} {\bibfield  {journal} {\bibinfo  {journal} {Nanomaterials}\ }\textbf {\bibinfo {volume} {12}},\ \bibinfo {pages} {1929} (\bibinfo {year} {2022})}\BibitemShut {NoStop}%
\bibitem [{\citenamefont {Lin}\ \emph {et~al.}(2022)\citenamefont {Lin}, \citenamefont {Qian}, \citenamefont {Dai}, \citenamefont {Sun},\ and\ \citenamefont {Wang}}]{lin2022regulation}%
  \BibitemOpen
  \bibfield  {author} {\bibinfo {author} {\bibfnamefont {Q.-L.}\ \bibnamefont {Lin}}, \bibinfo {author} {\bibfnamefont {Z.-F.}\ \bibnamefont {Qian}}, \bibinfo {author} {\bibfnamefont {X.-Y.}\ \bibnamefont {Dai}}, \bibinfo {author} {\bibfnamefont {Y.-L.}\ \bibnamefont {Sun}},\ and\ \bibinfo {author} {\bibfnamefont {R.-H.}\ \bibnamefont {Wang}},\ }\bibfield  {title} {\bibinfo {title} {Regulation of electronic structure of monolayer mos$_2$ by pressure},\ }\href@noop {} {\bibfield  {journal} {\bibinfo  {journal} {Rare Metals}\ }\textbf {\bibinfo {volume} {41}},\ \bibinfo {pages} {1761} (\bibinfo {year} {2022})}\BibitemShut {NoStop}%
\bibitem [{\citenamefont {Wang}\ \emph {et~al.}(2022)\citenamefont {Wang}, \citenamefont {Li}, \citenamefont {Wang}, \citenamefont {Zhao},\ and\ \citenamefont {Zhuo}}]{wang2022first}%
  \BibitemOpen
  \bibfield  {author} {\bibinfo {author} {\bibfnamefont {C.}~\bibnamefont {Wang}}, \bibinfo {author} {\bibfnamefont {S.}~\bibnamefont {Li}}, \bibinfo {author} {\bibfnamefont {S.}~\bibnamefont {Wang}}, \bibinfo {author} {\bibfnamefont {P.}~\bibnamefont {Zhao}},\ and\ \bibinfo {author} {\bibfnamefont {R.}~\bibnamefont {Zhuo}},\ }\bibfield  {title} {\bibinfo {title} {First principles study of the effect of uniaxial strain on monolayer mos$_2$},\ }\href@noop {} {\bibfield  {journal} {\bibinfo  {journal} {Physica E: Low-dimensional Systems and Nanostructures}\ }\textbf {\bibinfo {volume} {144}},\ \bibinfo {pages} {115401} (\bibinfo {year} {2022})}\BibitemShut {NoStop}%
\bibitem [{\citenamefont {Le}\ \emph {et~al.}(2020)\citenamefont {Le}, \citenamefont {Chihaia}, \citenamefont {Pham-Ho} \emph {et~al.}}]{le2020electronic}%
  \BibitemOpen
  \bibfield  {author} {\bibinfo {author} {\bibfnamefont {O.~K.}\ \bibnamefont {Le}}, \bibinfo {author} {\bibfnamefont {V.}~\bibnamefont {Chihaia}}, \bibinfo {author} {\bibfnamefont {M.-P.}\ \bibnamefont {Pham-Ho}}, \emph {et~al.},\ }\bibfield  {title} {\bibinfo {title} {Electronic and optical properties of monolayer mos$_2$ under the influence of polyethyleneimine adsorption and pressure},\ }\href@noop {} {\bibfield  {journal} {\bibinfo  {journal} {RSC advances}\ }\textbf {\bibinfo {volume} {10}},\ \bibinfo {pages} {4201} (\bibinfo {year} {2020})}\BibitemShut {NoStop}%
\bibitem [{\citenamefont {Chen}\ \emph {et~al.}(2023)\citenamefont {Chen}, \citenamefont {Lu}, \citenamefont {Kong}, \citenamefont {Tao}, \citenamefont {Ma}, \citenamefont {Liu}, \citenamefont {Lu}, \citenamefont {Li}, \citenamefont {Wu}, \citenamefont {Duan} \emph {et~al.}}]{chen2023mobility}%
  \BibitemOpen
  \bibfield  {author} {\bibinfo {author} {\bibfnamefont {Y.}~\bibnamefont {Chen}}, \bibinfo {author} {\bibfnamefont {D.}~\bibnamefont {Lu}}, \bibinfo {author} {\bibfnamefont {L.}~\bibnamefont {Kong}}, \bibinfo {author} {\bibfnamefont {Q.}~\bibnamefont {Tao}}, \bibinfo {author} {\bibfnamefont {L.}~\bibnamefont {Ma}}, \bibinfo {author} {\bibfnamefont {L.}~\bibnamefont {Liu}}, \bibinfo {author} {\bibfnamefont {Z.}~\bibnamefont {Lu}}, \bibinfo {author} {\bibfnamefont {Z.}~\bibnamefont {Li}}, \bibinfo {author} {\bibfnamefont {R.}~\bibnamefont {Wu}}, \bibinfo {author} {\bibfnamefont {X.}~\bibnamefont {Duan}}, \emph {et~al.},\ }\bibfield  {title} {\bibinfo {title} {Mobility enhancement of strained mos$_2$ transistor on flat substrate},\ }\href@noop {} {\bibfield  {journal} {\bibinfo  {journal} {ACS nano}\ }\textbf {\bibinfo {volume} {17}},\ \bibinfo {pages} {14954} (\bibinfo {year} {2023})}\BibitemShut {NoStop}%
\bibitem [{\citenamefont {Yang}\ \emph {et~al.}(2014)\citenamefont {Yang}, \citenamefont {Cui}, \citenamefont {Zhang}, \citenamefont {Wang}, \citenamefont {Shen}, \citenamefont {Zeng}, \citenamefont {Dayeh}, \citenamefont {Feng},\ and\ \citenamefont {Xiang}}]{yang2014lattice}%
  \BibitemOpen
  \bibfield  {author} {\bibinfo {author} {\bibfnamefont {L.}~\bibnamefont {Yang}}, \bibinfo {author} {\bibfnamefont {X.}~\bibnamefont {Cui}}, \bibinfo {author} {\bibfnamefont {J.}~\bibnamefont {Zhang}}, \bibinfo {author} {\bibfnamefont {K.}~\bibnamefont {Wang}}, \bibinfo {author} {\bibfnamefont {M.}~\bibnamefont {Shen}}, \bibinfo {author} {\bibfnamefont {S.}~\bibnamefont {Zeng}}, \bibinfo {author} {\bibfnamefont {S.~A.}\ \bibnamefont {Dayeh}}, \bibinfo {author} {\bibfnamefont {L.}~\bibnamefont {Feng}},\ and\ \bibinfo {author} {\bibfnamefont {B.}~\bibnamefont {Xiang}},\ }\bibfield  {title} {\bibinfo {title} {Lattice strain effects on the optical properties of mos$_2$ nanosheets},\ }\href@noop {} {\bibfield  {journal} {\bibinfo  {journal} {Scientific reports}\ }\textbf {\bibinfo {volume} {4}},\ \bibinfo {pages} {5649} (\bibinfo {year} {2014})}\BibitemShut {NoStop}%
\bibitem [{\citenamefont {Guo}(2016)}]{guo2016biaxial}%
  \BibitemOpen
  \bibfield  {author} {\bibinfo {author} {\bibfnamefont {S.-D.}\ \bibnamefont {Guo}},\ }\bibfield  {title} {\bibinfo {title} {Biaxial strain tuned thermoelectric properties in monolayer ptse$_2$},\ }\href@noop {} {\bibfield  {journal} {\bibinfo  {journal} {Journal of Materials Chemistry C}\ }\textbf {\bibinfo {volume} {4}},\ \bibinfo {pages} {9366} (\bibinfo {year} {2016})}\BibitemShut {NoStop}%
\bibitem [{\citenamefont {Ge}\ \emph {et~al.}(2023)\citenamefont {Ge}, \citenamefont {Zhou}, \citenamefont {Sun},\ and\ \citenamefont {Chen}}]{ge2023first}%
  \BibitemOpen
  \bibfield  {author} {\bibinfo {author} {\bibfnamefont {X.}~\bibnamefont {Ge}}, \bibinfo {author} {\bibfnamefont {X.}~\bibnamefont {Zhou}}, \bibinfo {author} {\bibfnamefont {D.}~\bibnamefont {Sun}},\ and\ \bibinfo {author} {\bibfnamefont {X.}~\bibnamefont {Chen}},\ }\bibfield  {title} {\bibinfo {title} {First-principles study of structural and electronic properties of monolayer ptx$_2$ and janus ptxy (x, y= s, se, and te) via strain engineering},\ }\href@noop {} {\bibfield  {journal} {\bibinfo  {journal} {ACS omega}\ }\textbf {\bibinfo {volume} {8}},\ \bibinfo {pages} {5715} (\bibinfo {year} {2023})}\BibitemShut {NoStop}%
\bibitem [{\citenamefont {Wang}\ \emph {et~al.}(2023)\citenamefont {Wang}, \citenamefont {Jing}, \citenamefont {Duan}, \citenamefont {Liu}, \citenamefont {Kang}, \citenamefont {Xu}, \citenamefont {Chen}, \citenamefont {Xia}, \citenamefont {Chang}, \citenamefont {Zhao} \emph {et~al.}}]{wang20232d}%
  \BibitemOpen
  \bibfield  {author} {\bibinfo {author} {\bibfnamefont {Z.}~\bibnamefont {Wang}}, \bibinfo {author} {\bibfnamefont {X.}~\bibnamefont {Jing}}, \bibinfo {author} {\bibfnamefont {S.}~\bibnamefont {Duan}}, \bibinfo {author} {\bibfnamefont {C.}~\bibnamefont {Liu}}, \bibinfo {author} {\bibfnamefont {D.}~\bibnamefont {Kang}}, \bibinfo {author} {\bibfnamefont {X.}~\bibnamefont {Xu}}, \bibinfo {author} {\bibfnamefont {J.}~\bibnamefont {Chen}}, \bibinfo {author} {\bibfnamefont {Y.}~\bibnamefont {Xia}}, \bibinfo {author} {\bibfnamefont {B.}~\bibnamefont {Chang}}, \bibinfo {author} {\bibfnamefont {C.}~\bibnamefont {Zhao}}, \emph {et~al.},\ }\bibfield  {title} {\bibinfo {title} {2d ptse$_2$ enabled wireless wearable gas monitoring circuits with distinctive strain-enhanced performance},\ }\href@noop {} {\bibfield  {journal} {\bibinfo  {journal} {ACS nano}\ } (\bibinfo {year} {2023})}\BibitemShut {NoStop}%
\bibitem [{\citenamefont {Li}\ \emph {et~al.}(2016)\citenamefont {Li}, \citenamefont {Li},\ and\ \citenamefont {Zeng}}]{li2016tuning}%
  \BibitemOpen
  \bibfield  {author} {\bibinfo {author} {\bibfnamefont {P.}~\bibnamefont {Li}}, \bibinfo {author} {\bibfnamefont {L.}~\bibnamefont {Li}},\ and\ \bibinfo {author} {\bibfnamefont {X.~C.}\ \bibnamefont {Zeng}},\ }\bibfield  {title} {\bibinfo {title} {Tuning the electronic properties of monolayer and bilayer ptse$_2$ via strain engineering},\ }\href@noop {} {\bibfield  {journal} {\bibinfo  {journal} {Journal of Materials Chemistry C}\ }\textbf {\bibinfo {volume} {4}},\ \bibinfo {pages} {3106} (\bibinfo {year} {2016})}\BibitemShut {NoStop}%
\bibitem [{\citenamefont {Zhang}\ \emph {et~al.}(2017{\natexlab{a}})\citenamefont {Zhang}, \citenamefont {Qin}, \citenamefont {Huang},\ and\ \citenamefont {Zhang}}]{zhang2017mechanism}%
  \BibitemOpen
  \bibfield  {author} {\bibinfo {author} {\bibfnamefont {W.}~\bibnamefont {Zhang}}, \bibinfo {author} {\bibfnamefont {J.}~\bibnamefont {Qin}}, \bibinfo {author} {\bibfnamefont {Z.}~\bibnamefont {Huang}},\ and\ \bibinfo {author} {\bibfnamefont {W.}~\bibnamefont {Zhang}},\ }\bibfield  {title} {\bibinfo {title} {The mechanism of layer number and strain dependent bandgap of 2d crystal ptse$_2$},\ }\href@noop {} {\bibfield  {journal} {\bibinfo  {journal} {Journal of Applied Physics}\ }\textbf {\bibinfo {volume} {122}} (\bibinfo {year} {2017}{\natexlab{a}})}\BibitemShut {NoStop}%
\bibitem [{\citenamefont {Han}\ \emph {et~al.}(2022)\citenamefont {Han}, \citenamefont {Gao}, \citenamefont {Zhou}, \citenamefont {Hou}, \citenamefont {Jia}, \citenamefont {Cao}, \citenamefont {Duan},\ and\ \citenamefont {Lu}}]{han2022deep}%
  \BibitemOpen
  \bibfield  {author} {\bibinfo {author} {\bibfnamefont {Y.}~\bibnamefont {Han}}, \bibinfo {author} {\bibfnamefont {L.}~\bibnamefont {Gao}}, \bibinfo {author} {\bibfnamefont {J.}~\bibnamefont {Zhou}}, \bibinfo {author} {\bibfnamefont {Y.}~\bibnamefont {Hou}}, \bibinfo {author} {\bibfnamefont {Y.}~\bibnamefont {Jia}}, \bibinfo {author} {\bibfnamefont {K.}~\bibnamefont {Cao}}, \bibinfo {author} {\bibfnamefont {K.}~\bibnamefont {Duan}},\ and\ \bibinfo {author} {\bibfnamefont {Y.}~\bibnamefont {Lu}},\ }\bibfield  {title} {\bibinfo {title} {Deep elastic strain engineering of 2d materials and their twisted bilayers},\ }\href@noop {} {\bibfield  {journal} {\bibinfo  {journal} {ACS Applied Materials \& Interfaces}\ }\textbf {\bibinfo {volume} {14}},\ \bibinfo {pages} {8655} (\bibinfo {year} {2022})}\BibitemShut {NoStop}%
\bibitem [{\citenamefont {Hong}(2019)}]{hong2019tunable}%
  \BibitemOpen
  \bibfield  {author} {\bibinfo {author} {\bibfnamefont {Z.}~\bibnamefont {Hong}},\ }\bibfield  {title} {\bibinfo {title} {Tunable structure and electronic properties of multilayer ptse$_2$},\ }in\ \href@noop {} {\emph {\bibinfo {booktitle} {Journal of Physics: Conference Series}}},\ Vol.\ \bibinfo {volume} {1411}\ (\bibinfo {organization} {IOP Publishing},\ \bibinfo {year} {2019})\ p.\ \bibinfo {pages} {012019}\BibitemShut {NoStop}%
\bibitem [{\citenamefont {Conley}\ \emph {et~al.}(2013)\citenamefont {Conley}, \citenamefont {Wang}, \citenamefont {Ziegler}, \citenamefont {Haglund~Jr}, \citenamefont {Pantelides},\ and\ \citenamefont {Bolotin}}]{conley2013bandgap}%
  \BibitemOpen
  \bibfield  {author} {\bibinfo {author} {\bibfnamefont {H.~J.}\ \bibnamefont {Conley}}, \bibinfo {author} {\bibfnamefont {B.}~\bibnamefont {Wang}}, \bibinfo {author} {\bibfnamefont {J.~I.}\ \bibnamefont {Ziegler}}, \bibinfo {author} {\bibfnamefont {R.~F.}\ \bibnamefont {Haglund~Jr}}, \bibinfo {author} {\bibfnamefont {S.~T.}\ \bibnamefont {Pantelides}},\ and\ \bibinfo {author} {\bibfnamefont {K.~I.}\ \bibnamefont {Bolotin}},\ }\bibfield  {title} {\bibinfo {title} {Bandgap engineering of strained monolayer and bilayer mos$_2$},\ }\href@noop {} {\bibfield  {journal} {\bibinfo  {journal} {Nano letters}\ }\textbf {\bibinfo {volume} {13}},\ \bibinfo {pages} {3626} (\bibinfo {year} {2013})}\BibitemShut {NoStop}%
\bibitem [{\citenamefont {Armstrong}\ \emph {et~al.}(2023)\citenamefont {Armstrong}, \citenamefont {McKenna},\ and\ \citenamefont {Wang}}]{armstrong2023directional}%
  \BibitemOpen
  \bibfield  {author} {\bibinfo {author} {\bibfnamefont {A.}~\bibnamefont {Armstrong}}, \bibinfo {author} {\bibfnamefont {K.~P.}\ \bibnamefont {McKenna}},\ and\ \bibinfo {author} {\bibfnamefont {Y.}~\bibnamefont {Wang}},\ }\bibfield  {title} {\bibinfo {title} {Directional dependence of band gap modulation via uniaxial strain in mos$_2$ and tis$_3$},\ }\href@noop {} {\bibfield  {journal} {\bibinfo  {journal} {Nanotechnology}\ }\textbf {\bibinfo {volume} {35}},\ \bibinfo {pages} {015704} (\bibinfo {year} {2023})}\BibitemShut {NoStop}%
\bibitem [{\citenamefont {Maniadaki}\ \emph {et~al.}(2016)\citenamefont {Maniadaki}, \citenamefont {Kopidakis},\ and\ \citenamefont {Remediakis}}]{maniadaki2016strain}%
  \BibitemOpen
  \bibfield  {author} {\bibinfo {author} {\bibfnamefont {A.~E.}\ \bibnamefont {Maniadaki}}, \bibinfo {author} {\bibfnamefont {G.}~\bibnamefont {Kopidakis}},\ and\ \bibinfo {author} {\bibfnamefont {I.~N.}\ \bibnamefont {Remediakis}},\ }\bibfield  {title} {\bibinfo {title} {Strain engineering of electronic properties of transition metal dichalcogenide monolayers},\ }\href@noop {} {\bibfield  {journal} {\bibinfo  {journal} {Solid State Communications}\ }\textbf {\bibinfo {volume} {227}},\ \bibinfo {pages} {33} (\bibinfo {year} {2016})}\BibitemShut {NoStop}%
\bibitem [{\citenamefont {Sharma}\ and\ \citenamefont {Singh}(2022)}]{sharma2022tuning}%
  \BibitemOpen
  \bibfield  {author} {\bibinfo {author} {\bibfnamefont {M.}~\bibnamefont {Sharma}}\ and\ \bibinfo {author} {\bibfnamefont {R.}~\bibnamefont {Singh}},\ }\bibfield  {title} {\bibinfo {title} {Tuning electronic properties of pentagonal pdse$_2$ monolayer by applying external strain},\ }\href@noop {} {\bibfield  {journal} {\bibinfo  {journal} {Indian Journal of Physics}\ }\textbf {\bibinfo {volume} {96}},\ \bibinfo {pages} {1037} (\bibinfo {year} {2022})}\BibitemShut {NoStop}%
\bibitem [{\citenamefont {Su}\ \emph {et~al.}(2021)\citenamefont {Su}, \citenamefont {Wang}, \citenamefont {Wong}, \citenamefont {Wang}, \citenamefont {Huang}, \citenamefont {Sheng}, \citenamefont {Tang}, \citenamefont {Chou}, \citenamefont {Chou}, \citenamefont {Jeng} \emph {et~al.}}]{su2021thermally}%
  \BibitemOpen
  \bibfield  {author} {\bibinfo {author} {\bibfnamefont {T.-Y.}\ \bibnamefont {Su}}, \bibinfo {author} {\bibfnamefont {T.-H.}\ \bibnamefont {Wang}}, \bibinfo {author} {\bibfnamefont {D.~P.}\ \bibnamefont {Wong}}, \bibinfo {author} {\bibfnamefont {Y.-C.}\ \bibnamefont {Wang}}, \bibinfo {author} {\bibfnamefont {A.}~\bibnamefont {Huang}}, \bibinfo {author} {\bibfnamefont {Y.-C.}\ \bibnamefont {Sheng}}, \bibinfo {author} {\bibfnamefont {S.-Y.}\ \bibnamefont {Tang}}, \bibinfo {author} {\bibfnamefont {T.-C.}\ \bibnamefont {Chou}}, \bibinfo {author} {\bibfnamefont {T.-L.}\ \bibnamefont {Chou}}, \bibinfo {author} {\bibfnamefont {H.-T.}\ \bibnamefont {Jeng}}, \emph {et~al.},\ }\bibfield  {title} {\bibinfo {title} {Thermally strain-induced band gap opening on platinum diselenide-layered films: A promising two-dimensional material with excellent thermoelectric performance},\ }\href@noop {} {\bibfield  {journal} {\bibinfo  {journal} {Chemistry of Materials}\ }\textbf {\bibinfo {volume} {33}},\ \bibinfo {pages} {3490}
  (\bibinfo {year} {2021})}\BibitemShut {NoStop}%
\bibitem [{\citenamefont {Giannozzi}\ \emph {et~al.}(2009)\citenamefont {Giannozzi}, \citenamefont {Baroni}, \citenamefont {Bonini}, \citenamefont {Calandra}, \citenamefont {Car}, \citenamefont {Cavazzoni}, \citenamefont {Ceresoli}, \citenamefont {Chiarotti}, \citenamefont {Cococcioni}, \citenamefont {Dabo} \emph {et~al.}}]{giannozzi2009quantum}%
  \BibitemOpen
  \bibfield  {author} {\bibinfo {author} {\bibfnamefont {P.}~\bibnamefont {Giannozzi}}, \bibinfo {author} {\bibfnamefont {S.}~\bibnamefont {Baroni}}, \bibinfo {author} {\bibfnamefont {N.}~\bibnamefont {Bonini}}, \bibinfo {author} {\bibfnamefont {M.}~\bibnamefont {Calandra}}, \bibinfo {author} {\bibfnamefont {R.}~\bibnamefont {Car}}, \bibinfo {author} {\bibfnamefont {C.}~\bibnamefont {Cavazzoni}}, \bibinfo {author} {\bibfnamefont {D.}~\bibnamefont {Ceresoli}}, \bibinfo {author} {\bibfnamefont {G.~L.}\ \bibnamefont {Chiarotti}}, \bibinfo {author} {\bibfnamefont {M.}~\bibnamefont {Cococcioni}}, \bibinfo {author} {\bibfnamefont {I.}~\bibnamefont {Dabo}}, \emph {et~al.},\ }\bibfield  {title} {\bibinfo {title} {Quantum espresso: a modular and open-source software project for quantum simulations of materials},\ }\href@noop {} {\bibfield  {journal} {\bibinfo  {journal} {Journal of physics: Condensed matter}\ }\textbf {\bibinfo {volume} {21}},\ \bibinfo {pages} {395502} (\bibinfo {year} {2009})}\BibitemShut {NoStop}%
\bibitem [{\citenamefont {Bl{\"o}chl}(1994)}]{blochl1994projector}%
  \BibitemOpen
  \bibfield  {author} {\bibinfo {author} {\bibfnamefont {P.~E.}\ \bibnamefont {Bl{\"o}chl}},\ }\bibfield  {title} {\bibinfo {title} {Projector augmented-wave method},\ }\href@noop {} {\bibfield  {journal} {\bibinfo  {journal} {Physical review B}\ }\textbf {\bibinfo {volume} {50}},\ \bibinfo {pages} {17953} (\bibinfo {year} {1994})}\BibitemShut {NoStop}%
\bibitem [{\citenamefont {Feynman}(1939)}]{feynman1939forces}%
  \BibitemOpen
  \bibfield  {author} {\bibinfo {author} {\bibfnamefont {R.~P.}\ \bibnamefont {Feynman}},\ }\bibfield  {title} {\bibinfo {title} {Forces in molecules},\ }\href@noop {} {\bibfield  {journal} {\bibinfo  {journal} {Physical review}\ }\textbf {\bibinfo {volume} {56}},\ \bibinfo {pages} {340} (\bibinfo {year} {1939})}\BibitemShut {NoStop}%
\bibitem [{\citenamefont {Momma}\ and\ \citenamefont {Izumi}(2011)}]{momma2011vesta}%
  \BibitemOpen
  \bibfield  {author} {\bibinfo {author} {\bibfnamefont {K.}~\bibnamefont {Momma}}\ and\ \bibinfo {author} {\bibfnamefont {F.}~\bibnamefont {Izumi}},\ }\bibfield  {title} {\bibinfo {title} {Vesta 3 for three-dimensional visualization of crystal, volumetric and morphology data},\ }\href@noop {} {\bibfield  {journal} {\bibinfo  {journal} {Journal of applied crystallography}\ }\textbf {\bibinfo {volume} {44}},\ \bibinfo {pages} {1272} (\bibinfo {year} {2011})}\BibitemShut {NoStop}%
\bibitem [{\citenamefont {Perdew}\ \emph {et~al.}(1996)\citenamefont {Perdew}, \citenamefont {Burke},\ and\ \citenamefont {Ernzerhof}}]{perdew1996generalized}%
  \BibitemOpen
  \bibfield  {author} {\bibinfo {author} {\bibfnamefont {J.~P.}\ \bibnamefont {Perdew}}, \bibinfo {author} {\bibfnamefont {K.}~\bibnamefont {Burke}},\ and\ \bibinfo {author} {\bibfnamefont {M.}~\bibnamefont {Ernzerhof}},\ }\bibfield  {title} {\bibinfo {title} {Generalized gradient approximation made simple},\ }\href@noop {} {\bibfield  {journal} {\bibinfo  {journal} {Physical review letters}\ }\textbf {\bibinfo {volume} {77}},\ \bibinfo {pages} {3865} (\bibinfo {year} {1996})}\BibitemShut {NoStop}%
\bibitem [{\citenamefont {Giannozzi}\ \emph {et~al.}(2017)\citenamefont {Giannozzi}, \citenamefont {Andreussi}, \citenamefont {Brumme}, \citenamefont {Bunau}, \citenamefont {Nardelli}, \citenamefont {Calandra}, \citenamefont {Car}, \citenamefont {Cavazzoni}, \citenamefont {Ceresoli}, \citenamefont {Cococcioni} \emph {et~al.}}]{giannozzi2017advanced}%
  \BibitemOpen
  \bibfield  {author} {\bibinfo {author} {\bibfnamefont {P.}~\bibnamefont {Giannozzi}}, \bibinfo {author} {\bibfnamefont {O.}~\bibnamefont {Andreussi}}, \bibinfo {author} {\bibfnamefont {T.}~\bibnamefont {Brumme}}, \bibinfo {author} {\bibfnamefont {O.}~\bibnamefont {Bunau}}, \bibinfo {author} {\bibfnamefont {M.~B.}\ \bibnamefont {Nardelli}}, \bibinfo {author} {\bibfnamefont {M.}~\bibnamefont {Calandra}}, \bibinfo {author} {\bibfnamefont {R.}~\bibnamefont {Car}}, \bibinfo {author} {\bibfnamefont {C.}~\bibnamefont {Cavazzoni}}, \bibinfo {author} {\bibfnamefont {D.}~\bibnamefont {Ceresoli}}, \bibinfo {author} {\bibfnamefont {M.}~\bibnamefont {Cococcioni}}, \emph {et~al.},\ }\bibfield  {title} {\bibinfo {title} {Advanced capabilities for materials modelling with quantum espresso},\ }\href@noop {} {\bibfield  {journal} {\bibinfo  {journal} {Journal of physics: Condensed matter}\ }\textbf {\bibinfo {volume} {29}},\ \bibinfo {pages} {465901} (\bibinfo {year} {2017})}\BibitemShut {NoStop}%
\bibitem [{\citenamefont {Zhang}\ \emph {et~al.}(2017{\natexlab{b}})\citenamefont {Zhang}, \citenamefont {Yan}, \citenamefont {Zhang}, \citenamefont {Huang}, \citenamefont {Arita}, \citenamefont {Sun}, \citenamefont {Duan}, \citenamefont {Wu},\ and\ \citenamefont {Zhou}}]{zhang2017experimental}%
  \BibitemOpen
  \bibfield  {author} {\bibinfo {author} {\bibfnamefont {K.}~\bibnamefont {Zhang}}, \bibinfo {author} {\bibfnamefont {M.}~\bibnamefont {Yan}}, \bibinfo {author} {\bibfnamefont {H.}~\bibnamefont {Zhang}}, \bibinfo {author} {\bibfnamefont {H.}~\bibnamefont {Huang}}, \bibinfo {author} {\bibfnamefont {M.}~\bibnamefont {Arita}}, \bibinfo {author} {\bibfnamefont {Z.}~\bibnamefont {Sun}}, \bibinfo {author} {\bibfnamefont {W.}~\bibnamefont {Duan}}, \bibinfo {author} {\bibfnamefont {Y.}~\bibnamefont {Wu}},\ and\ \bibinfo {author} {\bibfnamefont {S.}~\bibnamefont {Zhou}},\ }\bibfield  {title} {\bibinfo {title} {Experimental evidence for type-ii dirac semimetal in ptse$_2$},\ }\href@noop {} {\bibfield  {journal} {\bibinfo  {journal} {Physical Review B}\ }\textbf {\bibinfo {volume} {96}},\ \bibinfo {pages} {125102} (\bibinfo {year} {2017}{\natexlab{b}})}\BibitemShut {NoStop}%
\bibitem [{\citenamefont {Absor}\ \emph {et~al.}(2020)\citenamefont {Absor}, \citenamefont {Santoso}, \citenamefont {Yamaguchi},\ and\ \citenamefont {Ishii}}]{absor2020spin}%
  \BibitemOpen
  \bibfield  {author} {\bibinfo {author} {\bibfnamefont {M.~A.~U.}\ \bibnamefont {Absor}}, \bibinfo {author} {\bibfnamefont {I.}~\bibnamefont {Santoso}}, \bibinfo {author} {\bibfnamefont {N.}~\bibnamefont {Yamaguchi}},\ and\ \bibinfo {author} {\bibfnamefont {F.}~\bibnamefont {Ishii}},\ }\bibfield  {title} {\bibinfo {title} {Spin splitting with persistent spin textures induced by the line defect in the 1 t phase of monolayer transition metal dichalcogenides},\ }\href@noop {} {\bibfield  {journal} {\bibinfo  {journal} {Physical Review B}\ }\textbf {\bibinfo {volume} {101}},\ \bibinfo {pages} {155410} (\bibinfo {year} {2020})}\BibitemShut {NoStop}%
\bibitem [{\citenamefont {Lei}\ \emph {et~al.}(2017)\citenamefont {Lei}, \citenamefont {Liu}, \citenamefont {Huang},\ and\ \citenamefont {Zhou}}]{lei2017comparative}%
  \BibitemOpen
  \bibfield  {author} {\bibinfo {author} {\bibfnamefont {J.-Q.}\ \bibnamefont {Lei}}, \bibinfo {author} {\bibfnamefont {K.}~\bibnamefont {Liu}}, \bibinfo {author} {\bibfnamefont {S.}~\bibnamefont {Huang}},\ and\ \bibinfo {author} {\bibfnamefont {X.-L.}\ \bibnamefont {Zhou}},\ }\bibfield  {title} {\bibinfo {title} {The comparative study on bulk-ptse 2 and 2d 1-layer-ptse 2 under high pressure via first-principle calculations},\ }\href@noop {} {\bibfield  {journal} {\bibinfo  {journal} {Theoretical Chemistry Accounts}\ }\textbf {\bibinfo {volume} {136}},\ \bibinfo {pages} {1} (\bibinfo {year} {2017})}\BibitemShut {NoStop}%
\bibitem [{\citenamefont {Villaos}\ \emph {et~al.}(2019)\citenamefont {Villaos}, \citenamefont {Crisostomo}, \citenamefont {Huang}, \citenamefont {Huang}, \citenamefont {Padama}, \citenamefont {Albao}, \citenamefont {Lin},\ and\ \citenamefont {Chuang}}]{villaos2019thickness}%
  \BibitemOpen
  \bibfield  {author} {\bibinfo {author} {\bibfnamefont {R.~A.~B.}\ \bibnamefont {Villaos}}, \bibinfo {author} {\bibfnamefont {C.~P.}\ \bibnamefont {Crisostomo}}, \bibinfo {author} {\bibfnamefont {Z.-Q.}\ \bibnamefont {Huang}}, \bibinfo {author} {\bibfnamefont {S.-M.}\ \bibnamefont {Huang}}, \bibinfo {author} {\bibfnamefont {A.~A.~B.}\ \bibnamefont {Padama}}, \bibinfo {author} {\bibfnamefont {M.~A.}\ \bibnamefont {Albao}}, \bibinfo {author} {\bibfnamefont {H.}~\bibnamefont {Lin}},\ and\ \bibinfo {author} {\bibfnamefont {F.-C.}\ \bibnamefont {Chuang}},\ }\bibfield  {title} {\bibinfo {title} {Thickness dependent electronic properties of pt dichalcogenides},\ }\href@noop {} {\bibfield  {journal} {\bibinfo  {journal} {npj 2D Materials and Applications}\ }\textbf {\bibinfo {volume} {3}},\ \bibinfo {pages} {2} (\bibinfo {year} {2019})}\BibitemShut {NoStop}%
\bibitem [{\citenamefont {Gronvold}\ \emph {et~al.}(1960)\citenamefont {Gronvold}, \citenamefont {Haraldsen},\ and\ \citenamefont {Kjekshus}}]{gronvold1960sulfides}%
  \BibitemOpen
  \bibfield  {author} {\bibinfo {author} {\bibfnamefont {F.}~\bibnamefont {Gronvold}}, \bibinfo {author} {\bibfnamefont {H.}~\bibnamefont {Haraldsen}},\ and\ \bibinfo {author} {\bibfnamefont {A.}~\bibnamefont {Kjekshus}},\ }\bibfield  {title} {\bibinfo {title} {On the sulfides, selenides and tellurides of platinum},\ }\href@noop {} {\bibfield  {journal} {\bibinfo  {journal} {Acta Chem. Scand}\ }\textbf {\bibinfo {volume} {14}},\ \bibinfo {pages} {1879} (\bibinfo {year} {1960})}\BibitemShut {NoStop}%
\bibitem [{\citenamefont {Bucko}\ \emph {et~al.}(2010)\citenamefont {Bucko}, \citenamefont {Hafner}, \citenamefont {Leb{\`e}gue},\ and\ \citenamefont {Angy{\'a}n}}]{bucko2010improved}%
  \BibitemOpen
  \bibfield  {author} {\bibinfo {author} {\bibfnamefont {T.}~\bibnamefont {Bucko}}, \bibinfo {author} {\bibfnamefont {J.}~\bibnamefont {Hafner}}, \bibinfo {author} {\bibfnamefont {S.}~\bibnamefont {Leb{\`e}gue}},\ and\ \bibinfo {author} {\bibfnamefont {J.~G.}\ \bibnamefont {Angy{\'a}n}},\ }\bibfield  {title} {\bibinfo {title} {Improved description of the structure of molecular and layered crystals: ab initio dft calculations with van der waals corrections},\ }\href@noop {} {\bibfield  {journal} {\bibinfo  {journal} {The Journal of Physical Chemistry A}\ }\textbf {\bibinfo {volume} {114}},\ \bibinfo {pages} {11814} (\bibinfo {year} {2010})}\BibitemShut {NoStop}%
\bibitem [{\citenamefont {Fang}\ \emph {et~al.}(2019)\citenamefont {Fang}, \citenamefont {Liang}, \citenamefont {Feng},\ and\ \citenamefont {Luo}}]{fang2019structural}%
  \BibitemOpen
  \bibfield  {author} {\bibinfo {author} {\bibfnamefont {L.}~\bibnamefont {Fang}}, \bibinfo {author} {\bibfnamefont {W.}~\bibnamefont {Liang}}, \bibinfo {author} {\bibfnamefont {Q.}~\bibnamefont {Feng}},\ and\ \bibinfo {author} {\bibfnamefont {S.-N.}\ \bibnamefont {Luo}},\ }\bibfield  {title} {\bibinfo {title} {Structural engineering of bilayer ptse$_2$ thin films: a first-principles study},\ }\href@noop {} {\bibfield  {journal} {\bibinfo  {journal} {Journal of Physics: Condensed Matter}\ }\textbf {\bibinfo {volume} {31}},\ \bibinfo {pages} {455001} (\bibinfo {year} {2019})}\BibitemShut {NoStop}%
\bibitem [{\citenamefont {Shawkat}\ \emph {et~al.}(2020)\citenamefont {Shawkat}, \citenamefont {Gil}, \citenamefont {Han}, \citenamefont {Ko}, \citenamefont {Wang}, \citenamefont {Dev}, \citenamefont {Kwon}, \citenamefont {Lee}, \citenamefont {Oh}, \citenamefont {Chung} \emph {et~al.}}]{shawkat2020thickness}%
  \BibitemOpen
  \bibfield  {author} {\bibinfo {author} {\bibfnamefont {M.~S.}\ \bibnamefont {Shawkat}}, \bibinfo {author} {\bibfnamefont {J.}~\bibnamefont {Gil}}, \bibinfo {author} {\bibfnamefont {S.~S.}\ \bibnamefont {Han}}, \bibinfo {author} {\bibfnamefont {T.-J.}\ \bibnamefont {Ko}}, \bibinfo {author} {\bibfnamefont {M.}~\bibnamefont {Wang}}, \bibinfo {author} {\bibfnamefont {D.}~\bibnamefont {Dev}}, \bibinfo {author} {\bibfnamefont {J.}~\bibnamefont {Kwon}}, \bibinfo {author} {\bibfnamefont {G.-H.}\ \bibnamefont {Lee}}, \bibinfo {author} {\bibfnamefont {K.~H.}\ \bibnamefont {Oh}}, \bibinfo {author} {\bibfnamefont {H.-S.}\ \bibnamefont {Chung}}, \emph {et~al.},\ }\bibfield  {title} {\bibinfo {title} {Thickness-independent semiconducting-to-metallic conversion in wafer-scale two-dimensional ptse$_2$ layers by plasma-driven chalcogen defect engineering},\ }\href@noop {} {\bibfield  {journal} {\bibinfo  {journal} {ACS applied materials \& interfaces}\ }\textbf {\bibinfo {volume} {12}},\ \bibinfo {pages} {14341} (\bibinfo
  {year} {2020})}\BibitemShut {NoStop}%
\bibitem [{\citenamefont {Liu}\ \emph {et~al.}(2018)\citenamefont {Liu}, \citenamefont {Lazzaroni},\ and\ \citenamefont {Di~Valentin}}]{liu2018nature}%
  \BibitemOpen
  \bibfield  {author} {\bibinfo {author} {\bibfnamefont {H.}~\bibnamefont {Liu}}, \bibinfo {author} {\bibfnamefont {P.}~\bibnamefont {Lazzaroni}},\ and\ \bibinfo {author} {\bibfnamefont {C.}~\bibnamefont {Di~Valentin}},\ }\bibfield  {title} {\bibinfo {title} {Nature of excitons in bidimensional wse$_2$ by hybrid density functional theory calculations},\ }\href@noop {} {\bibfield  {journal} {\bibinfo  {journal} {Nanomaterials}\ }\textbf {\bibinfo {volume} {8}},\ \bibinfo {pages} {481} (\bibinfo {year} {2018})}\BibitemShut {NoStop}%
\bibitem [{\citenamefont {Chang}\ \emph {et~al.}(2014)\citenamefont {Chang}, \citenamefont {Register},\ and\ \citenamefont {Banerjee}}]{chang2014ballistic}%
  \BibitemOpen
  \bibfield  {author} {\bibinfo {author} {\bibfnamefont {J.}~\bibnamefont {Chang}}, \bibinfo {author} {\bibfnamefont {L.~F.}\ \bibnamefont {Register}},\ and\ \bibinfo {author} {\bibfnamefont {S.~K.}\ \bibnamefont {Banerjee}},\ }\bibfield  {title} {\bibinfo {title} {Ballistic performance comparison of monolayer transition metal dichalcogenide mx$_2$ (m= mo, w; x= s, se, te) metal-oxide-semiconductor field effect transistors},\ }\href@noop {} {\bibfield  {journal} {\bibinfo  {journal} {Journal of Applied Physics}\ }\textbf {\bibinfo {volume} {115}} (\bibinfo {year} {2014})}\BibitemShut {NoStop}%
\bibitem [{\citenamefont {Allah}(2015)}]{allah2015regroupement}%
  \BibitemOpen
  \bibfield  {author} {\bibinfo {author} {\bibfnamefont {A.~F.}\ \bibnamefont {Allah}},\ }\emph {\bibinfo {title} {Regroupement de techniques de caract{\'e}risation de mat{\'e}riaux destin{\'e}s {\`a} l’{\'e}nergie solaire pour optimisation et mesures industrielles}},\ \href@noop {} {Ph.D. thesis},\ \bibinfo  {school} {Universit{\'e} Paris Sud-Paris XI} (\bibinfo {year} {2015})\BibitemShut {NoStop}%
\bibitem [{\citenamefont {Yoffe}(1993)}]{yoffe1993low}%
  \BibitemOpen
  \bibfield  {author} {\bibinfo {author} {\bibfnamefont {A.~D.}\ \bibnamefont {Yoffe}},\ }\bibfield  {title} {\bibinfo {title} {Low-dimensional systems: quantum size effects and electronic properties of semiconductor microcrystallites (zero-dimensional systems) and some quasi-two-dimensional systems},\ }\href@noop {} {\bibfield  {journal} {\bibinfo  {journal} {Advances in Physics}\ }\textbf {\bibinfo {volume} {42}},\ \bibinfo {pages} {173} (\bibinfo {year} {1993})}\BibitemShut {NoStop}%
\bibitem [{\citenamefont {Zhuang}\ and\ \citenamefont {Hennig}(2013)}]{zhuang2013computational}%
  \BibitemOpen
  \bibfield  {author} {\bibinfo {author} {\bibfnamefont {H.~L.}\ \bibnamefont {Zhuang}}\ and\ \bibinfo {author} {\bibfnamefont {R.~G.}\ \bibnamefont {Hennig}},\ }\bibfield  {title} {\bibinfo {title} {Computational search for single-layer transition-metal dichalcogenide photocatalysts},\ }\href@noop {} {\bibfield  {journal} {\bibinfo  {journal} {The Journal of Physical Chemistry C}\ }\textbf {\bibinfo {volume} {117}},\ \bibinfo {pages} {20440} (\bibinfo {year} {2013})}\BibitemShut {NoStop}%
\bibitem [{\citenamefont {Kou}\ \emph {et~al.}(2013)\citenamefont {Kou}, \citenamefont {Yan}, \citenamefont {Hu}, \citenamefont {Wu}, \citenamefont {Wehling}, \citenamefont {Felser}, \citenamefont {Chen},\ and\ \citenamefont {Frauenheim}}]{kou2013graphene}%
  \BibitemOpen
  \bibfield  {author} {\bibinfo {author} {\bibfnamefont {L.}~\bibnamefont {Kou}}, \bibinfo {author} {\bibfnamefont {B.}~\bibnamefont {Yan}}, \bibinfo {author} {\bibfnamefont {F.}~\bibnamefont {Hu}}, \bibinfo {author} {\bibfnamefont {S.-C.}\ \bibnamefont {Wu}}, \bibinfo {author} {\bibfnamefont {T.~O.}\ \bibnamefont {Wehling}}, \bibinfo {author} {\bibfnamefont {C.}~\bibnamefont {Felser}}, \bibinfo {author} {\bibfnamefont {C.}~\bibnamefont {Chen}},\ and\ \bibinfo {author} {\bibfnamefont {T.}~\bibnamefont {Frauenheim}},\ }\bibfield  {title} {\bibinfo {title} {Graphene-based topological insulator with an intrinsic bulk band gap above room temperature},\ }\href@noop {} {\bibfield  {journal} {\bibinfo  {journal} {Nano letters}\ }\textbf {\bibinfo {volume} {13}},\ \bibinfo {pages} {6251} (\bibinfo {year} {2013})}\BibitemShut {NoStop}%
\bibitem [{\citenamefont {Kou}\ \emph {et~al.}(2014)\citenamefont {Kou}, \citenamefont {Wu}, \citenamefont {Felser}, \citenamefont {Frauenheim}, \citenamefont {Chen},\ and\ \citenamefont {Yan}}]{kou2014robust}%
  \BibitemOpen
  \bibfield  {author} {\bibinfo {author} {\bibfnamefont {L.}~\bibnamefont {Kou}}, \bibinfo {author} {\bibfnamefont {S.-C.}\ \bibnamefont {Wu}}, \bibinfo {author} {\bibfnamefont {C.}~\bibnamefont {Felser}}, \bibinfo {author} {\bibfnamefont {T.}~\bibnamefont {Frauenheim}}, \bibinfo {author} {\bibfnamefont {C.}~\bibnamefont {Chen}},\ and\ \bibinfo {author} {\bibfnamefont {B.}~\bibnamefont {Yan}},\ }\bibfield  {title} {\bibinfo {title} {Robust 2d topological insulators in van der waals heterostructures},\ }\href@noop {} {\bibfield  {journal} {\bibinfo  {journal} {ACS nano}\ }\textbf {\bibinfo {volume} {8}},\ \bibinfo {pages} {10448} (\bibinfo {year} {2014})}\BibitemShut {NoStop}%
\bibitem [{\citenamefont {Hamada}\ and\ \citenamefont {Otani}(2010)}]{hamada2010comparative}%
  \BibitemOpen
  \bibfield  {author} {\bibinfo {author} {\bibfnamefont {I.}~\bibnamefont {Hamada}}\ and\ \bibinfo {author} {\bibfnamefont {M.}~\bibnamefont {Otani}},\ }\bibfield  {title} {\bibinfo {title} {Comparative van der waals density-functional study of graphene on metal surfaces},\ }\href@noop {} {\bibfield  {journal} {\bibinfo  {journal} {Physical Review B}\ }\textbf {\bibinfo {volume} {82}},\ \bibinfo {pages} {153412} (\bibinfo {year} {2010})}\BibitemShut {NoStop}%
\bibitem [{\citenamefont {Berland}\ \emph {et~al.}(2011)\citenamefont {Berland}, \citenamefont {Borck},\ and\ \citenamefont {Hyldgaard}}]{berland2011van}%
  \BibitemOpen
  \bibfield  {author} {\bibinfo {author} {\bibfnamefont {K.}~\bibnamefont {Berland}}, \bibinfo {author} {\bibfnamefont {{\O}.}~\bibnamefont {Borck}},\ and\ \bibinfo {author} {\bibfnamefont {P.}~\bibnamefont {Hyldgaard}},\ }\bibfield  {title} {\bibinfo {title} {van der waals density functional calculations of binding in molecular crystals},\ }\href@noop {} {\bibfield  {journal} {\bibinfo  {journal} {Computer Physics Communications}\ }\textbf {\bibinfo {volume} {182}},\ \bibinfo {pages} {1800} (\bibinfo {year} {2011})}\BibitemShut {NoStop}%
\bibitem [{\citenamefont {Antunes}\ \emph {et~al.}(2018)\citenamefont {Antunes}, \citenamefont {Vaiss}, \citenamefont {Tavares}, \citenamefont {Capaz},\ and\ \citenamefont {Leit{\~a}o}}]{antunes2018van}%
  \BibitemOpen
  \bibfield  {author} {\bibinfo {author} {\bibfnamefont {F.~P.~N.}\ \bibnamefont {Antunes}}, \bibinfo {author} {\bibfnamefont {V.~S.}\ \bibnamefont {Vaiss}}, \bibinfo {author} {\bibfnamefont {S.~R.}\ \bibnamefont {Tavares}}, \bibinfo {author} {\bibfnamefont {R.~B.}\ \bibnamefont {Capaz}},\ and\ \bibinfo {author} {\bibfnamefont {A.~A.}\ \bibnamefont {Leit{\~a}o}},\ }\bibfield  {title} {\bibinfo {title} {Van der waals interactions and the properties of graphite and 2h-, 3r-and 1t-mos2: A comparative study},\ }\href@noop {} {\bibfield  {journal} {\bibinfo  {journal} {Computational Materials Science}\ }\textbf {\bibinfo {volume} {152}},\ \bibinfo {pages} {146} (\bibinfo {year} {2018})}\BibitemShut {NoStop}%
\bibitem [{\citenamefont {Abdukayumov}\ \emph {et~al.}(2024)\citenamefont {Abdukayumov}, \citenamefont {Mi{\v{c}}ica}, \citenamefont {Ibrahim}, \citenamefont {Voj{\'a}{\v{c}}ek}, \citenamefont {Vergnaud}, \citenamefont {Marty}, \citenamefont {Veuillen}, \citenamefont {Mallet}, \citenamefont {de~Moraes}, \citenamefont {Dosenovic} \emph {et~al.}}]{abdukayumov2024atomic}%
  \BibitemOpen
  \bibfield  {author} {\bibinfo {author} {\bibfnamefont {K.}~\bibnamefont {Abdukayumov}}, \bibinfo {author} {\bibfnamefont {M.}~\bibnamefont {Mi{\v{c}}ica}}, \bibinfo {author} {\bibfnamefont {F.}~\bibnamefont {Ibrahim}}, \bibinfo {author} {\bibfnamefont {L.}~\bibnamefont {Voj{\'a}{\v{c}}ek}}, \bibinfo {author} {\bibfnamefont {C.}~\bibnamefont {Vergnaud}}, \bibinfo {author} {\bibfnamefont {A.}~\bibnamefont {Marty}}, \bibinfo {author} {\bibfnamefont {J.-Y.}\ \bibnamefont {Veuillen}}, \bibinfo {author} {\bibfnamefont {P.}~\bibnamefont {Mallet}}, \bibinfo {author} {\bibfnamefont {I.~G.}\ \bibnamefont {de~Moraes}}, \bibinfo {author} {\bibfnamefont {D.}~\bibnamefont {Dosenovic}}, \emph {et~al.},\ }\bibfield  {title} {\bibinfo {title} {Atomic-layer controlled transition from inverse rashba--edelstein effect to inverse spin hall effect in 2d ptse2 probed by thz spintronic emission},\ }\href@noop {} {\bibfield  {journal} {\bibinfo  {journal} {Advanced Materials}\ }\textbf {\bibinfo {volume} {36}},\ \bibinfo {pages}
  {2304243} (\bibinfo {year} {2024})}\BibitemShut {NoStop}%
\end{thebibliography}%

\end{document}